\documentclass[12pt]{article}

\pdfoutput=1

\usepackage[utf8]{inputenc}

\usepackage[top=80pt,bottom=85pt,left=85pt,right=85pt]{geometry}
\usepackage{amssymb}
\usepackage{amsmath}
\usepackage{nccmath}
\usepackage{graphicx,xcolor,subfigure}

\makeatletter
\renewcommand{\p@subfigure}{}
\renewcommand{\@thesubfigure}{(\alph{subfigure})\hskip\subfiglabelskip}
\makeatother
\usepackage{setspace}
\usepackage{sectsty}
\usepackage[debug,pageanchor=false]{hyperref}
\definecolor{link}{rgb}{.8,.15,.1}
\hypersetup{colorlinks=true,linkcolor=link,citecolor=link,urlcolor=link,linktocpage}
\usepackage[nosort]{cite}
\usepackage{tikz}

\newcommand{\rr}{\mathbb{R}}
\newcommand{\cc}{\mathbb{C}}

\newcommand{\zz}{\mathbb{Z}}

\numberwithin{equation}{section}
\numberwithin{figure}{section}
\numberwithin{table}{section}

\def\CP{\mathbb{CP}}
\def\bF{\mathbb{F}}
\def\bC{\mathbb{C}}
\def\bD{\mathbb{D}}
\def\bZ{\mathbb{Z}}

\def\Spin{\mathrm{Spin}}
\def\SU{\mathrm{SU}}

\def\tr{\mathop{\mathrm{tr}}\nolimits}
\def\Om#1{\mathrm{O}#1{}_-}
\def\Op#1{\mathrm{O}#1{}_+}

\def\divisor#1{#1}

\def\inc#1{\vcenter{\hbox{\includegraphics[scale=.3]{#1}}}}
\def\incF#1{\vcenter{\hbox{\includegraphics[scale=.4]{#1}}}}

\let\hat\widehat

\newcommand{\cN}{\mathcal{N}}
\newcommand{\bea}{\begin{align}}
\newcommand{\eea}{\end{align}}
\newcommand{\be}{\begin{equation}}
\newcommand{\ee}{\end{equation}}
\newcommand{\bit}{\begin{itemize}}
\newcommand{\eit}{\end{itemize}}
\newcommand{\ben}{\begin{enumerate}}
\newcommand{\een}{\end{enumerate}}

\newcommand{\fg}{\mathfrak{g}}

\newcommand{\su}{\mathfrak{su}}
\newcommand{\so}{\mathfrak{so}}
\renewcommand{\sp}{\mathfrak{sp}}
\newcommand{\fe}{\mathfrak{e}}

\newcommand{\half}{\frac{1}{2}}

\def\rep#1{\mathsf{#1}}

\newcommand{\fk}{\mathfrak{k}}

\begin{document}

	\begin{titlepage}

	\begin{flushright}
	IPMU-18-0001 \\ UCSB Math 2018-16
	\end{flushright}
	\begin{center}

	\vskip .5in %.3in 
	\noindent

	{\Large \bf{The frozen phase of F-theory}}

	\bigskip\medskip

	Lakshya Bhardwaj,$^1$ David R. Morrison,$^2$ Yuji Tachikawa,$^3$ Alessandro Tomasiello$^4$\\

	\bigskip\medskip
	{\small
	$^1$ Perimeter Institute for Theoretical Physics, Waterloo, Ontario, Canada N2L 2Y5\\ 
	$^2$ Departments of Mathematics and Physics, University of California,\\ 
	Santa Barbara, Santa Barbara, CA 93106, USA
	\\
	$^3$ IPMU, University of Tokyo, Kashiwa, Chiba 277-8583, Japan
	\\	
%	\vspace{.3cm}%{.1cm}
	$^4$ Dipartimento di Fisica, Universit\`a di Milano--Bicocca, \\ Piazza della Scienza 3, I-20126 Milano, Italy, \\ and  INFN, sezione di Milano--Bicocca

	}

	\vskip .5cm %.3cm
	{\small \tt lbhardwaj@pitp.ca,  drm@math.ucsb.edu, \\
	yuji.tachikawa@ipmu.jp, alessandro.tomasiello@unimib.it}
	\vskip .9cm %.6cm
	     	{\bf Abstract }
	\vskip .1in
	\end{center}

	\noindent
	We study the interpretation of O7$_+$-planes in F-theory, mainly in the context of the six-dimensional models.
	In particular, we study how to assign  gauge algebras and matter content to seven-branes and their intersections, and the implication of  anomaly cancellation in our construction, generalizing earlier analyses without any O7$_+$-planes.
	By including O7$_+$-planes we can realize 6d superconformal field theories hitherto unobtainable in F-theory, such as those with hypermultiplets in the symmetric representation of $\mathfrak{su}$.
	We also examine a couple of compact models.
These reproduce some famous perturbative models, and in some cases enhance their gauge symmetries non-perturbatively.

	\noindent

	\bigskip
	
	\begin{center}
	--- \textsl{dedicated to the memory of Joe Polchinski} ---
	\end{center}
	
	\bigskip

	\vfill
	\eject

	\end{titlepage}

	\setcounter{tocdepth}{2}
	\tableofcontents
	\eject
	
\section{Introduction} % (fold)
\label{sec:intro}

F-theory \cite{Vafa:1996xn,Morrison:1996na,Morrison:1996pp} is a geometrical way to describe non-perturbative backgrounds of type IIB string theory, whose transition functions include S-duality in addition to the more usual symmetries.  
Supersymmetric backgrounds of F-theory describe a spacetime which includes the base of an elliptic Calabi--Yau variety, with a variable axio-dilaton field whose value is specified by the elliptic fibration.
The degeneration loci of the fibration, called the irreducible components of the discriminant locus, are interpreted as seven-branes on which various gauge algebras are realized.
Among these, one finds as particular examples the ordinary D7-branes and O7-planes of perturbative IIB theory.

The perturbative definition of O-planes, however, allows for several different variants.\footnote{%
That the Chan-Paton indices can carry $\mathfrak{u}$, $\mathfrak{so}$ and $\mathfrak{sp}$ indices was originally pointed out by Schwarz in \cite{Schwarz:1982md} and that they were the only possibilities was soon showed by Marcus and Sagnotti in \cite{Marcus:1982fr}, both in 1982;
see also Sec.~1.3 of Schwarz's review \cite{Schwarz:1982jn}.
That the choice of $\mathfrak{so}$ and $\mathfrak{sp}$ is reflected in the sign of the RR-charge of the O9-plane was already essentially noticed in the seminal paper by Green and Schwarz on the anomaly cancellation in Type I superstring theory \cite{Green:1984sg} in 1984.
That one can have a consistent $T^2/\bZ_2$ compactification of type IIB theory with three O7$_-$s and one O7$_+$ was originally noted in \cite{Bianchi:1991eu} in 1991.
}
In particular, we have two types of O7-planes called the O7$_-$-plane and the O7$_+$-plane, whose charge in units where a (full) D7-brane has charge 1  equals $-4$ and $+4$, respectively.
As was pointed out in the early days of F-theory, the one reproduced in conventional F-theory is the O7$_-$-plane \cite{Sen:1996vd}.
At a fixed total D7-charge, an object with O7$_+$ allows for fewer deformations than an object with O7$_-$.
For example, an O7$_-$ with 8 D7s on top, with total charge 4, can be deformed in various ways by pulling the D7s away, while a single O7$_+$ with the same charge does not allow for such a possibility.
The F-theory description of the latter should hence involve a divisor which for some reason cannot be deformed. This was analyzed and called a \emph{frozen} singularity in \cite{Witten:1997bs}, where this was also discussed in several dual frames.
This phenomenon was then further investigated in \cite{deBoer:2001px}.

Thus it was known for a long time that F-theory includes O7$_+$-planes but they were basically ignored in the vast existing literature on the compactifications of F-theory.
One motivation for revisiting this issue at present rests in the classifications of six-dimensional superconformal theories (SCFTs).
In a series of works initiated in \cite{Heckman:2013pva}, and in particular in \cite{Heckman:2015bfa}, it was shown that \emph{almost} all known 6d SCFTs at that time and a lot more were realizable using 6d compactifications of F-theory.
(For a recent comprehensive review, see \cite{Heckman:2018jxk}.)
However, if one compares this classification against the known examples constructed using massive IIA brane constructions \cite{Brunner:1997gk,Brunner:1997gf,Hanany:1997gh,Hanany:1999sj} and the purely-field theoretical analyses \cite{Danielsson:1997kt,Bhardwaj:2015xxa}, 
one recognizes that there are indeed cases not realized by conventional F-theory constructions.

A typical feature of these cases is that their massive IIA brane construction involves O8$_+$s.
By a T-duality, this is mapped to a IIB brane construction involving O7$_+$s.
This motivated us to look at F-theory compactifications to six-dimensions in the presence of O7$_+$s.

At this point, it is natural to worry if there could be frozen singularities other than O7$_+$-planes which have not been studied in conventional F-theory.
This question was settled, at least for supersymmetric seven-branes, in a recent  re-analysis of 7-branes in F-theory \cite{Tachikawa:2015wka} which concluded that the O7$_+$ is in fact
the only type of frozen singularity in F-theory.\footnote{%
There are  various other less-studied types of higher-codimension singularities one can incorporate in F-theory, such as the ones used by Garc\'\i a-Etxebarria and Regalado \cite{Garcia-Etxebarria:2016erx} to construct 4d $\mathcal{N}{=}3$ SCFTs.
Frozen versions of singularities also occur in M-theory \cite{Witten:1997bs,deBoer:2001px}, where they play an important role in M5-brane fractionation \cite{deBoer:2001px,Ohmori:2015pua,Mekareeya:2017sqh}.  }
Therefore, the only ingredient missing in conventional F-theory compactifications to six-dimensions is the inclusion of O7$_+$-planes, 
and indeed including them we find F-theory realizations of `missing' 6d SCFTs,
as we will see later in the paper.\footnote{We will find F-theory realizations
for certain examples, but we defer a general treatment of the classification
problem formulated in \cite{Heckman:2013pva,Heckman:2015bfa} to future work.}

Once we are convinced that O7$_+$-planes can be included in the F-theory construction,
we realize that we need to revisit every part of the standard F-theory machinery, such as 
the assignment of the gauge algebras and of the matter content to the components of the discriminant and to their intersections,
and the way the 6d anomalies cancel via the Green--Schwarz-West-Sagnotti effect \cite{Green:1984bx,Sagnotti:1992qw}, derived geometrically for F-theory by Sadov in \cite{Sadov:1996zm}.
This paper is the authors' first attempt to provide such generalizations.

One unexpected consequence of the introduction of O7$_+$-planes is the following.
To appreciate it, let us first recall the situation \emph{without} O7$_+$-planes.
In a conventional F-theory compactification without O7$_+$-plane, once one is given the geometry of the elliptically-fibered Calabi--Yau, 
there is a standard method to assign a unique set of gauge algebras and matter content to the geometry.
In particular, under this standard assignment, each simple factor in the gauge algebra is associated to a single component of the discriminant divisor, and each component has at most one simple factor of gauge algebras associated to it.
This choice corresponds to having zero holonomies of the gauge fields on these divisors themselves.
We have the option of turning on the non-trivial gauge configurations, including the effects often called the T-branes \cite{Cecotti:2010bp}, but we also have the standard option of not turning them on at all.

\emph{With} O7$_+$-planes, however, we will often be forced to have at least some nontrivial gauge configurations on some of the components.
More precisely, we even lose the concept of a unique, standard  assignment of gauge algebras and matter content, since we do not even have a natural origin in the space of the all possible holonomies.
Because of this, we often have multiple simple factors of gauge algebras on a single component of the discriminant locus, and also a single simple factor of gauge algebra shared across multiple components, as we will see later.

Unfortunately, at present, we do not have any algorithmic method to find consistent assignments given an elliptic Calabi--Yau and a specification of where the O7$_+$-planes are;
we do not even have a method to tell if there are any consistent assignments at all.
Therefore we are forced to rely on consistency checks via anomaly cancellation
and dualities to backgrounds that are better understood.

The rest of the paper is organized as follows.
In Sec.~\ref{sec:pert}, we study the properties of O7$_+$-planes in the context of F-theory, using string theory and M-theory dualities.
This will let us figure out how to assign gauge algebras and matter content.
In Sec.~\ref{sec:anom}, we study the anomaly cancellation of F-theory models with O7$_+$-planes.
We will see that the analysis of Sadov \cite{Sadov:1996zm} can be naturally generalized by introducing a divisor which represents where O7$_+$-planes lie.
Then in Sec.~\ref{sec:scfts}, we discuss some 6d SCFTs which can be realized only with O7$_+$-planes in F-theory construction,
and 
in Sec.~\ref{sec:comp}, we study the massless spectrum of a couple of compact six-dimensional models with O7$_+$-planes.

In Appendix~\ref{sec:8d}, we review the 8d compactifications with O7$_+$-planes, which is simpler than the 6d examples discussed in the main text.
Finally in Appendix~\ref{app:branes}, we give an alternative derivation, using intersecting brane models, of the spectrum of some compact models discussed in Sec.~\ref{sec:comp}.

% section intro (end)

\section{Frozen seven-branes and their properties} % (fold)
\label{sec:pert}

In this section, we use perturbative string techniques to obtain some properties of frozen singularities.

We start in section \ref{sub:o} with a lightning review of O-planes.
We then discuss the basics of O7$_+$-planes in F-theory in Sec.~\ref{sub:bas},
and in Sec.~\ref{sub:int} we study the physics at individual intersection points of O7$_+$-planes and other seven-branes.
To prepare ourselves for the analysis of an O7$_+$-plane which intersects with more than one seven-brane,
we then need to have short digressions, on the T-duals of NS5- and D6-branes in Sec.~\ref{sub:ns5} and on the phenomenon of shared gauge algebras in Sec.~\ref{sub:shared}.
We then come back to the case with O7$_+$-planes in Sec.~\ref{sub:o7ns5}.
In the final subsection \ref{sub:smooth}, we see that with O7$_+$-planes a shrunken divisor does not necessarily signify any singularity in the low energy physics.

\subsection{Basics of orientifold planes} % (fold)
\label{sub:o}

Let us start by a quick review of the basics of the orientifolds.\footnote{%
A good review of the basics can also be found in \cite{Dabholkar:1997zd}.
More detailed and rigorous analysis of perturbative orientifolds were done e.g.~in \cite{Distler:2009ri,Gao:2010ava}, but we stick to the traditional, ad hoc approach in this paper.
The name \emph{orientifold} itself was introduced in \cite{Dai:1989ua} by Dai, Leigh and Polchinski.
The concept of the orientifold goes back further in history, see e.g.~\cite{Sagnotti:1987tw,Horava:1989vt} and references therein.
}
\paragraph{Action on the closed strings:}
An orientifold is usually defined as a $\zz_2$ symmetry $\Pi$ that includes world-sheet parity $\Omega$.
It can also include a spacetime involution $\sigma$.
It is often necessary to also include an extra factor $(-)^{F_{\rm L}}$ (where $F_{\rm L}$ is the left-moving spacetime fermion number) so that $\Pi^2$ acts as the identity.
If locally $\sigma$ is the reflection of $9-p$ coordinates, so that the \emph{orientifold plane} O$p$ (the fixed locus of $\sigma$)\footnote{We will also consider actions that include translations and thus have no fixed locus as in \eqref{eq:shift-o}; the conclusions in (\ref{eq:o-action}) below also apply.} has (spatial) dimension $p$, one needs to include $(-)^{F_{\rm L}}$ if $p=2,3$ mod 4.\footnote{%
To check this, one first uses the fact that a reflection $R_I$ of the $I$-th spatial coordinate acts by $\Gamma_I$ on the 10d Majorana spinor, which satisfies $(\Gamma_I)^2=+1$.
Therefore, $R_{I_1\cdots I_p}^2 =1$ or $(-1)^{F_L+F_R}$ depending on whether $p=0,1$ or $2,3$ mod $4$, respectively.
Then one compensates this $(-1)^{F_L+F_R}$ by the fact that $\Omega(-1)^{F_L}\Omega=(-1)^{F_R}$ and therefore $(\Omega(-1)^{F_L})^2=(-1)^{F_L+F_R}$.}
To summarize, locally the orientifold action is 
\begin{equation}
\begin{array}{ccccccccccccccc}
 \mathrm{O}9 &\mathrm{O}8 &\mathrm{O}7 &\mathrm{O}6 &\mathrm{O}5 & \cdots \\
 \hline
\Omega & \Omega R_9 & \Omega R_8 R_9(-1)^{F_L}  & \Omega R_7 R_8 R_9(-1)^{F_L}\vphantom{\Big[}  & \Omega R_6 R_7 R_8 R_9 & \cdots
\end{array},\label{eq:o-action}
\end{equation} 
where $R_p$ denotes a reflection of the $p$-th coordinate.
This specifies the orientifold's action on closed strings.
In this paper, we will be interested in particular in O7s, with O6s and O8s making occasional appearances.

\paragraph{Action on the open strings:}
In presence of open strings, one also needs to decide its action on the Chan--Paton matrix $\lambda$, which appears in a superposition $\sum_{i,j}\lambda_{ij}|ij\rangle$ 	of the states $|ij \rangle$, that in turn can be interpreted as going from the $i$-th to the $j$-th brane in a stack (omitting other quantum numbers).
Since the world-sheet parity $\Omega$ reverses orientation, it acts by transposing $\lambda$, but it may also mix the states with a change of basis $M$: namely, $\lambda \to M \lambda^t M^{-1}$.
Imposing that this action is an involution leads to the condition that
\begin{equation}\label{eq:sign}
	M^{-1}M^t= \mp 1\,.
\end{equation} 
This sign choice leads to two different types of O-plane, which we call O$p_\pm$.\footnote{%
\label{foot:wpm}In \cite{Witten:1997bs} and other older papers,  O$p_\pm$-planes are called planes of type ${\cal O}^\mp$, with the opposite sign.
We stick to the more modern conventions which are now standard.} 

\paragraph{The RR-charge:}
The RR charge can be computed through a one-loop computation, which contains $-\tr M^{-1}M^t$ in its M\"obius strip contribution (see for example the reviews \cite{Dabholkar:1997zd,Angelantonj:2002ct}).
In the end one concludes that the charge is 
$\pm 2^{p-5}$ that of a full D$p$-brane\footnote{%
Naively the fractional charge of the O$p$-plane for $p\le 4$ contradicts the Dirac quantization.
For a resolution, see \cite{Tachikawa:2018njr}.
}: explicitly,
\begin{equation}\label{eq:charge}
\begin{array}{c|cccccccc}
	p& 9&8&7&6&5&4&3& \cdots \\
	\hline
	\pm 2^{p-5} & \pm 16 & \pm 8 & \pm4 & \pm 2 & \pm 1 & \pm \frac12 & \pm \frac14 & \cdots
\end{array}
\end{equation}
Thus, the O$p_-$ has negative charge and the O$p_+$ has positive charge, as the name implies.

\paragraph{The gauge group:}
The gauge group is also influenced by the sign (\ref{eq:sign}).
If a stack of $N$ D$p$-branes is parallel to the O$p$-plane but not on top of it, the action will relate the strings ending on them to strings ending on an image stack in a different locus; the gauge group will be the usual ${\rm U}(N)$.
On the other hand, if the stack is on top of the O$p$-plane, the action will relate the open string states to themselves, projecting out some of them.
To read off the gauge group, we can consider the gauge field states $\lambda_{ij}\alpha_{-1/2}^\mu |0;ij\rangle$.
Since $\Omega\alpha_{-1/2} \Omega= - \alpha_{-1/2}$, the surviving states will be those with Chan--Paton factors $\lambda$ such that $\lambda = -M \lambda^t M^{-1}$.
If the sign in (\ref{eq:sign}) is $-1$, $M$ is antisymmetric; by a change of basis ($\lambda \to C^{-1} \lambda C$, $M \to C M C^t$) it can be chosen to be of the form $J\equiv (\begin{smallmatrix}0 & 1_N \\ -1_N & 0 \end{smallmatrix})$, and thus $\lambda$ will be in the $\mathfrak{sp}_N$ algebra.\footnote{%
We follow the standard convention that $\mathfrak{sp}_1=\mathfrak{su}_2$.
}
If on the other hand the sign in (\ref{eq:sign}) is $+1$, then $M$ can be chosen to be $1_{2N}$, and $\lambda \in \mathfrak{so}_{2N}$.

Summarizing, the choice (\ref{eq:sign}) leads to two different orientifolds: 
\begin{itemize}
\item O$p_-$, with $\mathfrak{so}_{2N}$ gauge algebra and charge $-2^{p-5}$, and 
\item O$p_+$, with $\mathfrak{sp}_{N}$ gauge algebra and charge $+2^{p-5}$.
\end{itemize}

\paragraph{D$q$-branes intersecting O$p$-planes:}
More generally, if we also have a stack of D$q$-branes which intersect our O$p$, there are subtle signs \cite{Gimon:1996rq} coming from the fact that the strings from the O$p$- to the D$q$-branes needed to be  expanded to both integer and half-integer modes.
In flat space (and vanishing $B$ field), the number $\#\text{ND}$ of Neumann--Dirichlet directions (the number of directions transverse to the D$p$ and parallel to the D$q$, or vice versa) has to be a multiple of 4, for unbroken supersymmetry.
The result for the gauge algebra on the D$q$-branes is then as follows:\footnote{\label{foot:red}The fields on the D$q$ stack get mapped to fields on another point of the stack,
unless the D$q$ stack is completely embedded in the O$p$-plane.
A priori this only restricts the behavior as a function of the coordinates of the gauge field, which would then locally remain of $\mathfrak{u}(2m)$ type.
However, in situations where the divisor wrapped by the stack is compact, in most applications we want to keep only the zero-modes of the gauge field under its equation of motion, and this restricts the gauge group as in (\ref{eq:OpDq}).} 
\begin{equation}
\begin{array}{l|cc}
& {\rm O}p_+ & {\rm O}p_- \\
\hline
\# \text{ND}=0,8& \text{symplectic} & \text{orthogonal} \\
\# \text{ND}=4& \text{orthogonal} & \text{symplectic}
\end{array}\quad .\label{eq:OpDq}
\end{equation}

\paragraph{T-duality:}
Let us next discuss the T-duality of orientifolds, since we often need to perform T-duality of the setup on $S^1/\bZ_2$ where two fixed points support O$p$-planes, possibly of different types.
Two most straightforward cases are when both fixed points have $\Om{p}$ or both fixed points have $\Op{p}$.
The T-dual is then simply $\Om{(p+1)}$ or $\Op{(p+1)}$ wrapped around $S^1$.

\begin{figure}[ht]
	\centering
		\includegraphics[width=12cm]{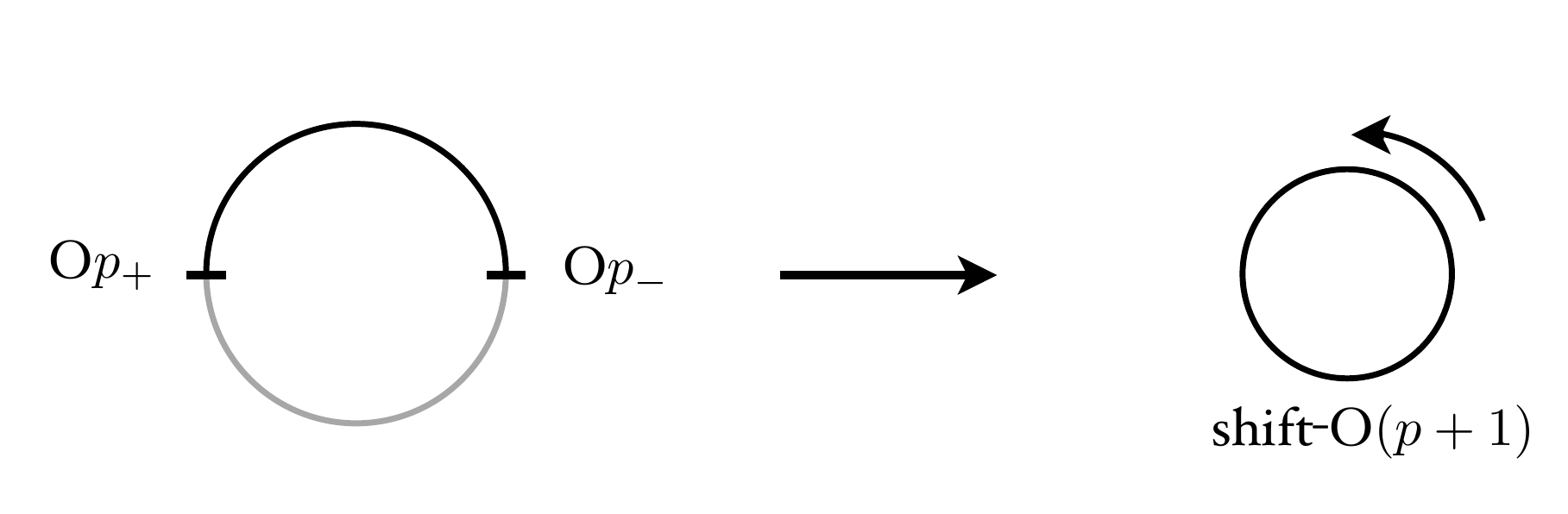}
	\caption{A model with two O$p$-planes with opposite sign is turned by T-duality.}
	\label{fig:t-shift}
\end{figure}

When one fixed point has $\Om{p}$ and the other fixed point has $\Op{p}$,
the T-dual is known to be a \emph{shift-orientifold}, namely an orientifold whose spacetime action $\sigma$ not only flips the coordinates transverse to the orientifold, but also translates a circle by half its radius
\begin{equation}
	\sigma: (x_{p+1},x_{p+2},\cdots,x_9)\sim   \left(x_{p+1}+\frac R2,-x_{p+2},\cdots,-x_9\right)\,.
\label{eq:shift-o}
\end{equation}
See Fig.~\ref{fig:t-shift} for a pictorial representation.
Note that this action fixes no point.

The derivation of this fact can be found e.g.~in \cite[p.~41]{Witten:1997bs} or \cite{Hanany:2001iy}.
A rough argument goes as follows.
We start from the shift-orientifold background \eqref{eq:shift-o},
and T-dualize the $x_{p+1}$ direction.
Its T-dual should be a compactification on $S^1/\bZ_2$.
Therefore this should result in a combination of two O$p$-planes at two fixed points.
The original shift-orientifold background did not have any D$(p+1)$-charge.
Therefore, in the T-dual, we should have zero D$p$-plane charge.
This is only possible if one fixed point is $\Om{p}$ and the other is $\Op{p}$.

Another intuitive argument is as follows.
The shift operator $s: x_{p+1}\to x_{p+1}+R/2$ can be thought of as $e^{i \frac R2\widehat p}$, where $\widehat p$ is the momentum operator.
Its T-dual is $\tilde s= e^{i \frac 1{2R}\widehat w}$, where $\widehat w$ is the ``winding operator'', which measures the length of the string.
$\tilde s$ gives $1$ on strings of total length zero, such as those that begin and end on the same O$p$, but it gives $-1$ on the strings that begin and end on different O$p$'s, signaling the fact that the two have different signs.

\paragraph{Other types of orientifolds:}
It is also known that there are $\tilde{\mathrm{O}}p_\pm$-planes when $p\le 6$,
distinguished from the more ordinary O$p_\pm$-planes by the RR-torsion flux.
As we will not use them heavily, we will not discuss them further.

% subsection o (end)

\subsection{Frozen divisors in F-theory} % (fold)
\label{sub:bas}

Our main interest lies in seven-branes in Type IIB theory and F-theory.
An ordinary O7$_-$ without any D7-branes on top is known to lift to two $I_1$ divisors, due to quantum effects \cite{Sen:1996vd}.
Similarly, with $n<4$ D7-branes on top, the F-theory realization is given by $(n+2)$ $I_1$ divisors.
With at least $4$ D7-branes,  it is interpreted in F-theory as an $I^*_{n-4}$ divisor (where $n$ is the number of D7-branes).
Since string theory also has O7$_+$-planes, it is natural to ask how these are described in F-theory.

First of all, from (\ref{eq:charge}) we see that O7$_\pm$ have charge equal to that of $\pm 4$ full D7-branes.
So an O7$_+$ has the same charge and tension as an O7$_-$ with 8 full D7-branes on top.
In F-theory, they will give rise to the same monodromy \cite{Witten:1997bs,Landsteiner:1997ei}; we expect both to be described by an $I^*_4$ divisor.
However, the O7$_-$ with  $8$ D7 gives rise to an $\mathfrak{so}_{16}$ gauge algebra, while the O7$_+$ gives rise to none.
A related difference is that the O7$_-$ with  $8$ D7 can be deformed by pulling the D7s away (which corresponds in F-theory to a complex structure deformation), while the O7$_+$ cannot.
Thus an O7$_+$ is described by a $I^*_4$ singularity which for some reason cannot be deformed; we will call this a \emph{frozen} singularity, and denote it by $\widehat{I}^*_4$.

More generally, an O7$_-$ with  $n$ D7s has the same charge and tension as an O7$_+$ with $(n-8)$ D7s; both are described by an $I^*_{n-4}$ singularity, but in the latter case the gauge algebra is $\mathfrak{sp}_{n-4}$ rather than $\mathfrak{so}_{2n}$, and the deformations are correspondingly reduced.
In this case too we say that the singularity is frozen, and we denote by $\widehat I^*_{n-4}$.

To be more expicit, an F-theory vacuum is typically described by the 
``Weierstrass coefficients'' $f$ and $g$ which are sections of the 
line bundles $\mathcal{O}_B(-4K_B)$ and $\mathcal{O}_B(-6K_B)$ on
the F-theory base $B$, and which lead to the equation
\begin{equation}
y^2 = x^3 + f x + g
\end{equation}
for the total space of the elliptic fibration.  Along a divisor $D$ with
a $\widehat I^*_{n-4}$ singularity, $f$ vanishes to order $2$, $g$ vanishes
to order $3$, and the equation $4f^3+27g^2$ of the discriminant locus
vanishes to order $(n-8)+10$, for a configuration with $n-8$ D7-branes
on top of an O7$_+$.  Although the ``freezing'' mechanism is not understood,
it must prevent any deformation which lowers the order of vanishing of
either $f$ or $g$ at all, or which lowers the order of vanishing of $4f^3+27g^2$
below $10$.  

Note that the Weierstrass coefficients are accompanied by periods of
type IIB two-forms over appropriate two-cycles in $B$; for compactifications
to 6d, the complex moduli provided by Weierstrass coefficients are
paired with these periods of two-cycles to provide the two complex
scalars in a hypermultiplet.  In particular, by activating a vev represented
by one of these two-form periods we may disturb the gauge group
assigned to a divisor without changing the geometry of the divisor (which
would have required a change of complex modulus).  Such deformations
are often described in the language of T-branes \cite{Cecotti:2010bp}, for which a number
of geometric tools have been developed 
\cite{Anderson:2013rka,Collinucci:2014qfa,Anderson:2017rpr}.

As an exercise in using the rule \eqref{eq:OpDq}, let us consider D3-branes embedded in the worldvolume of O7$_\pm$.
Since $\#\text{ND}=4$, the gauge group on the embedded D3-branes is $\mathfrak{so}$ for O7$_+$ and $\mathfrak{sp}$ for O7$_-$.
In particular, the smallest gauge algebra allowed is $\mathfrak{so}_{1}$ and $\mathfrak{sp}_{1}$, with
one and two Chan-Paton indices, respectively.
A bulk D3-brane has two Chan-Paton indices.
Therefore, a bulk D3-brane can fractionate into two separate objects on O7$_+$ but not on O7$_-$.
These D3-branes can be considered as point-like instantons of the gauge fields on O7$_\pm$,
and therefore the D3-charges of the minimal-charge instanton on O7$_\pm$ differ by a factor of 2.
This fact becomes important in the anomaly analysis in Sec.~\ref{sec:anomcancel}.

% subsection bas (end)

\subsection{Intersections: perturbative analysis} % (fold)
\label{sub:int}

As mentioned in the introduction, O7$_+$s are the only frozen F-theory singularities \cite{Tachikawa:2015wka}.
As our main interest lies in the compactification to 6d, we now want to understand their behavior when they intersect other singularities, namely, how they modify the gauge algebras of neighboring divisors and the matter representations at intersections with them.
We will do so by using perturbative string techniques, and dualities.

Some readers might want to study the simpler situation in 8d  summarized in Appendix~\ref{sec:8d}, before considering the more interesting but complicated examples of 6d compactifications discussed here.

\subsubsection{$\widehat I^*$--$I$ intersection} % (fold)
\label{ssub:i*i}

Let us now start working out what happens when the frozen divisors intersect ordinary divisors.  
We will begin with the intersections of frozen $\widehat I^*_n$ with $I_m$ divisors.

Let us first recall what this intersection gives in the unfrozen case, i.e.~an $I^*$--$I$ intersection.
The intersection with the $I^*$ induces on the $I$ a so-called ``Tate'' monodromy, a nontrivial automorphism of the gauge algebra that reduces it \cite{Bershadsky:1996nh}.\footnote{This is not to be confused with the ``Kodaira'' monodromy, describing how the geometry changes when one goes around a singular divisor.}
This is expressed by saying that the divisor is \emph{non-split}, and denoted by a superscript ${}^{\rm ns}$.
Its effect on the gauge algebra is that it reduces from $\mathfrak{u}_{2m}$ to $\mathfrak{sp}_{m}$.
We summarize this situation by writing
\begin{equation}\label{eq:I*I}
\begin{array}{cc}
	\mathfrak{so}_{2n+8} & \mathfrak{sp}_{m}\\
	I^*_n & I^{\rm ns}_{2m}\,.
\end{array}
\end{equation}
As a warm-up, let us also see how it is reproduced by orientifolds.
Consider an intersection of an O7$_- + (n+4)$ D7 along directions 01256789 with $m$ full D7s along directions 03456789.
From (\ref{eq:OpDq}) we see again that the gauge algebra on the $m$ D7s is reduced to $\mathfrak{sp}_{m}$; see also footnote \ref{foot:red}.
We thus recover (\ref{eq:I*I}).
Notice that the spacetime action of the orientifold projection can be interpreted as the Tate monodromy we mentioned above.

\begin{figure}[ht]
	\centering
		\includegraphics[width=6cm]{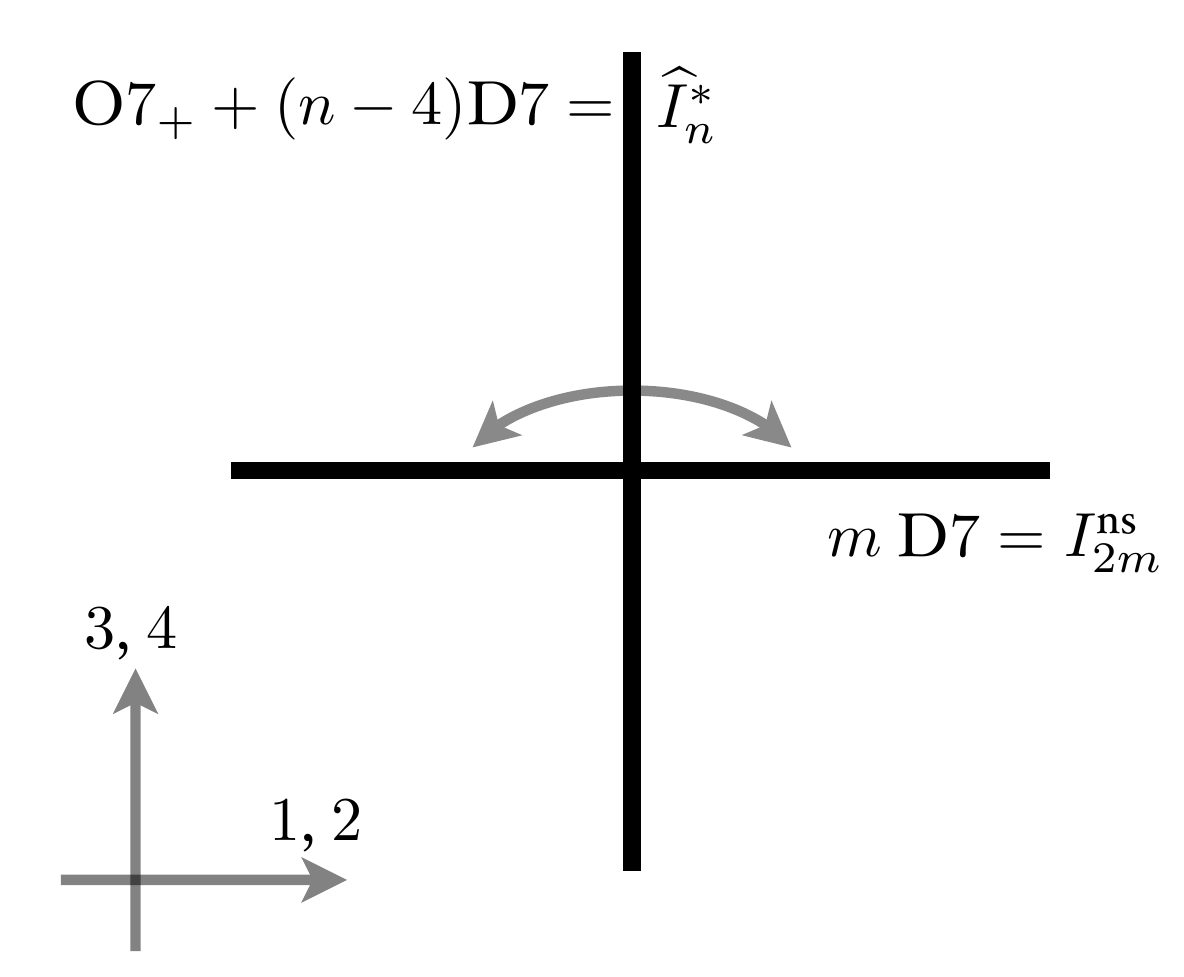}
	\caption{An O7--D7 intersection, interpreted in F-theory as an intersection between an $\widehat I^*_{n+4}$ and an $I^{\rm ns}_{2m}$.}
	\label{fig:O7-D7}
\end{figure}

We can similarly work out what happens if the $I^*$ divisor is replaced by its frozen $\widehat I^*$ counterpart: the configuration now involves an O7$_+ + (n-4)$ D7s, and $2m$ transverse D7s (see Fig.~\ref{fig:O7-D7}, where only directions 6789 are depicted).
Looking again at (\ref{eq:OpDq}), we see that the gauge algebra on the $m$ D7s is reduced this time to $\mathfrak{so}_{2m}$.
We conclude
\begin{equation}\label{eq:hI*I}
\begin{array}{cc}
	\mathfrak{sp}_{n-4} & \mathfrak{so}_{2m}\\
	\widehat I^*_n & I^{\rm ns}_{2m}\,.
\end{array}
\end{equation}
Thus, an $I^{\rm ns}$ divisor intersecting a frozen divisor has an $\mathfrak{so}$ gauge algebra, rather than an $\mathfrak{sp}$ gauge algebra.
In both cases (\ref{eq:I*I}) and (\ref{eq:hI*I}) there is a bifundamental at the intersection, due to the strings from one set of branes to the other.

% subsubsection i*i (end)

\subsubsection{$I^*$--$I^*$, $I^*$--$\widehat I^*$, $\widehat I^*$--$\widehat I^*$ intersections} % (fold)
\label{ssub:i*i*}

We will now consider intersections between two $I^*$ divisors, both frozen and unfrozen.
We will see that using perturbative O7s we will have only partial success in understanding the full possibilities.
This will lead us in Sec.~\ref{sub:o7ns5} to consider T-dual configurations.

\paragraph{$I^*$--$I^*$ intersection:}
Let us again start by recalling what F-theory gives in the ordinary unfrozen case.
The intersection of two $I^*$ divisors actually falls outside Kodaira's classification.
To cure this, one can blow-up the base; this reveals a new divisor that touches both $I^*$'s, and that behaves like in (\ref{eq:I*I}):
\begin{equation}\label{eq:Dtflow}
\begin{array}{ccc}
	\mathfrak{so}_{2k+8} & & \mathfrak{so}_{2\ell+8}\\
	I^*_k & \bullet & I^*_\ell
\end{array}
\qquad \leftarrow \qquad 
\begin{array}{ccc}
	\mathfrak{so}_{2k+8} &  \mathfrak{sp}_{(k+\ell)/2} & \mathfrak{so}_{2\ell+8}\\
	I^*_k & I^{\rm ns}_{k+\ell} & I^*_\ell 
\end{array}\,
\end{equation}
where we assumed $k+\ell$ to be even,
and the $\bullet$ denotes the bad singularity that we blew up.
Physically, it signals a six-dimensional superconformal sector which is sometimes called \emph{$\bD_{k+4}$--$\bD_{\ell+4}$ conformal matter};\footnote{%
\label{foot:Dcm}In fact this superconformal theory depends only on $k+\ell$
and has $\so(2k+2\ell+16)$ flavor symmetry.
Thus we will simply call it $\bD_{k+\ell+8}$ conformal matter in what follows.
We use the blackboard letter $\bD$ since the notation $D_i$  denotes an $i$-th divisor in this paper.
One can also define $\bD_{2n}$ as the 6d superconformal theory which has a one-dimensional tensor branch on which it becomes an $\mathfrak{sp}_{n-4}$ theory with $4n$ fundamentals with at least $\mathfrak{so}_{4n}$ flavor symmetry.
For example, then, the $\bD_8$ theory is the E-string theory.
} the blow-up represents moving along its tensor branch, namely the part of its moduli space where we give a vev to the scalar in the tensor multiplet.

Let us now try to engineer an $I^*$--$I^*$ intersection using O7s.
The most natural generalization of Fig.~\ref{fig:O7-D7} consists of two O7s that intersect transversally.
This can be achieved by an orientifold projection in flat space that has more than one generator of the type we recalled in (\ref{eq:o-action}).
For an intersection of two O7s, locally one takes the two generators
\begin{equation}\label{eq:2O7}
	\Omega R_6 R_7 (-1)^{F_L} \, ,\qquad \Omega R_8 R_9(-1)^{F_L}\,.
\end{equation}
We can see that in this situation there is an O7 on the locus $x^6=x^7=0$, and another on the locus $x^8=x^9=0$.
(Notice that one is then also quotienting by their product $R_6 R_7 R_8 R_9$, so that at the intersection between the O7s there is in fact also a $\zz_2$ orbifold singularity.) 
Choosing the $\pm$ type of these two orientifold planes affects their charge and their action on Chan--Paton indices in the way we reviewed earlier; we will see shortly what their combined effect amounts to.

Another ingredient is that the projection on the closed $\zz_2$-twisted sector is reversed if two orientifolds of different type intersect \cite{Polchinski:1996ry}.
This comes about by considering the exchange of closed strings between two crosscaps, one from one O7 and another from another O7.
The sign of this diagram is reversed when two orientifolds are of different type, and the modular transformation of this diagram determines the orientifolding projection on the closed string $\zz_2$ twisted sector.
In the end, one finds that an O7$_-$--O$7_+$ intersection has a six-dimensional tensor multiplet, while O7$_-$--O$7_-$ or O7$_+$--O$7_+$ intersection has a hypermultiplet:
\begin{equation}\label{eq:Oint}
\begin{array}{c|cc}
&{\rm O}7_- & {\rm O}7_+\\
\hline
{\rm O}7_-& \text{hyper} & \text{tensor} \\
{\rm O}7_+ & \text{tensor} & \text{hyper}
\end{array} \quad . 
\end{equation}

As we mentioned, if D-branes are present, they will now feel the effect of both projections.
Consider for example choosing both planes to be O7$_-$, with $k+4$ and $\ell+4$ D7s present on the $x^6=x^7=0$ and $x^8=x^9=0$ loci respectively.
The first set of D7s, say, would be projected to $\mathfrak{so}_{2k+8}$ by the O7$_-$ parallel to it; but, recalling (\ref{eq:OpDq}), it would also be projected to $\mathfrak{sp}_{k+4}$ by the O7$_-$ transverse to it.
This means that it actually gets projected to the intersection of the two, $\mathfrak{u}_{k+4}$.
In the language of F-theory branes, this gives
\begin{equation}\label{eq:I*I*}
	\begin{array}{ccc}
		\mathfrak{u}_{k+4} & & \mathfrak{u}_{\ell+4}\\
		I^*_k  & \cdot & I^*_\ell
	\end{array}\,,
\end{equation}
where the $\cdot$ now represents the hypermultiplet found in (\ref{eq:Oint}).\footnote{%
A warning is in order. The orientifold projection leaves the gauge algebra $\mathfrak{u}$ on $I^*$,
but the $\mathfrak{u}_1$ part usually gets Higgsed and becomes massive by the Green-Schwarz mechanism, each $\mathfrak{u}_1$ eating a neutral hypermultiplet.
This point was carefully analyzed in \cite[Sec.~2]{Berkooz:1996iz}.
In our case, the diagonal $\mathfrak{u}_1$ of $\mathfrak{u}_{k+4}$ and $\mathfrak{u}_{\ell+4}$
will be gone.
In a compact model, we usually expect every $\mathfrak{u}_1$ part to be eliminated in this manner,
agreeing with the usual expectation that only the $\mathfrak{su}$ algebras are realized on the 7-branes,
not the $\mathfrak{u}$ algebras.
}
This hypermultiplet is neutral under $\mathfrak{u}_{k+4}\oplus\mathfrak{u}_{\ell+4}$. The presence of this neutral hypermultiplet signals that the configuration (\ref{eq:I*I*}) is obtained by moving along a particular direction in the \emph{Higgs} branch of $\bD_{k+4}$--$\bD_{\ell+4}$ conformal matter whose tensor branch was depicted in (\ref{eq:Dtflow}). This particular direction in the Higgs branch is parametrized by vevs of the neutral hypermultiplet in (\ref{eq:I*I*}). Another well-known direction in the Higgs branch, distinct from the one represented by (\ref{eq:I*I*}), is provided by brane recombination, where the two $I^*$ divisors merge.

\paragraph{$\widehat I^*$--$\widehat I^*$ intersection:}
For an O7$_+$--O7$_+$ projection, for the same reason we get 
\begin{equation}\label{eq:hI*hI*}
	\begin{array}{ccc}
		\mathfrak{u}_{k-4} & & \mathfrak{u}_{\ell-4}\\
		\widehat I^*_k  & \cdot & \widehat I^*_\ell
	\end{array}\,.
\end{equation}
In analogy with our discussion below (\ref{eq:I*I*}), it is natural to think that this is the Higgsing of a ``frozen conformal matter'' 
\begin{equation}\label{eq:hI*cm}
	\begin{array}{ccc}
		\mathfrak{sp}_{k-4} & & \mathfrak{sp}_{\ell-4}\\
		\widehat I^*_k  & \bullet & \widehat I^*_\ell
	\end{array}\,,
\end{equation}
and that upon blowing up (moving along the tensor branch) an $I^{\rm ns}_{k+\ell}$ with $\so(k+\ell)$ gauge algebra would be created, which would behave as in (\ref{eq:hI*I}).
We will see later that this expectation is borne out.  

\paragraph{$\widehat I^*$--$ I^*$ intersection:}
For an O7$_+$--O7$_-$ intersection, on each set of D7s the two projections will be of the same type.
For example, on the D7s on the O7$_-$, we have $\lambda= -M_1 \lambda^t M_1^{-1} =-M_2 \lambda M_2^{-1}$, with both $M_i$ symmetric.
We can make $M_1=1$ as in section \ref{sub:o}; with the residual freedom in change of basis we can diagonalize $M_2$, but a priori it could have any number of positive and negative eigenvalues.
If we also impose that the D7s can move off the O7$_-$, we obtain that $M_2=\left(\begin{smallmatrix}
1_{\ell+4} & 0 \\ 0 & -1_{\ell+4}\end{smallmatrix}\right)$, and the gauge symmetry is 
$\mathfrak{so}_{\ell+4}\oplus \mathfrak{so}_{\ell+4}$.
Similar considerations apply to the O$7_+ + (k-4)$D7s; hence we get
\begin{equation}\label{eq:I*hI*}
	\begin{array}{ccc}
		\mathfrak{sp}_{k/2-2}\oplus \mathfrak{sp}_{k/2-2} & & 
\mathfrak{so}_{\ell+4} \oplus \mathfrak{so}_{\ell+4}\\
		\widehat I^*_k  & \circ & I^*_\ell
	\end{array}\,
\end{equation}
where we assumed $k$ to be even.
Notice that in this case there is no neutral hypermultiplet at the origin, according to (\ref{eq:Oint}); we have included the symbol $\circ$ to mark this.
So in this case we do not expect this configuration to be a Higgsing of a conformal one. This might look surprising, but it will become clearer in section \ref{sub:smooth} below, where we will see an alternative realization of the same setup (in the case $k=\ell$ is even). 

% subsubsection i*i* (end)

% subsection int (end)

\subsection{NS5- and D6-branes} % (fold)
\label{sub:ns5}

To go beyond the results in section \ref{ssub:i*i*}, we will need to consider configurations which are dual to IIA in presence of NS5-branes.
To set the stage, in this subsection we will discuss a situation without orientifolds.

We consider IIA on $\rr^9\times S^1$; let us say the $S^1$ corresponds to direction 4, and has periodicity $R$.
Let us have a single NS5 whose worldvolume is in directions 056789, localized at $x^\alpha = x^4=0$, $\alpha=1,2,3$.
T-dualizing it along direction 4 turns it into an Euclidean Taub--NUT geometry.
The space transverse to the NS5 is $\rr^3\times S^1$; T-duality turns the $H$ flux of the NS5 into a Chern class that signals the $S^1$ is now Hopf-fibred over the $S^2$s at $x^\alpha x^\alpha=r^2$.
The inverse images of these $S^2$s are thus copies of $S^3$.
These shrink smoothly at $x^\alpha=0$, so that locally around this point the  fibration is $S^1\hookrightarrow \rr^3 \to \rr^4$.
One way to realize this fibration in coordinates is
\begin{align}
	\mathbb{H}\cong \cc^2 &\to \rr^3 \\
	q={\binom{z}{w}} &\mapsto x^\alpha= q^\dagger \sigma^\alpha q  
\end{align}
where $\sigma^\alpha$ are the Pauli matrices.
So
\begin{equation}\label{eq:xzw}
	x^1+ i x^2  = z w \ ,\qquad x^3 = |z|^2 - |w|^2 \ .
\end{equation}

\begin{figure}[ht]
	\centering
		\includegraphics[width=7cm]{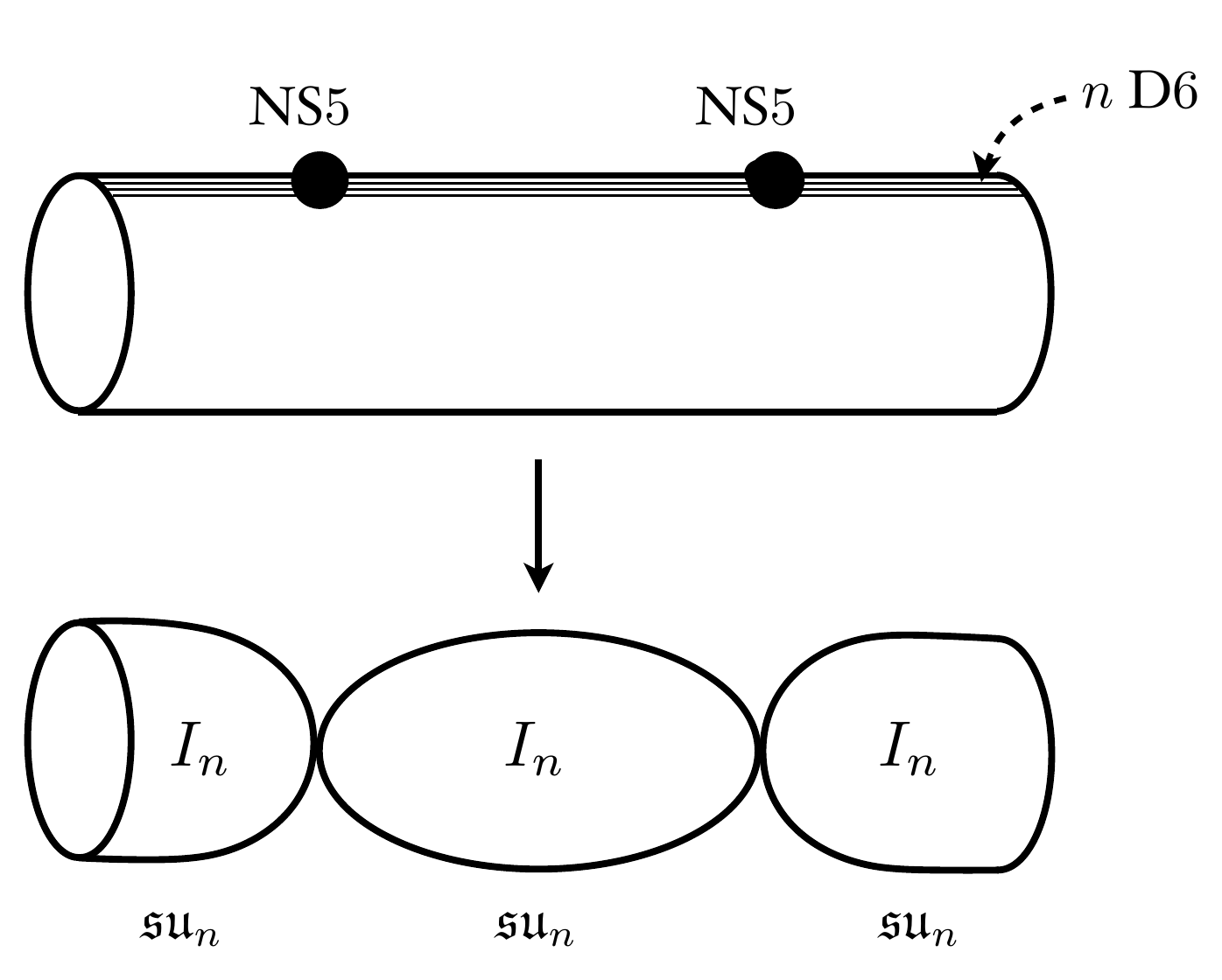}
	\caption{\small NS5-branes, D6-branes, and T-duality. The compact and noncompact directions of the cylinder are called respectively directions 4 and 3 in the text.}
	\label{fig:NS5-D6}
\end{figure}

If we have several NS5s localized at several positions in the 3 direction ($x^3=x^3_i$, $x^1=x^2=x^4=0$), T-duality turns the geometry into a multi-Taub--NUT geometry where the $S^1$ shrinks at the $x^3=x^3_i$.
The inverse image under the $S^1$ fibration of a path between two of these points is an $S^2$.
We represent this in Fig.~\ref{fig:NS5-D6}.

Let us now suppose some D6s are also stretched along the 0356789 directions.
First let us imagine there are $n$ D6s stretched along the entire 3 axis, 
i.e.~when $n$ D6s are placed at $x^1=x^2=x^4=0$.
Under T-duality along direction 4, they will turn into $n$ D7s.
More precisely, as Fig.~\ref{fig:NS5-D6} suggests, they will turn into a sequence of D7s wrapping the various $S^2$ on the Taub--NUT with multiplicity $n$.
What the picture does not show is that these $S^2$s are holomorphic cycles.
Locally around an NS5 at $x^\alpha=x^4=0$, for example, the locus $x^\alpha =0$ is turned into $x^1= x^2= 0$.
From (\ref{eq:xzw}) we see this to be  $zw=0$, which is the union of the curve $z=0$ and of $w=0$.
In F-theory terms, this is a chain of intersecting $I_n$ curves.

In the presence of a Romans mass, parameterized conventionally by an integer $2\pi F_0\equiv n_0\neq 0$,
the number of D6s ending on an NS5 from the left minus the number of D6s from the right is $n_0$.
Focusing on an NS5 on which a D6 ends from the right and does not continue to the left, 
we see again from (\ref{eq:xzw}) that T-duality turns it into the single curve $z=0$.
This would be one of the $S^2$s in Fig.~\ref{fig:NS5-D6}.
We then have a chain of intersecting curves supporting $I_{n}$, $I_{n+n_0}$, $I_{n+2n_0}$, \ldots.

Another possible generalization is to move the D6s in the $x^4$ direction, so that there is now a stack of $n_j$ D6s at $x^4=x^4_j$.
On the IIB side, this corresponds to Wilson lines for the gauge field on the D7s.

% subsection ns5 (end)

\subsection{Shared gauge algebras} % (fold)
\label{sub:shared}

From the setup of Fig.~\ref{fig:NS5-D6}, we can also wonder what happens if we move only some of the D6s away from the NS5s in direction 4; say from an initial stack of $n$ D6s we move $m$ to the position $x^4=x^4_0$.
These D6s recombine: they no longer end on the NS5s.
In field theory, this corresponds to a partial Higgsing 
\begin{equation}\label{eq:shared}
	\mathfrak{su}_{n}\oplus \mathfrak{su}_{n}\ \to\  \mathfrak{su}_{n-m}\oplus \mathfrak{su}_m\oplus \mathfrak{su}_{n-m}\ 
\end{equation} 
where the $\mathfrak{su}_m$  at the middle is the diagonal subalgebra of two copies of $\mathfrak{su}_m\subset \mathfrak{su}_n$.

Since the displacement has happened along the 4 direction, it is not immediately apparent on the IIB side: the T-dual still consists of two stacks of $m+n$ D7-branes meeting at a point, as in section \ref{sub:ns5}.
The only consequence of the displacement is the presence of a Wilson line: there is a worldvolume gauge field with non-zero holonomy, $a=\frac{x^4_0}{l_s^2}{\rm diag}(0,\ldots,0,1,\ldots,1)d\tilde x^4$.
Since direction $\tilde 4$ shrinks at the intersection point, on both D7s there is a worldvolume $da=f$ field strength proportional to a $\delta$-function supported on the intersection point.

\begin{figure}[ht]
	\centering
		\includegraphics[width=6cm]{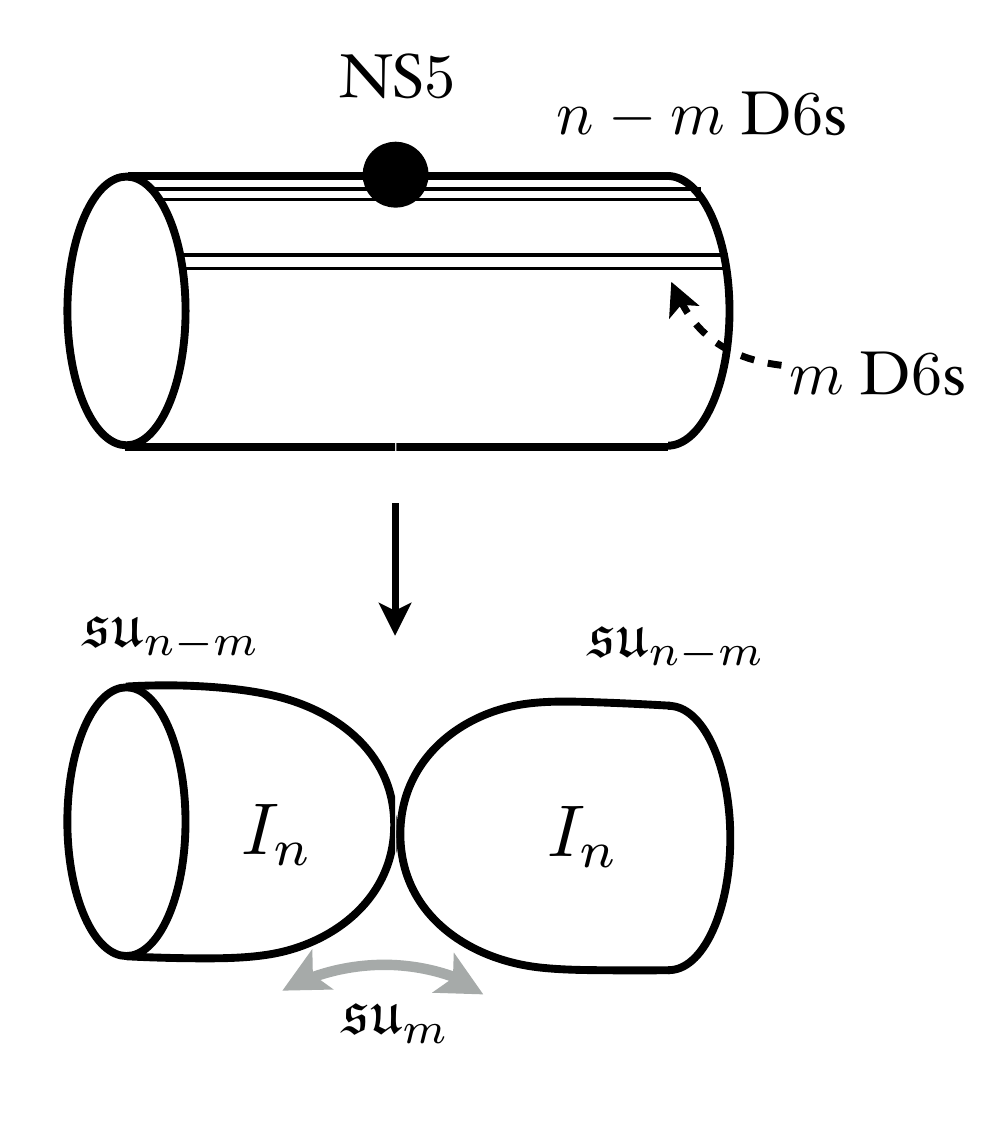}
	\caption{\small On the IIA side, we can move $m$ of the D6s off the NS5s and make them recombine.
On the IIB side, this corresponds to a gauge algebra $\mathfrak{su}_m$ that is shared between two curves meeting at a point.
We denote this with a double-sided arrow.}
	\label{fig:shared}
\end{figure}

By comparing with the IIA picture, we conclude that a Wilson line can partially break the gauge algebra on two intersecting D7s, as in (\ref{eq:shared}): part of the gauge algebra can \emph{recombine}.
The $\mathfrak{su}_m$ algebra is now shared between the two intersecting divisors; this is summarized in Fig.~\ref{fig:shared}.
In what follows, we fill find other examples of such shared gauge algebras.

If we move all the D6s off the NS5 (i.e.~if $n=m$), only the shared gauge algebra is present.
In this case, one might be puzzled by the fact that on the IIB side the Wilson line is now proportional to the identity.
This would not seem to cause a Higgsing, while from the IIA picture it is clear that it does, since the D6s are away from the NS5.

To clarify this point, we need to identify the T-dual of the NS5 position in IIB.
Since the NS5 position in IIA is shifted by a diffeomorphism in the $x^4$ position, its T-dual should be shifted in IIB by a gauge transformation for the NS-NS two-form field, namely $B\to B + d \Lambda$, for $\Lambda$ a one-form.
In fact this one-form was identified in \cite[Sec.~2.2]{Witten:2009xu} explicitly.
More generally we conclude that, in the intersection between two curves ${\cal C}_1$, ${\cal C}_2$, there is a shared gauge algebra if on either curve there is an eigenvalue $a_i$ of the Wilson line $\alpha$ on the curves that does \emph{not} match with the pullback of $\Lambda$ at large distance from the intersection:\footnote{To see more clearly what (\ref{eq:aiLambda}) gives, our $\Lambda$ in (\ref{eq:aiLambda}) is equal to a number $\tilde x$ (the dual of the NS5 displacement) times the $\Lambda$ in \cite[(2.3)]{Witten:2009xu}. Going at large distance from the intersection, the pullback $\Lambda$ will just look like $\tilde x d \theta$, and it makes sense to compare it with the $a_i$.} 
\begin{equation}\label{eq:aiLambda}
	a_i \neq \Lambda|_{{\cal C}_1}\ 
	\text{or}\
	a_i \neq \Lambda|_{{\cal C}_2}\ .
\end{equation}

In F-theory language, we could consider a deformation of the Weierstrass
coefficients which ``recombined'' two branes, i.e., smoothed the two
divisors out into a single divisor.  If instead of this deformation,
the corresponding periods of two-forms are activated, the gauge theory
will recombine without any change in the geometry.

% subsection shared (end)

\subsection{Intersections: via T-duality} % (fold)
\label{sub:o7ns5}
Having made a detour in the last two subsections, we now reintroduce O-planes in our story.

First we need to review the behavior of NS5s in presence of orientifolds.
Like any other brane, any NS5 must come with a mirror image under the orientifold action.
Each copy is usually called a \emph{half-brane} to emphasize that it can become \emph{full} if the two copies are brought to the O-plane.
It turns out \cite{Evans:1997hk} that when this is done the two half-NS5s can be separated again: this time along the O-plane worldvolume, while staying on it.
When this happens, the orientifold type changes between the two half-NS5s.

The situation relevant for our purposes consists in having an O6 defined by a reflection inverting directions 124, and for example two half-NS5s at two values of $x^3$.
(Thus the O6-plane and the half-NS5s are stretched along the same directions as the D6 and NS5 in the previous subsection.) If the O6 is taken to be an O6$_-$ outside the two half-NS5s, its type will change to O6$_+$ inside.
This leads to a sequence of gauge algebras
\begin{equation}\label{eq:so-sp-so}
	\mathfrak{so}_{2n+8}\,, \ \mathfrak{sp}_n\,, \ \mathfrak{so}_{2n+8}\ .
\end{equation}

\begin{figure}[ht]
	\centering
		\subfigure[\label{fig:O7--}]{\includegraphics[width=7cm]{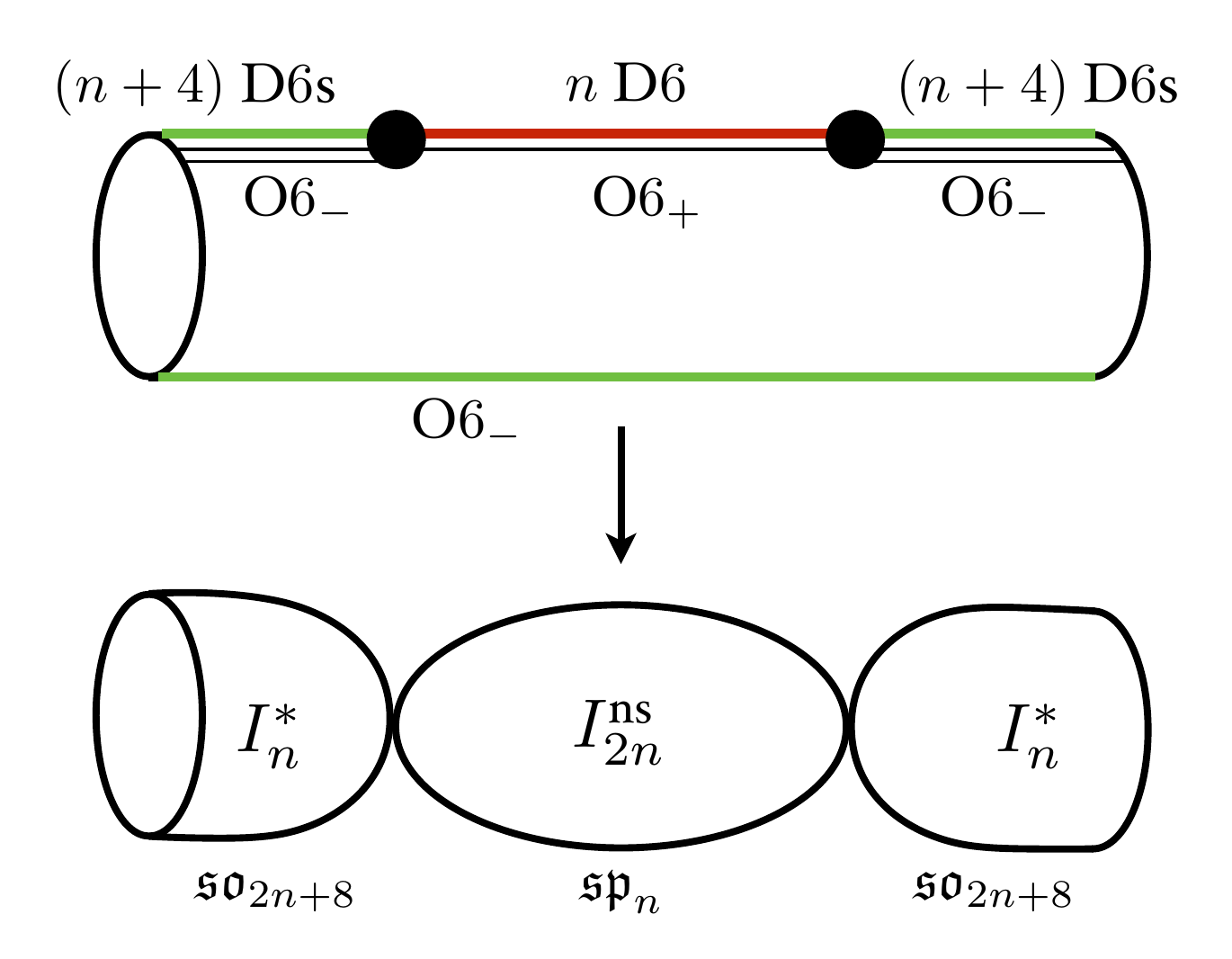}}\hspace{1cm}
		\subfigure[\label{fig:O7-+}]{\includegraphics[width=7cm]{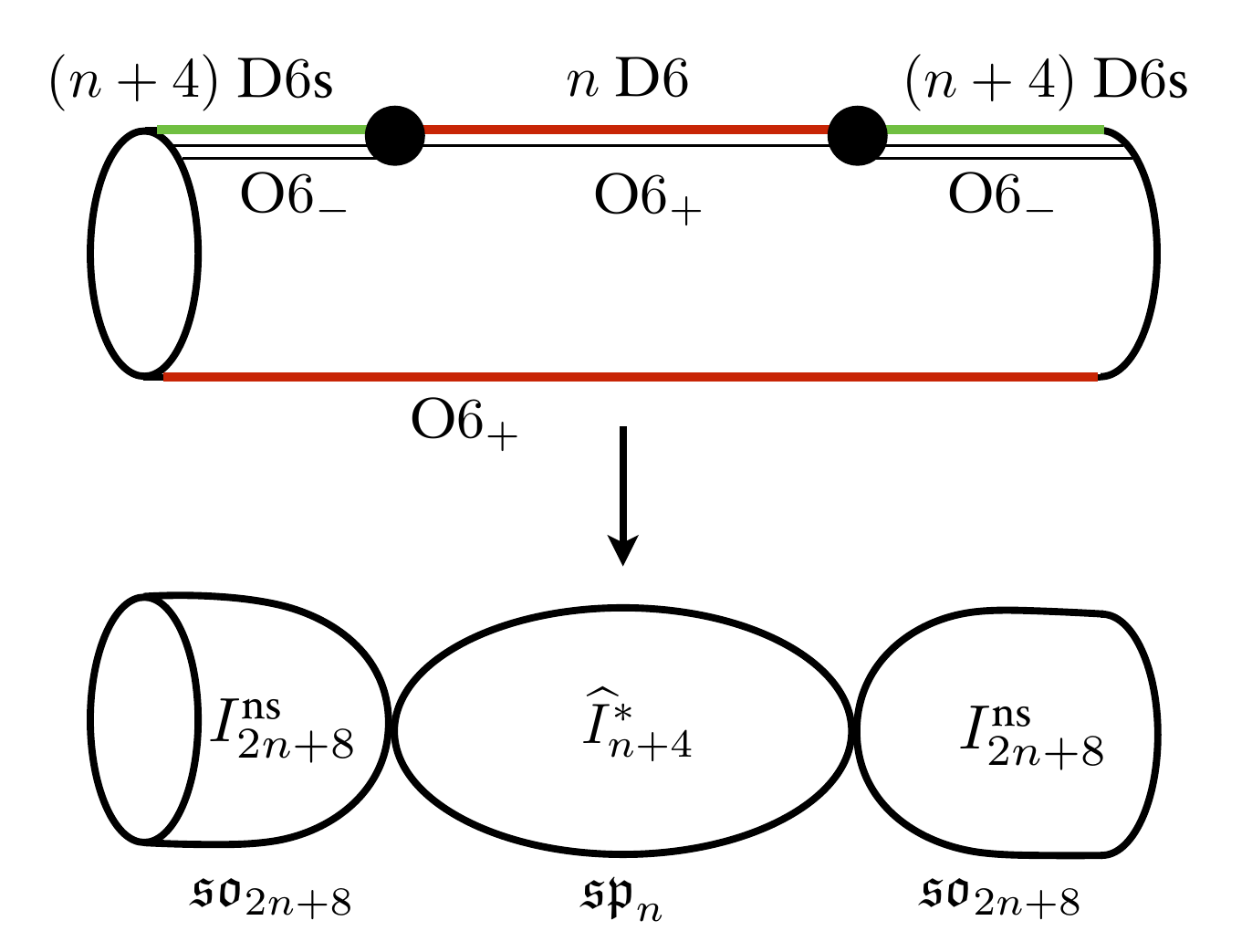}}
	\caption{\small Two configurations with O7$_\pm$-planes, and their T-duals. The dots now represent half-NS5s}
	\label{fig:TO7}
\end{figure}

Actually, since direction 4 is compact, a reflection involving 124 will have a fixed point both at $x^4=0$ and at $x^4=R/2$, the opposite locus on the circle.
The O6-plane on that locus can be of both O6$_-$ and O6$_+$ type.
We show both those cases in Fig.~\ref{fig:TO7}.
In both cases the gauge algebras are still as in (\ref{eq:so-sp-so}), since the difference with the case of Fig.~\ref{fig:O7--} happens in a region where no D6s are present.

Upon T-duality, we again find a chain of curves.
To see what type of curves we have, we need to use the rules reviewed in section \ref{sub:o}; see in particular Fig.~\ref{fig:t-shift}.
We learn from there that an orientifold with O6$_\pm$-planes on both sides of a circle gets T-dualized to an orientifold with an O7$_\pm$-plane, while a circle which has an O6$_+$ on one side and an O6$_-$ on the other gets T-dualized to a shift-orientifold.
This is another realization of Tate monodromy, which we discussed at the beginning of section \ref{ssub:i*i}.

Thus, in the case of Fig.~\ref{fig:O7--}, after T-duality we end up with a curve $I^{\rm ns}_{2n}$ between two ordinary $I^*_n$ curves.
This is familiar from (\ref{eq:Dtflow}) with $m=n$, and is in agreement with the sequence of gauge algebras (\ref{eq:so-sp-so}) we found in IIA.

In the case of Fig.~\ref{fig:O7-+}, we have a frozen $\widehat I^{*}_{n+4}$ curve touching two $I^{\rm ns}_{2n+8}$ ones.
The presence of the frozen singularity alters the usual F-theory rules: from the IIA picture, we see that as expected an $\widehat I^{*}_{n+4}$ curve supports an $\mathfrak{sp}_n$ gauge algebra; moreover, we also see that an $I^{\rm ns}_{2m}$ touching a frozen curve supports an $\mathfrak{so}_{2m}$.
This can be generalized to 
\begin{equation}
	\begin{array}{ccc}
		\mathfrak{sp}_{k-4} &  \mathfrak{so}_{k+\ell} & \mathfrak{sp}_{\ell-4}\\
		\widehat I^*_k & I^{\rm ns}_{k+\ell} & \widehat I^*_\ell 
	\end{array}\,
\end{equation}
(with $k=n+4$).
This is the theory on the tensor branch of (\ref{eq:hI*cm}), thus realizing the expectation discussed there.

\begin{figure}[ht]
	\centering
\subfigure[\label{fig:O6higgs-nong}]{\includegraphics[width=5cm]{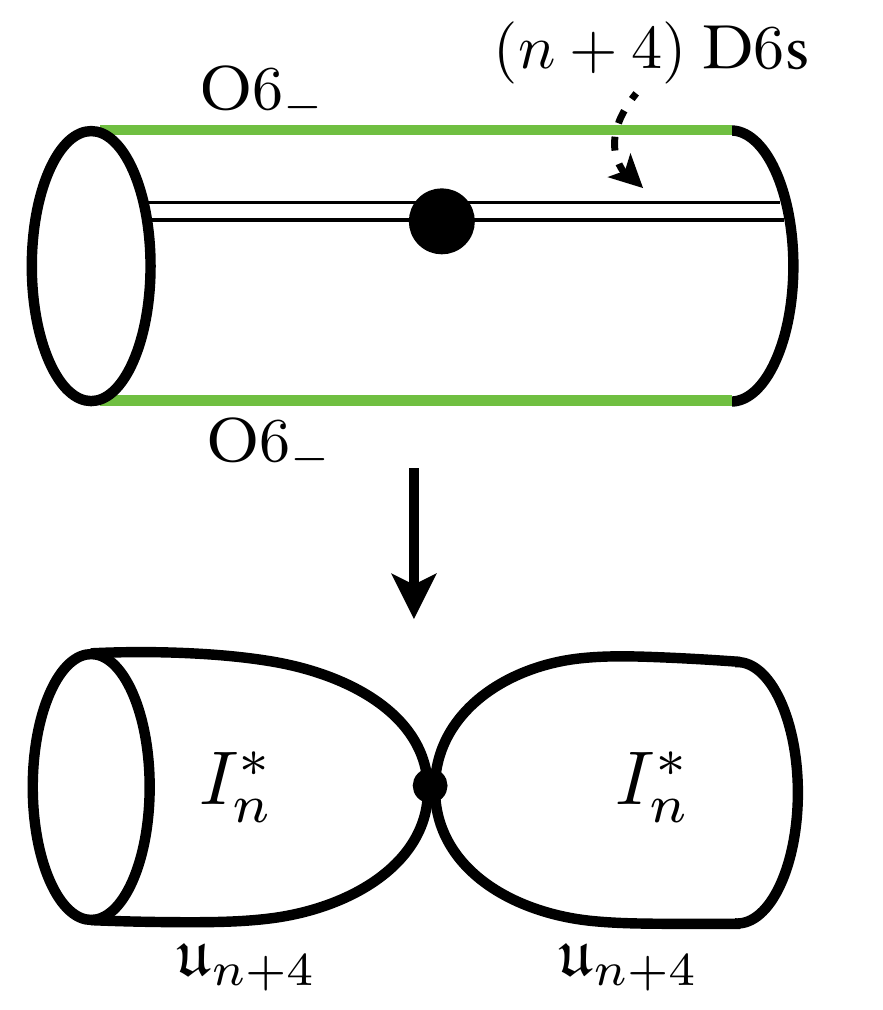}}	
\subfigure[\label{fig:O6higgs-rec}]{\includegraphics[width=5cm]{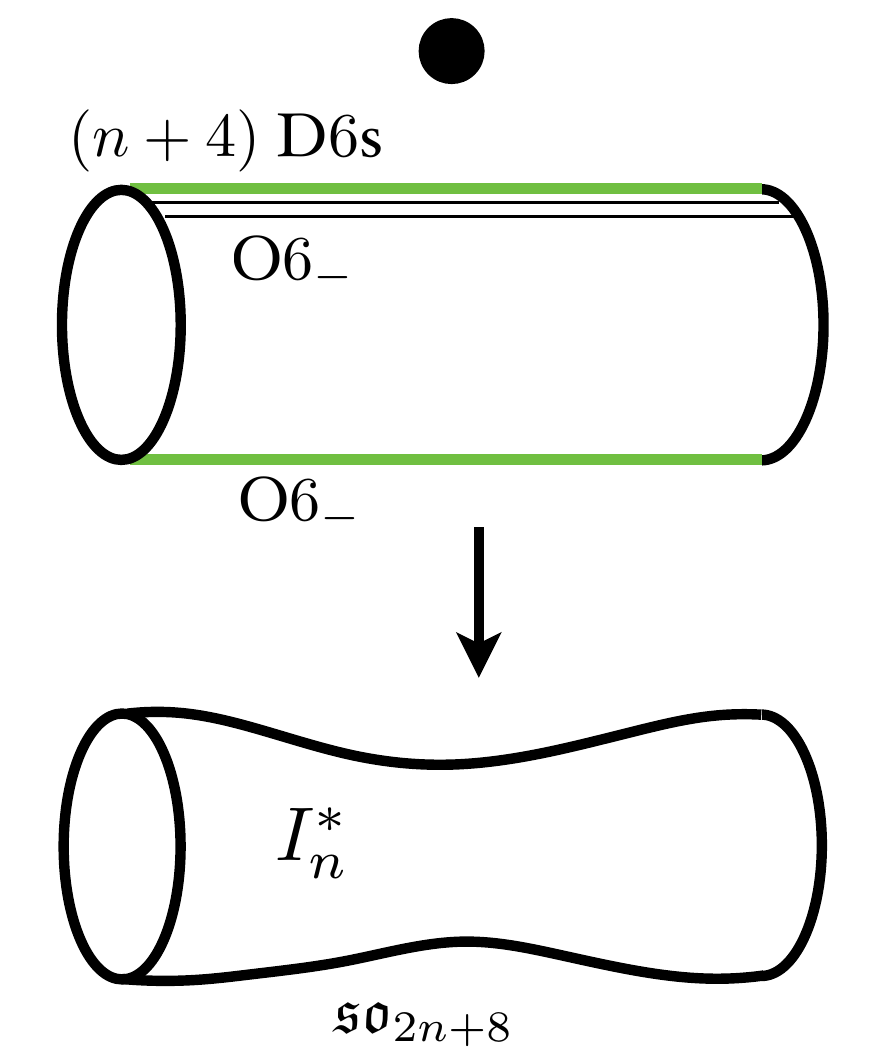}}	
	\caption{\small Two different ways of Higgsing $\bD_{m+4}$--$\bD_{n+4}$ conformal matter. \subref{fig:O6higgs-nong} reproduces (\ref{eq:I*I*}); \subref{fig:O6higgs-rec} corresponds to brane recombination.}
	\label{fig:O6higgs}
\end{figure}

If we put the half-NS5s back on top of each other, we recover a full NS5. We can now split it again by moving the two halves along the periodic 4 direction, together with some of the D6s, or by moving them in another direction, so that the degeneration induced by T-dualizing the NS5s no longer happens on the O6--D6 system. These two new configurations represent respectively the Higgsing in (\ref{eq:I*I*}), and the one mentioned below it involving brane recombination. These two possibilities are depicted in Fig.~\ref{fig:O6higgs}.

The setup of this section can also be decorated by adding $m$ D6-branes at the bottom orientifold plane; this would add a gauge algebra $\mathfrak{so}_{2m}$ to Fig.~\ref{fig:O7--}, and $\mathfrak{sp}_{m}$ to Fig.~\ref{fig:O7-+}.
On the F-theory side, this would correspond to the presence of a Wilson line, and to a  gauge algebra that is shared among the three curves, in the language of section \ref{sub:shared}.
Again, this can be realized through the T-brane-like phenomena of activating
the two-form-period partner of a geometric deformation.

% subsection o7ns5 (end)

\subsection{Smooth transitions} % (fold)
\label{sub:smooth}

In the chains of curves considered so far, shrinking one or more of the curves leads to some strongly coupled physics.
This is clear from the IIA picture, where it corresponds to making two or more NS5-branes coincide.
In an effective field theory description, this often manifests itself in a gauge coupling becoming infinite.
The positions of the NS5s parameterize the tensor branch of a six-dimensional effective theory; these situations correspond to non-generic loci of the tensor branch.

For example, in the situations depicted in figures \ref{fig:NS5-D6} and \ref{fig:TO7}, there is a one-dimensional tensor branch, parameterized by a 6d tensor multiplet whose scalar $\phi$ corresponds to the distance between the two NS5s, and which in the 6d theory also plays the role of the inverse square of the gauge coupling.
At the origin $\phi=0$, the gauge coupling diverges.
At this strong coupling point it is expected that a CFT arises, describing two coincident NS5s on top of a D6 stack.

\begin{figure}[ht]
	\centering
		\includegraphics[width=14cm]{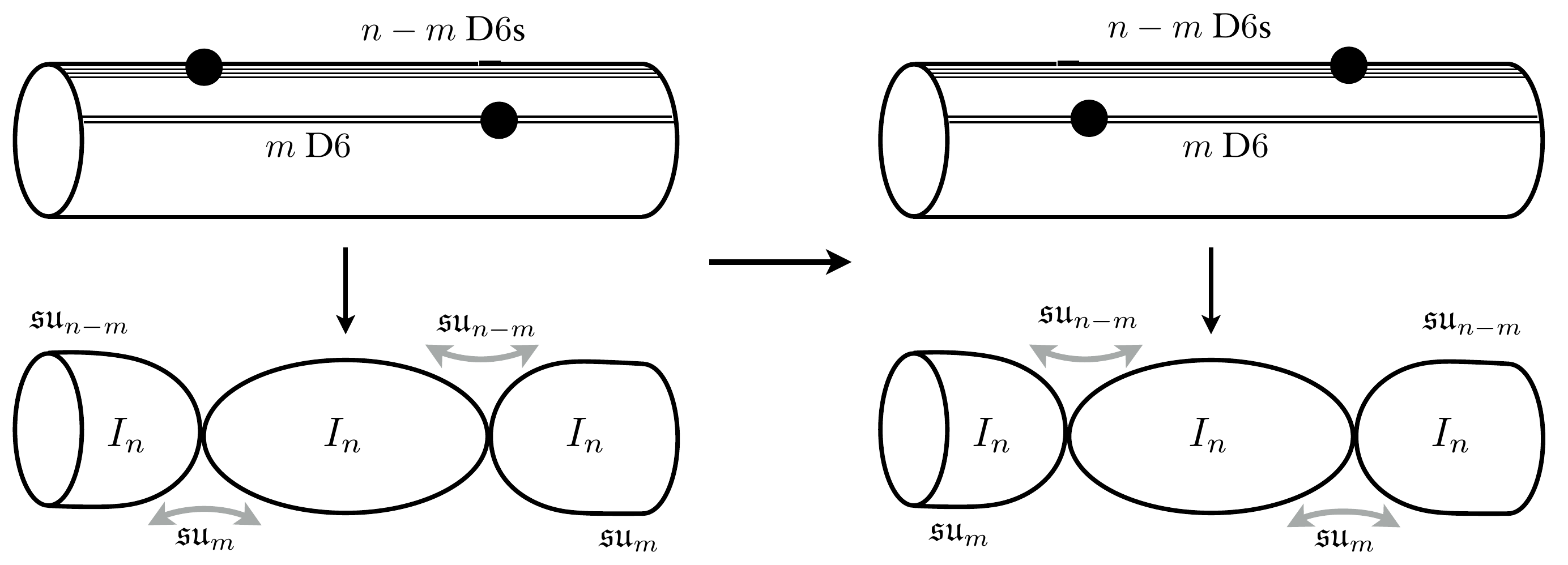}
	\caption{\small A smooth transition, in IIA and in F-theory.}
	\label{fig:smooth}
\end{figure}

However, on the IIA side we can also consider placing the NS5s at different values of $x^9$ (the compact direction).
In this case, bringing the NS5s at the same value of $x^9$ does not actually put them on top of each other; 
now we do not expect strong coupling physics at the origin $\phi=0$ of the tensor branch.
A first example not involving orientifolds is shown in Fig.~\ref{fig:smooth}.
In this case without frozen seven-branes, we can of course put all NS5-branes on the same stack of D6-branes so that this smooth transition does not happen.

\begin{figure}[ht]
	\centering
		\includegraphics[width=14cm]{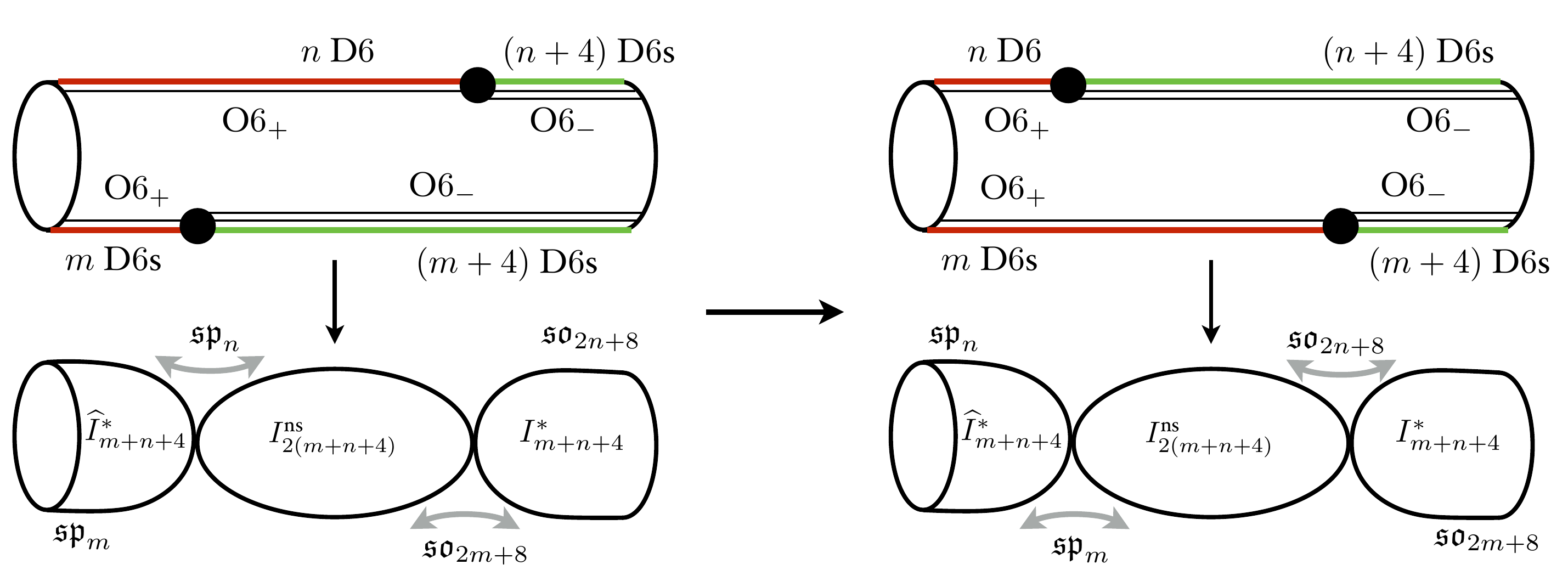}
	\caption{\small A configuration that produces a curve touching both an $I^*$ and and an $\widehat I^*$.
The gauge algebras $\mathfrak{sp}_n$ and $\mathfrak{so}_{2m+8}$ are shared between the first two and the second two curves respectively.}
	\label{fig:smooth-O7}
\end{figure}

When we start involving orientifolds, we can engineer more interesting situations.
The example in Fig.~\ref{fig:smooth-O7} has a non-split $I^{\rm ns}_{2n}$ touching both a frozen and a non-frozen $I^*$.
In this case there is no way to put all NS5-branes on the same side of the O6-planes.
Note also that in both sides of the figure the overall gauge algebra remains the same, but the roles of localized and shared simple subalgebras  are exchanged.

When the two NS5s are aligned,
for $m=n$ we are in fact in the situation of (\ref{eq:I*hI*}), with $k=\ell=2n+4$.
This is in agreement with our observation made there (motivated by the absence of a hypermultiplet) that there is no conformal point at that intersection; in this case the transition is completely smooth, and there is no special point on the tensor branch. 

In 6d compactifications of F-theory,  we are accustomed to getting conformal theories when a divisor shrinks. 
One reason for this is that one can engineer string states from D3-branes, and these strings become tensionless when we shrink a curve. In the situations of Fig.~\ref{fig:smooth} and \ref{fig:smooth-O7}, in fact we cannot wrap a D3-brane on the middle curve: this is made clear by T-dualizing back to IIA, where it would become a D2-brane, which can terminate on either one or the other half-NS5, but not on both.

The situation in Fig.~\ref{fig:smooth-O7} is a simple illustration of the fact mentioned in the introduction that in the presence of O7$_+$ we lose the notion of a canonical assignment of gauge algebras and matter content.
In this situation, this happens for two reasons.
First, we can only take $m$ D6-branes from bottom to top of the cylinder.
After doing that, we are still left with 4 D6-branes ending on half-NS5-brane.
This implies that there is no canonical `zero' for the Wilson lines.
Second, the half-NS5s are stuck at fixed values of $x^4$.
This implies that there are fixed non-zero periods of NS-NS 2-form potential on the curves.

% subsection smooth (end)

\subsection{Tangential intersections and O8-planes} % (fold)
\label{sub:tangential}

The discussion of  $I^*$--$I$ and $\widehat I^*$--$I$ intersections in section \ref{ssub:i*i} has an interesting exception, that occurs when the intersection is \emph{tangential}. We discuss it now because T-duality helps in the analysis, as we will now see.  

\begin{figure}[ht]
	\centering
		\includegraphics[width=14cm]{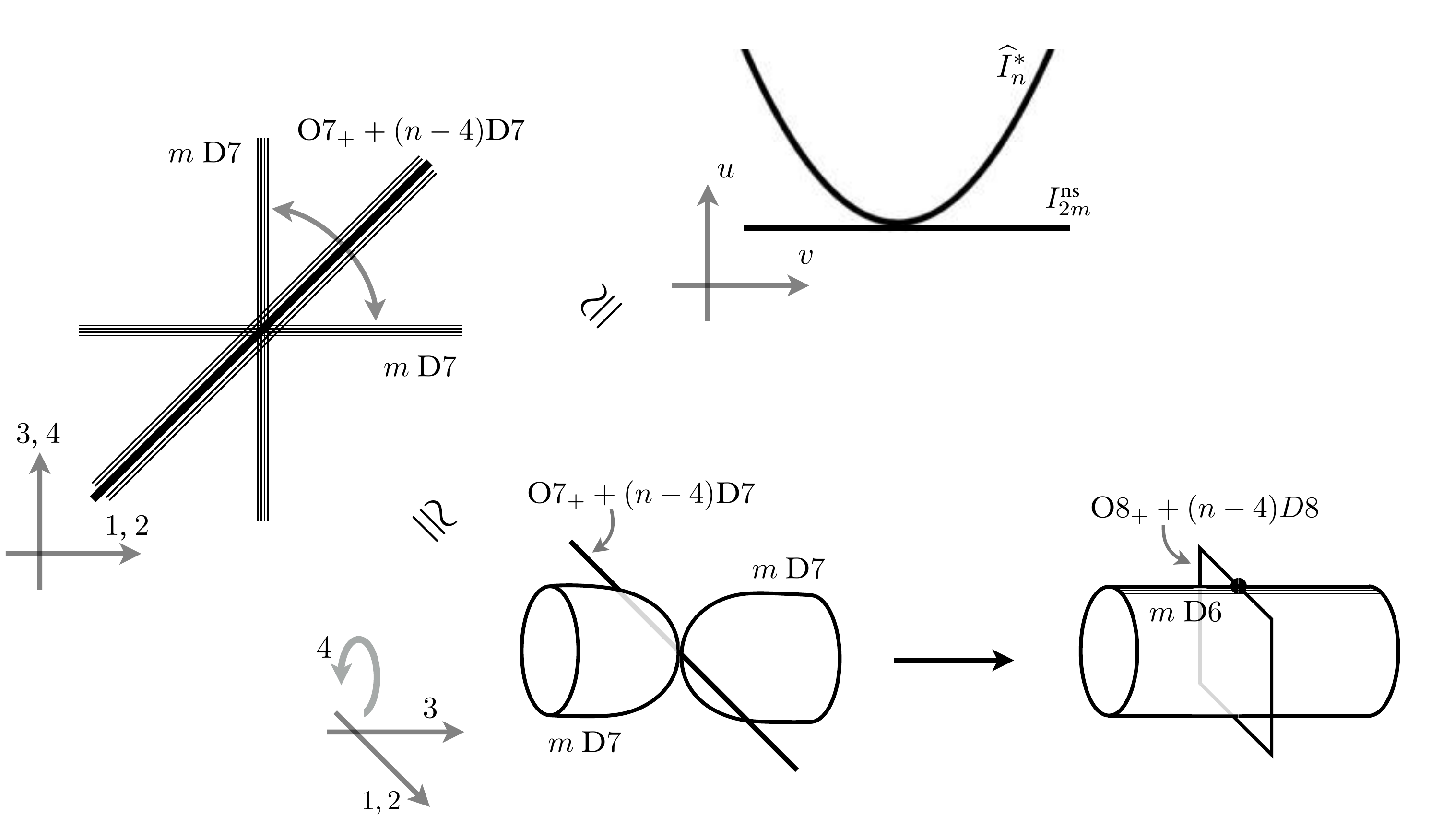}
	\caption{\small Various equivalent ways of seeing a tangential $\hat I^*$--$I$ intersection. As in recent figures, the dot on the bottom-right frame is a half-NS5.}
	\label{fig:tangential}
\end{figure}

We start by considering O7s and D7s that again share the directions 056789, but which are extended in the remaining directions in a more complicated fashion than in section \ref{ssub:i*i}. Define $z= x^1+ i x^2$, $w=x^3 + i x^4$, and let the orientifold act on the spacetime by $\sigma: z\leftrightarrow w$.
The O7$_\mp$ will then be on the locus $z=w$; place again $n\pm 4$ D7s on top of it.
Now also place $m$ half-D7s on the locus $z=0$; their $m$ images will be on the locus $w=0$.
In this case, the gauge fields on the D7s on $z=0$ will have a ${\rm U}(m)$ gauge field, which the O7 maps to a gauge field on the D7s on $w=0$.
To see why this is related to a tangential intersection, consider the invariant coordinates $v=z+w$, $u=zw$.
The configuration we are considering is then mapped to an O7$_\mp + (n\pm 4)$ D7s on the locus $v^2=4u$, and $m$ D7s on $u=0$.
These two loci intersect tangentially.
We can summarize this as follows:
\begin{equation}\label{eq:tangential}
	\begin{array}{ccc}
		\mathfrak{so}_{2n+8} & &\mathfrak{su}_{m}\\
		I^*_n &|| & I_{m}\,;
	\end{array}
	\qquad\qquad
	\begin{array}{ccc}
		\mathfrak{sp}_{n-4} & & \mathfrak{su}_{m}\\
		\widehat I^*_n & || &I_{m}\,,
	\end{array}
\end{equation}
where we have used $||$ to denote tangency as in \cite{Bhardwaj:2015oru}. This coordinate change is illustrated in the top part of Fig.~\ref{fig:tangential}, in the O7$_+$ case. 

An additional subtlety concerns the matter content in (\ref{eq:tangential}). One can in principle work this out directly in the original setup on the left of Fig.~\ref{fig:tangential}, but it is instructive to do it instead in a dual frame. First of all we change coordinates, using again (\ref{eq:xzw}); only this time we take $z=x^1+ i x^2$, $w=x^3+i x^4$ introduced earlier, and define new coordinates $\tilde x^1 + i \tilde x^2= z w$, $\tilde x^3 = |z|^2- |w|^2$, with a fourth periodic coordinate $e^{i\tilde x^4}= \frac{z \bar w}{\bar z w}$. We are once again rewriting $\mathbb{R}^4$ as a fibration $S^1 \hookrightarrow \mathbb{R}^4 \to \mathbb{R}^3$. The orientifold is now defined by the involution $\sigma: \tilde x^3 \to - \tilde x^3, \tilde x^4 \to - \tilde x^4$; the O7-plane then sits at $\tilde x^3=\tilde x^4=0$, while the D7s are on the locus $\tilde x^1=\tilde x^2=0$. (Notice that the $\tilde x^4$ circle shrinks at $\tilde x^3=0$.) If we now T-dualize along direction 4, we end up with an O8 at $\tilde x^3=0$ with a half-NS5 stuck on it, and with D6s crossing it. 

All this is depicted on the lower part of Fig.~\ref{fig:tangential}, again for the O8$_+$ case. At this point we can read off the matter content from a perturbative string computation similar to the one leading to (\ref{eq:OpDq}), as already done in \cite{Brunner:1997gf,Hanany:1997gh}; the result is that in the tangential intersection (\ref{eq:tangential}) the $\mathfrak{u}_m$ has a hypermultiplet in the antisymmetric in the unfrozen case, and in the symmetric in the frozen case.

We can deform a tangential intersection into two transverse intersections. 
This corresponds to giving a vev to the hypermultiplet in the antisymmetric or symmetric representation,
and breaks the gauge algebra to $\mathfrak{sp}$ or $\mathfrak{so}$ respectively.
We will study an explicit example in Sec.~\ref{sub:susu}.

% subsection tangential (end)

% section pert (end)

\section{Anomaly analysis} % (fold)
\label{sec:anom}

In this section we discuss the cancellation of one-loop anomalies and the Green--Schwarz contributions in 6d compactifications with frozen seven-branes.

\subsection{Anomaly cancellation with frozen singularities}
\label{sec:anomcancel}

A compactification of F-theory on an elliptically fibered Calabi--Yau threefold gives rise to an effective 6d gauge theory with $\cN=(1,0)$ supersymmetry at low energies.
When there are no frozen singularities present,
it is  possible to turn off the holonomies of gauge fields on stacks of seven-branes, and the periods of 2-form NS-NS and R-R potentials.
Then, each simple summand $\fg_i$ of the 6d gauge algebra is associated to a single irreducible component $D_i$ of the discriminant locus of the elliptic fibration, and can be determined from the knowledge of the type of singular fiber over $D_i$ along with the data of the monodromy of the elliptic fiber around $D_i$ \cite{Morrison:1996pp,Bershadsky:1996nh,Aspinwall:2000kf}.
The matter content \cite{Katz:1996xe,Grassi:2011hq} and the coupling of tensor multiplets \cite{Sadov:1996zm} is encoded in the intersection numbers of various divisors in the base of the elliptic fibration.
These data allow us to compute both the 1-loop contribution $I^8_\text{1-loop}$ to the anomaly polynomial, as well as the Green--Schwarz contribution $I^8_{GS}$ to the anomaly polynomial.
Combining these two, one finds that $I^8=I^8_\text{1-loop}+I^8_{GS}$ vanishes for any  smooth elliptically fibered Calabi--Yau threefold \cite{Grassi:2000we,Grassi:2011hq}.

Now let us include frozen singularities in the geometry.
In this situation, it is not always possible to tune the above mentioned holonomies to zero.
We do not have any canonical nonzero choice either.
Because of the nonzero holonomies, one is forced to consider situations in which simple summands of the 6d gauge algebra are realized on divisors which are positive linear combinations of irreducible components of discriminant locus.
We will call the divisors associated to simple summands of gauge algebra as \emph{gauge divisors}.

In this paper, we will not be able to list down all the possible 6d spectra that could result from a geometry, as that will require a systematic understanding of holonomies and fluxes in F-theory compactifications, which we do not have at present.
Therefore, we suppose that an assignment of gauge algebras on the components of the discriminant is given, and study  the Green--Schwarz contribution to the anomaly.
We follow the work of Sadov \cite{Sadov:1996zm} but we include the effects from the frozen singularities.

The 6d tensor multiplets descend from Kaluza-Klein reduction of the chiral 4-form $C^{(4)}$ of type IIB string theory.
To determine the coupling of 6d tensor multiplets, we need to look at two couplings of $C^{(4)}$ in ten-dimensional type IIB string theory,
namely the coupling to the gauge theory living on seven-branes and the coupling to gravity in the bulk.

\paragraph{Gauge Green--Schwarz terms:}
We start with the  coupling of the gauge fields to the RR 4-form field $C^{(4)}$.
When there are no O$7_+$-planes,
 the stack of seven-branes on $D_a$ has a ten-dimensional coupling given by
\begin{align}
\int C^{(4)} \nu(F_a)\divisor{D_a}\label{unfro}
\end{align}
where $F_a$ is the field strength valued in the ``Kodaira'' 8d gauge algebra $\fk_a$ on the $D_a$ component of the discriminant,
and $\nu(F_a)$ is the instanton number density\footnote{%
In the literature many different conventions have been used; 
$\mathop{\mathrm{tr}} F^2$ is defined variously as the trace
in the smallest nontrivial representation (e.g.~\cite{Bhardwaj:2015xxa}),
or in the adjoint representation divided by the dual Coxeter number (e.g.~\cite{Tachikawa:2018njr}),
or by twice the dual Coxeter number (e.g.~\cite{Grassi:2000we,Grassi:2011hq}),
with or without $(2\pi)^4$ in the denominator implicitly included.
We follow the physical convention introduced by Intriligator in \cite{Intriligator:2014eaa}, where the notation $c_2(F)$ was used.
This choice  has the virtue that the coefficient in the resulting anomaly polynomial of the term $\nu(F_a)\nu(F_b)$ have a direct physical meaning, i.e.~the Dirac pairing of two instanton-strings.
}, normalized so that it integrates to one on the standard BPST instanton embedded into $\fk_a$ with embedding index 1.
This normalization reflects the familiar fact that an instanton in the worldvolume of a seven-brane has D3-charge 1.

When the component $D_a$ carries an $\widehat I^*_{n+4}$ singularity, i.e.~when it corresponds to  an O$7_+$-plane with $n$ D7-branes on top, 
the local 8d gauge algebra  is $\fk_a=\sp_{n}$, and the ten-dimensional coupling is \begin{equation}
\int C^{(4)} \left(\frac12 \nu(F_a) \right)\divisor{D_a}.\label{fro}
\end{equation}
Note a factor-of-two difference in the coefficient between \eqref{unfro} and \eqref{fro}.
This is due to the fact that a bulk D3-brane can fractionate into two on O7$_+$, as reviewed in Sec.~\ref{sub:bas},
and the gauge instanton in the standard normalization corresponds to the D3-brane of minimal possible charge.

Let us now write the 6d gauge algebra in the form $\oplus_i \fg_i$ where $\fg_i$ is simple.
Each $\fg_i$ is shared on some  of the $D_a$;
we let $\mu_{i,a}=1$ or $0$ depending on whether $\fg_i$ is on $D_a$ or not.
An embedding $\rho_{i,a}:\fg_i \hookrightarrow \fk_a$ must exist whenever $\mu_{i,a}=1$, and otherwise we let $\rho_{i,a}$ be the zero map.
These embeddings have the properties
\begin{enumerate}
\item $\bigoplus_i \rho_{i,a}(\fg_i) \subset \fk_a$, and
\item $\fg_i$ is the diagonal in $\bigoplus_a \rho_{i,a}(\fg_i)$.
\end{enumerate}
The Green--Schwarz coupling for the gauge fields is given in terms
of the field strenghs $F_i$ valued in $\fg_i$ by 
\begin{equation}
\int C^{(4)}\sum_i \nu( F_{i})\divisor{\Sigma_i}
:= \sum_{i,a} \int C^{(4)}\left(\sum_i \mu_{i,a}o_{i,a} \nu( F_i)\right)\divisor{D_a}  
 \label{gaugecoupling}
\end{equation}
where we defined the $i$-th \emph{gauge divisor} to be \begin{equation}
\Sigma_i=\sum_a \mu_{i,a}o_{i,a}D_a,
\end{equation} and 
$o_{i,a}$ is the embedding index\footnote{%
The embedding indices we often encounter in this paper can be summarized in the following diagram:
\begin{equation}
\begin{tikzpicture}[baseline=(SP),scale=1,font=\footnotesize]
\node(UU) at (0,1) {$\mathfrak{u}_{2n}$};
\node(SP) at (1,0) {$\mathfrak{sp}_n$};
\node(SO) at (-1,0) {$\mathfrak{so}_{2n}$};
\node(U) at (0,-1) {$\mathfrak{u}_n$};
\draw[->] (U) -- node[midway,below left]{$\scriptstyle1$}  (SO);
\draw[->] (SO) -- node[midway,above left]{$\scriptstyle2$} (UU);
\draw[->] (U) -- node[midway,below right]{$\scriptstyle2$} (SP);
\draw[->] (SP) -- node[midway,above right]{$\scriptstyle1$} (UU);
\end{tikzpicture}
\end{equation}
where the numbers beside the arrow show the indices.
} of $\fg_i\subset \fk_a$,
multiplied by 1/2 when $\fk_a=\sp_{n}$ is supported on a frozen singularity.

Note that even when there is no ``sharing'' (so the gauge divisors are
$\Sigma_a=D_a$) and no O$7_+$-planes, $\fg_a$ could still be different from
$\fk_a$, due to the ``Tate monodromy'' phenomenon \cite{Bershadsky:1996nh}.

Before proceeding, we point out here that the inverse square of the gauge coupling of $\fg_i$ is given by $\sum_a \mu_{i,a} o_{i,a} A_a$ where $A_a$ is the area of $D_a$.
This follows from the fact that the scalar $A_a$ and the 2-form $\int_{D_a} C^{(4)}$ are the bosonic components of a single supermultiplet, and therefore Green-Schwarz coupling $\int C^{(4)} \sum_a\mu_{i,a}o_{i,a}\nu(F_i) D_a$
comes with the coupling $\int \sum_a A_a \mu_{i,a}o_{i,a} \tr F_{i}\wedge *F_i$.
This means in particular that when the gauge algebra $\fg_i$ is shared on multiple components, the gauge theory does \emph{not} become singular when a single component $D_a$ involved in the gauge divisor shrinks to zero size.

\paragraph{Gravitational Green--Schwarz terms:}
We turn our attention to the gravitational coupling.
When there are no O$7_+$s, the stack of seven-branes on $D_a$ has a ten-dimensional coupling to gravity given by
\begin{align}
\int C^{(4)}\left(\frac{N_a}{12}\; \frac{p_1(T)}{4} \right)\divisor{D_a}\label{zen}
\end{align}
where $N_a$ is the order of vanishing of discriminant $\Delta$ on $D_a$, 
$p_1(T)$ is the Pontryagin class of the tangent bundle of the worldvolume.
We also slightly abuse notation and use $D_a$ within  the integral to represent
 the two-form determined by the divisor.\footnote{%
The couplings \eqref{unfro} and \eqref{zen} follow in the case of $N_a$ D7-branes 
by starting from the coupling $(\sum_p C^{(p)}) \hat A(T)^{1/2} \tr e^{F}$ determined in \cite{Green:1996dd} 
and extracting the necessary parts, using $\hat A(T)|_4= -p_1(T)/24$ and $\tr e^{F} |_4 = -\nu(F)$.}
In particular, a D7-brane contributes  $N_a=1$ and an O$7_-$-plane contributes  $N_a=2$.

Now, the contribution of O$7_+$ to this gravitational coupling is opposite to that of O$7_-$;
the ``effective $N_a$'' is $-2$.
Since an $\hat I_n^*$ singularity corresponds to O$7_++(n-4)$D7-branes, 
its ``effective $N_a$'' is $-2+(n-4)=n-6$.
In comparison, $N_a$ of $I_n^*$ is $n+6$.
Hence, in the presence of O$7_+$ we need a correction term to the coupling, which be written as
\begin{align}
\int C^{(4)}\left(\left(\frac{N_a}{12}-s_a\right)\frac{p_1(T)}{4}\right)\divisor{D_a} \label{gravcoupling}
\end{align}
where $s_a=1$ when the curve $D_a$ carries an O$7_+$ and $s_a=0$ when it does not.

\paragraph{The  cancellation:}

Combining (\ref{gaugecoupling}) and (\ref{gravcoupling}), the full six-dimensional coupling relevant for the Green--Schwarz mechanism is
\be
\int_B C^{(4)}\left(-(K+F) \frac{p_1(T)}{4}+\sum_i \divisor{\Sigma_i}\nu( F_i)\right)\,,
\ee
where 
$C^{(4)}$ has two legs on the base $B$
and
 \begin{equation}
F=\sum_a s_a D_a
\end{equation} is the \emph{frozen divisor}, signifying the divisor along which we find the frozen singularities.
We have also used the condition for unbroken supersymmetry (the Calabi--Yau condition) to substitute the canonical divisor $K$ in place of $-\frac{1}{12}N_a\divisor{D_a}$.

The contribution to anomaly polynomial is then a square of the coefficient of $C^{(4)}$, with a factor of $1/2$ in front, to take into account that the RR 4-form field is self-dual:
\begin{align}
I^8_{GS}&=-\frac12 \left(
- (K+F) \frac{p_1(T)}4 +  \sum_i \Sigma_i \nu( F_i)
\right)^2.
\end{align}
It is a standard result (see e.g.~\cite{Erler:1993zy,Schwarz:1995zw,Seiberg:1996vs,Berkooz:1996iz,Sadov:1996zm}) that the one-loop anomaly of the 6d system 
is given by\footnote{%
Again there are various different normalizations in the literature.
We follow the  convention that $2\pi i I^8_\text{ours}$ yields the anomalous phase variation via the descent formalism;
in particular $I^8$ should have rational coefficients when expressed in terms of geometrically-defined characteristic classes.
The early paper by Erler \cite{Erler:1993zy} used $I^8_\text{Erler}=2\pi iI^8_\text{ours}$.
Another common convention during the early years of the second revolution, apparently introduced by Schwarz \cite{Schwarz:1995zw}, was to normalize $I^8$ to contain $(\tr R^2)^2$ with coefficient 1, for a model with one tensor multiplet. 
We have $I^8_\text{Schwarz}=16(2\pi)^4 I^8_\text{ours}$.
}
\begin{equation}
I^8_\text{1-loop}=\frac{9-n_T}{32}p_1(T)^2-\frac{N_i}4 \nu(F_i) p_1(T) + \frac{M_{ij}}2 \nu(F_i)\nu(F_j) \label{1-loop}
\end{equation} where $n_T$ is the number of tensor multiplets,
and $N_i$, $M_{ij}$ are some numerical coefficients,
assuming that the coefficient of $\tr R^4$ vanishes, i.e.
\begin{equation}
n_V-n_H-29n_T+273=0\,. \label{gr}
\end{equation}
At the end of this subsection,
we comment on how to obtain the numerical values $N_i$ and $M_{ij}$.

We see that cancellation of gauge and gauge-gravity anomalies  requires the following:
\begin{align}
N_i&=(K+F)\cdot \Sigma_i ,\label{cancelgg} &
M_{ij}&=\Sigma_i\cdot \Sigma_j. %\label{cancelg}
\end{align}
Here, $K\cdot \Sigma_i=K\cdot (\sum \mu_{i,a} o_{i,a} D_a)$ can be computed from the adjunction formula $2(g_a-1)=(K+D_a)\cdot D_a$, where $g_a$ is the genus of the curve $D_a$.

If the 6d theory contains dynamical gravity and satisfies (\ref{gr}), then we obtain the following condition as well, from the vanishing of the coefficient of $(\tr R^2)^2$:
\be
9-n_T=(K+F)^2 \label{cancelgrav}
\ee

This condition \eqref{cancelgrav} follows just from geometry,
as we now demonstrate.
If $D_a$ carries a frozen singularity, then the singular fiber over $D_a$ has Kodaira type $I^*_{n\ge4}$.
For these Kodaira fibers, it is known that $D_a\cdot(-2K-D_a)=0$ \cite{Grassi:2011hq}.
Moreover, any two distinct components $D_a$ and $D_b$ of $F$ must not intersect each other because two $I^*_{n\ge4}$ singularities cannot intersect each other(in the absence of conformal matter).
From these two facts, it follows that
\be
(K+F)^2=K^2+\sum_as_aD_a\cdot(2K+D_a)=K^2,
\ee
and the equality $K^2=9-n_T$.\footnote{%
Compactification of $C^{(4)}$ on a base $B$ produces $h^{1,1}(B)$ anti-symmetric 2-form potentials.
One of them goes into the supergravity multiplet and the remaining $h^{1,1}(B)-1$ go into tensor multiplets; hence $n_T=h^{1,1}(B)-1$.
Since $B$ is the base of Calabi--Yau, $h^{1,0}(B)=h^{2,0}(B)=0$ and it follows from Noether's formula that 
$K^2=10-h^{1,1}(B)=9-n_T$.}

By now, the cancellation of the Green-Schwarz anomaly and of the one-loop anomaly in the conventional F-theory compactification without O7$_+$ is well-established.
This allows us to read off $N_i$ and $M_{ij}$ for almost all the cases.
First, for $i\neq j$,
we have $M_{ij}=1$ for a bifundamental of $\mathfrak{su}$-$\mathfrak{su}$
or a half-bifundamental of $\mathfrak{so}$-$\mathfrak{sp}$.
To read off $N_i$ and $M_{ii}$ (without suming over $i$),
let us say that the given algebra is $\fg_i$ and the total set of hypermultiplets for $\fg_i$ is $\rho$.
One then looks up the pair of $(\fg_i,\rho)$ e.g.~in Eqs.~(2.10)--(2.14) of \cite{Kim:2016foj},
to find a conventional F-theory realization of the 6d gauge theory on a sphere of  self-intersection $-n$.
Then $M_{ii}=-n$ and $N_{i}=n-2$.
Essentially the only case not covered by this procedure is when $\fg_i=\su(n)$, with one $\rep{sym}$ and $n-8$ fundamentals.
For this, one first Higgses the hypers in $\rep{sym}$,
to give $\so(n)$ with $n-8$ fundamentals. 
This has a well-known anomaly polynomial, which can be determined in the method just described above.
Then one can convert it back to the anomaly polynomial of the original $\su(n)$ theory by using $\nu(\so(n))=2\nu(\su(n))$.

\subsection{Matter content with frozen singularities}
\paragraph{Transversal intersections:}
In the situation when there are no frozen singularities and each simple factor of gauge algebra $\fg_a$ is associated to a single irreducible component of discriminant locus $D_a$, Grassi and Morrison \cite{Grassi:2011hq} wrote down the matter content charged under $\fg_a$ in terms of intersection numbers of combinations of $D_a$ and $K$,
assuming that every intersection among $D_a$ and $D_b$ is transversal.
The geometry underlying the derivation of those formulas, analyzed in
the M-theory dual (and therefore on the Coulomb/tensor branch
of the theories), consists of finding the curves in the total space
upon which M2-branes can be wrapped, and finding the intersection numbers
of those curves with the divisors which represent the Cartan subgroup
of the original nonabelian gauge group, since those intersection numbers
specify gauge charges.  This was carried out in a number of works
\cite{Bershadsky:1996nh,Katz:1996xe,intriligator-morrison-seiberg,Aspinwall:2000kf,Morrison:2011mb}
which \cite{Grassi:2011hq} relied upon.

\begin{table}[htbp]
\begin{center}
\begin{tabular}{|c|c|c|} \hline
 $\mathfrak{g}_a$ & $\rho$ & Number of hypers in $\rho$ \\ \hline
 \hline
 $\mathfrak{su}_{2} $& $\rep{adj}$ &$\frac12(K'+\Sigma_i)\cdot \Sigma_i$\\
                   & $\rep{fund}$ &$(-8K'-2\Sigma_i)\cdot \Sigma_i$ \\ \hline
 $\mathfrak{su}_{3} $& $\rep{adj}$ &$\frac12(K'+\Sigma_i)\cdot \Sigma_i$\\
                   & $\rep{fund}$ &$(-9K'-3\Sigma_i)\cdot \Sigma_i$ \\ \hline
 $\mathfrak{su}_{n} $,& $\rep{adj}$ &$\frac12(K'+\Sigma_i)\cdot \Sigma_i$   \\
$n\ge4$         &      $\rep{fund}$ &$(-8K'-n\Sigma_i)\cdot \Sigma_i$ \\
     & $\rep{asym}^2$ &$-K'\cdot \Sigma_i$  \\
 \hline
 $\mathfrak{sp}_{n} $,&$\rep{adj}$ &$\frac12(K'+\Sigma_i)\cdot \Sigma_i$\\
     $n\ge2$          &$\rep{fund}$ &$(-8K'-2n\Sigma_i)\cdot \Sigma_i$\\
     &$\rep{asym}^2_{\text{irr}}$ &$\frac12(-{}K'+\Sigma_i)\cdot \Sigma_i$\\
\hline
$\mathfrak{so}_{7}$, & $\rep{adj}$ &$\frac12(K'+\Sigma_i)\cdot \Sigma_i$\\
$(-2K'-\Sigma_i)\cdot \Sigma_i\ge0$,       & $\rep{vect}$ &$\frac12(-3K'-\Sigma_i)\cdot \Sigma_i$\\
& $\rep{spin}$ & $(-4K'-2\Sigma_i)\cdot \Sigma_i$ \\ \hline
$\mathfrak{so}_{7}$, & $\rep{adj}$ &$\frac18(-2K-2F+\Sigma_i)\cdot \Sigma_i$\\
$(-2K'-\Sigma_i)\cdot \Sigma_i\le0$,       & $\rep{vect}$ &$\frac14(-16K'-7\Sigma_i)\cdot \Sigma_i$\\
& $\rep{sym}^2_{\text{irr}}$ & $\frac{1}{8}(2K'+\Sigma_i)\cdot \Sigma_i$ \\ \hline
$\mathfrak{so}_{n}$, & $\rep{adj}$ &$\frac12(K'+\Sigma_i)\cdot \Sigma_i$\\
$8\le n\le14$,       & $\rep{vect}$ &$\frac12({(4-n)}K'+(6-n)\Sigma_i)\cdot \Sigma_i$\\
& $\rep{spin}_*$ & $\frac1{\dim(\rep{spin}_*)}(-{32K'}-16\Sigma_i)\cdot \Sigma_i$ \\ \hline
$\mathfrak{so}_{n}$, & $\rep{adj}$ &$\frac12(K'+\Sigma_i)\cdot \Sigma_i$\\
$n\ge15$       & $\rep{vect}$ &$(-{4}K'-\frac n4 \Sigma_i)\cdot \Sigma_i$\\ \hline
 $\mathfrak{e}_6$ & $\rep{adj}$ &$\frac12(K'+\Sigma_i)\cdot \Sigma_i$\\
       & $\rep{27}$ &$(-3K'-2\Sigma_i)\cdot \Sigma_i$\\ \hline
 $\mathfrak{e}_7$ & $\rep{adj}$ &$\frac12(K'+\Sigma_i)\cdot \Sigma_i$\\
       & $\rep{56}$ &$\frac12(-4K'-3\Sigma_i)\cdot \Sigma_i$\\ \hline
 $\mathfrak{e}_8$ & $\rep{adj}$ &$\frac12(K'+\Sigma_i)\cdot \Sigma_i$\\ \hline
 $\mathfrak{f}_4$ & $\rep{adj}$ &$\frac12(K'+\Sigma_i)\cdot \Sigma_i$\\
       & $\rep{26}$ &$\frac12(-5K'-3\Sigma_i)\cdot \Sigma_i$\\ \hline
 $\mathfrak{g}_2$ & $\rep{adj}$ &$\frac12(K'+\Sigma_i)\cdot \Sigma_i$\\
        & $\rep{7}$ &$(-5K'-2\Sigma_i)\cdot \Sigma_i$\\ \hline
\end{tabular}
\end{center}
\smallskip
\caption{Number of hypermultiplets for each relevant representation of each simple gauge algebra when frozen singularities are present.
This includes the contribution of vector multiplet as a $-1$ hypermultiplet in adjoint. 
By definition, $K'=K+F$.
$\rep{spin}_*$ denotes the sum of number of hypers in two irreducible spinors $\rep{spin}_\pm$ for $\so_\text{even}$, and the number of hypers in the unique irreducible spinor for $\so_\text{odd}$. For $\so_8$, number of hypers in $\rep{spin}_+$ must equal number of hypers in $\rep{spin}_-$.
The two different proposals for $\so_{7}$ coincide when $(-2K'-\Sigma_i)\cdot \Sigma_i=0$.
For $\so_{n\ge15}$ we have a further constraint that $\Sigma_i\cdot(-2K'-\Sigma_i)=0$, and for $\fe_8$ we have a further constraint that $(6K'+5\Sigma_i)\cdot \Sigma_i=0$.}	\label{matter2}
\end{table}

Now we would like to understand the matter content
in the presence of the frozen singularities.
We do not have a geometric derivation for our proposed answer, since
the M-theory geometry of frozen singularities is not well understood.
However, as we have seen in detail, the effect of the frozen singularity in the anomaly contribution from the Green--Schwarz effect is summarized by the replacement of individual components $D_a$ by the gauge divisor $\Sigma_i$,
and of the canonical class $K$ by $K'=K+F$.
The one-loop contribution should then be able to exactly cancel this contribution.  We thus propose that the corect answer for the matter content
is to perform the same replacement in 
the results of \cite{Grassi:2011hq}.

We tabulate the results of this replacement, i.e., of our precise proposal
for matter content, in Table~\ref{matter2}.
A few comments on the table are in order:
\begin{itemize}
\item The number associated to adjoint representation in the table is $n_H^\rep{adj}-1$ where $n_H^\rep{adj}$ is the number of hypermultiplets charged in the adjoint representation.
The $-1$ incorporates the contribution to the anomaly of a vector multiplet, which indeed comes with the opposite sign with respect to an adjoint hypermultiplet.
\item For $\so_\text{even}$, the number of hypers in $\rep{spin}_*$ denotes the combined sum of the number of hypers in the two irreducible spinor representations $\rep{spin}_\pm$. For $\so_\text{odd}$, the number of hypers in $\rep{spin}_*$ denotes the number of hypers in the unique irreducible spinor representation.
\item For a generic $\so_\text{even}$ we can choose the number of hypers in $\rep{spin}_+$ and $\rep{spin}_-$ arbitrarily as long as their sum equals the number of hypers required in $\rep{spin}_*$. However, for $\so_8$, the number of hypers in $\rep{spin}_+$ must equal the number of hypers in $\rep{spin}_-$,
because there are two linearly-independent Casimirs of degree 4.
See \cite{Grassi:2011hq} for more details on this requirement.
\item 
The entry for $\so_{7}$ in our table contains a refinement over \cite{Grassi:2011hq}, in which only the spinor representation was considered.
But the coefficient of the spinor representation is negative whenever $(-2K'-\Sigma_i)\cdot \Sigma_i<0$.
In this case, a different representation with the same contribution to the anomaly needs to be used.
One finds $\rep{sym}^2_\text{irr}$ does the job.\footnote{%
Similar modifications are unnecessary for $\so_{n\ge 8}$.
Suppose $(2K'+\Sigma_i)\cdot \Sigma_i \ge 8$, so that we have at least one $\rep{sym}^2_\text{irr}$.
Combining this inequality with the inequalities that the number of vectors are non-negative and the number of adjoints are $\ge-1$, we obtain:
\begin{align}
(-2K'+ \Sigma_i)\cdot \Sigma_i&\ge-8,\\
\left(-4K'-\frac{n}{4} \Sigma_i\right)\cdot \Sigma_i&\ge0,\\
(2K'+ \Sigma_i)\cdot \Sigma_i&\ge8.
\end{align}
Combining the first and third inequalities, we find that $\Sigma_i^2\ge0$.
Combining the second and third inequalities, we find that $(8-n)\Sigma_i^2\ge32$.
These two together imply that $n<8$.}
\item For $\fg_a=\so_{n\ge15}$, we have the additional constraint $\Sigma_i\cdot(-2K'-\Sigma_i)=0$.
The physical meaning of this constraint is that the intersection points of $\Sigma_i$ and $-2K'-\Sigma_i$ carry spinor representations, but it is impossible to satisfy anomaly cancellation for $\so_{n\ge15}$ if we have hypermultiplets transforming as spinors.
There is a similar constraint for $\fe_8$ which states that $(6K'+5\Sigma_i)\cdot \Sigma_i=0$.
\end{itemize}

\paragraph{Tangential intersections:}

We know that this simple replacement cannot be the full story.
We saw at the end of Sec.~\ref{sub:tangential} that if a curve carrying frozen singularities intersects a curve carrying $I_n$ singularity tangentially, then it traps a hypermultiplet in the two-index symmetric representation of $\su_{n}$.
In light of this, for $\fg_i=\su_{n}$ we define $t_a$ to be the number of tangential intersections of $F$ with $D_a$.
Let $t_i=\sum_a \mu_{i,a}t_a$, in terms of which we write our modified proposal for $\su_{n}$ as
\begin{multline}
\rho = \left[\half(K+F+\Sigma_i)\cdot \Sigma_i-t_i\right] \rep{adj} + (-8K-8F-n\Sigma_i)\cdot \Sigma_i\:\rep{fund}\\
+ \left[(-K-F)\cdot \Sigma_i+t_i\right] \rep{asym}^2+t_i\:\rep{sym}^2\,.
\end{multline}
This still satisfies anomaly cancellation because $\sigma=-\rep{adj}+\:\rep{asym}^2+\rep{sym}^2$ has the property that tr$_\sigma F^2$ and tr$_\sigma F^4$ are both zero.
This proposal gives correct predictions for models which have a perturbative dual for which  the spectrum can be determined by other methods.

% section anom (end)

\section{Noncompact models}
\label{sec:scfts}
Now let us analyze how the anomaly cancellation works out in a few examples.
We are particularly interested in 6d SCFTs which supplement the lists
given in \cite{Heckman:2013pva,Heckman:2015bfa}.  As in
\cite{Heckman:2013pva,Heckman:2015bfa}, we expect to be able to realize the tensor
branch of a 6d SCFT by means of a contractible collection of curves
in the F-theory base, with the difference that we will now allow frozen
branes as well.

\subsection{$\so$-$\sp$ chains}
We first come back to the setup discussed in Sec.~\ref{sub:o7ns5}.
In the type IIA frame, we consider the following chain: \begin{equation}
\begin{array}{c|c|c|c}
\text{O6$_-$}&
\text{O6$_+$}&
\text{O6$_-$}&
\text{O6$_+$}\\
\text{$(n+4)$ D6s}&
\text{$n$ D6s}&
\text{$(n+4)$ D6s}&
\text{$n$ D6s}\\
\end{array}
\end{equation} separated by half-NS5-branes.
The leftmost and the rightmost stacks are semi-infinite.
This realizes the 6d quiver theory with the structure \begin{equation}
[\so_{2n+8}] \quad \sp_{n} \quad \so_{2n+8} \quad [\sp_{n}]
\end{equation}
where the bracketed algebras are flavor symmetries.

We perform a T-duality to bring this setup into F-theory.
The result depends on whether we have O6$_-$ or O6$_+$ on the other fixed locus, see Fig.~\ref{fig:TO7}.
The first case is a familiar setup without frozen singularities:
\begin{equation} \label{eq:43}
\begin{array}{cccc}
[\so_{2n+8}] & \sp_{n} & \so_{2n+8} & [\sp_{n}] \\
& 1 & 4 & \\
I^*_n  & I^\text{ns}_{2n} & I^*_n & I^\text{ns}_{2n} 
\end{array}
\end{equation}
where the first, the second, the third row shows the gauge algebra, the negative of the self-intersection number, and the singularity type, respectively.
Denoting the two $\CP^1$'s in the middle by $D_1$ and $D_2$, the Green--Schwarz contribution to the anomaly is 
\begin{equation}
-\frac12\left(  -\frac{p_1(T)}{4} \cdot K +( \nu(F_{\sp})  D_1 +\nu( F_{\so})  D_2) \right)^2.\label{linear1}
\end{equation}

In the second case we obtain a setup with frozen singularities:
\begin{equation} \label{eq:45}
\begin{array}{cccc}
[\so_{2n+8}] & \sp_{n} & \so_{2n+8} & [\sp_{n}] \\
& 4 & 1 & \\
 I^\text{ns}_{2n+8} & \hat I^*_{n+4} &  I^\text{ns}_{2n+8} & \hat I^*_{n+4}.
\end{array}
\end{equation}
Note that the gauge group, matter content, and flavor symmetry group
of \eqref{eq:45} are identical to those of \eqref{eq:43}: only the
F-theory realization is different.

Denoting the two middle $\CP^1$'s by $\tilde D_1$ and $\tilde D_2$ this time, and the canonical class by $\tilde K$ to distinguish it from the case above, the Green--Schwarz contribution is now 
\begin{equation}
-\frac12\left(   -\frac{p_1(T)}{4}\cdot (\tilde K+\tilde F) + (\nu( F_{\sp}) \cdot \frac12\tilde D_1 +\nu( F_{\so}) \cdot 2 \tilde D_2) \right)^2.\label{linear2}
\end{equation}
where the factor $1/2$ in front of $\tilde D_1$ is due to the fractionation of D3-branes on O7$_+$,
and the factor $2$ in front of $\tilde D_2$ is due to the embedding index 2 of $\so_{2n+8}\subset \su_{2n+8}$.
The frozen divisor $\tilde F$ is $\tilde D_1+\tilde D_3$, where $\tilde D_3$ is the noncompact divisor on the far right.

The terms with $\tr F^2_{\sp}$ and $\tr F^2_{\so}$ in the two expressions \eqref{linear1} and \eqref{linear2} should agree, since they cancel the same 1-loop anomalies.
Indeed, we can easily check that \begin{align}
\begin{pmatrix}
K\\
D_1\\
D_2\\
\end{pmatrix}\cdot
(D_1,D_2)
=
\begin{pmatrix}
\tilde K + \tilde F\\
\mfrac12\tilde D_1\\
2\tilde D_2
\end{pmatrix}\cdot
\left(\mfrac12 \tilde D_1,2\tilde D_2\right)
&=\begin{pmatrix}
-1 & 2 \\
-1 & 1 \\
1 & -4
\end{pmatrix}.
\end{align}
In addition, as observed earlier, $K^2 = (\tilde K + \tilde F)^2$.

\subsection{$\su$-$\su$ chains}
\label{sub:susu}

\begin{figure}
\[
\begin{array}{c|cc}
& \text{before} & \text{after}\\
\hline
\text{O}_-&\inc{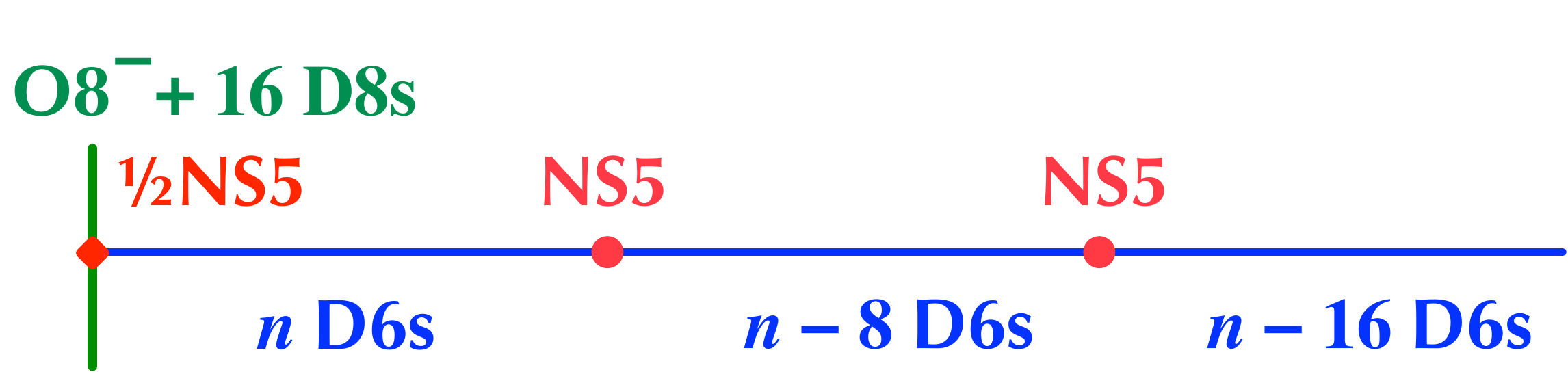}   & \inc{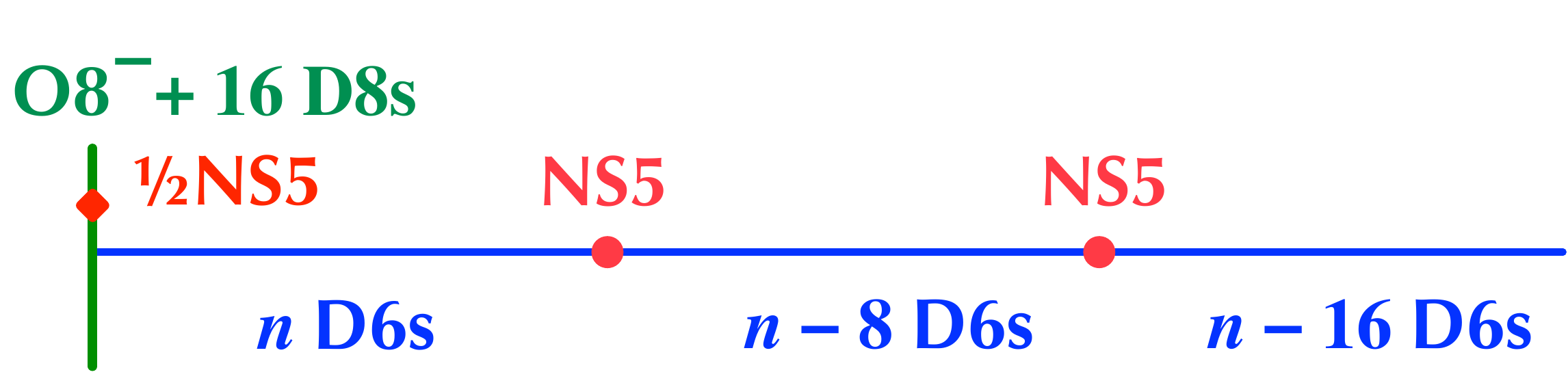} \\
\text{O}_+&\inc{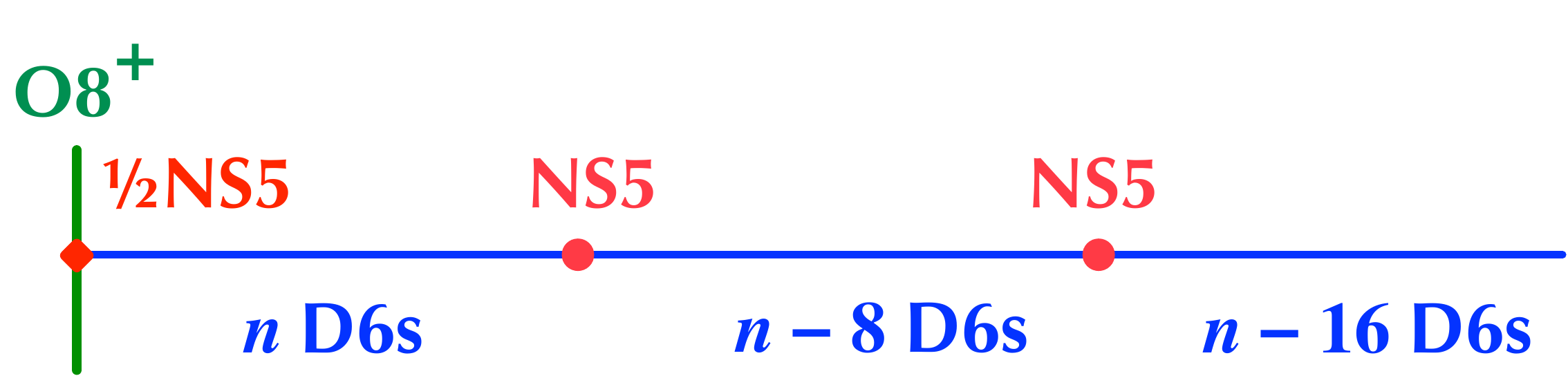}&  \inc{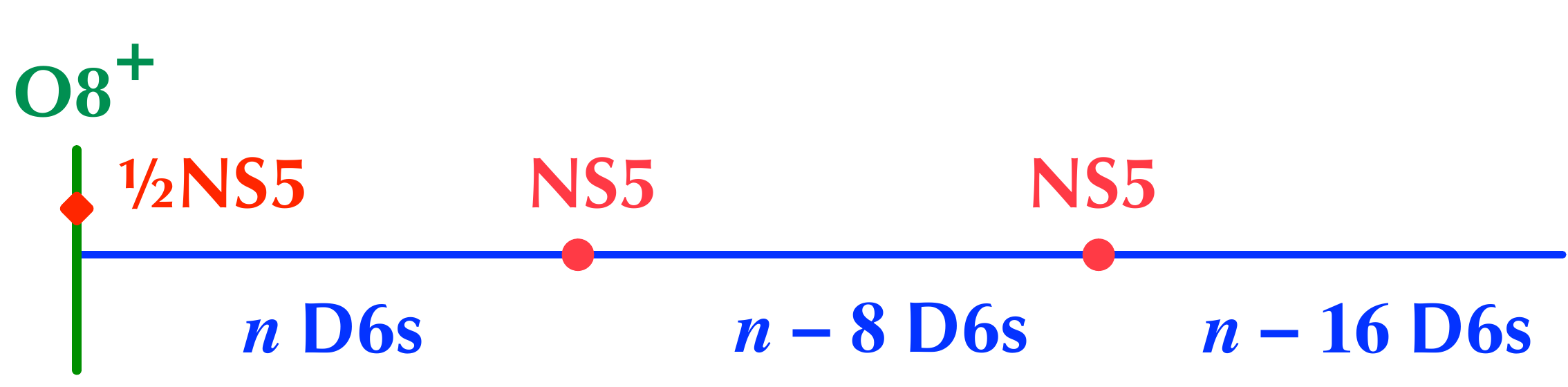} 
\end{array}
\]
\caption{Four type IIA configurations.\label{fubar}}

\bigskip

\[
\begin{array}{c|rr}
& \multicolumn{1}{c}{\text{before}} & \multicolumn{1}{c}{\text{after}}\\
\hline
\text{O}_-&  [\su_{16}] ,\su_{n} +\rep{asym}, \su_{n{-}8}, [\su_{n{-}16}]  & [\so_{32}] ,\sp_{n/2} , \su_{n{-}8}, [\su_{n{-}16}]  \\
\text{O}_+& \su_{n} +\rep{sym}, \su_{n{-}8}, [\su_{n{-}16}] &   \so_{n} , \su_{n{-}8}, [\su_{n{-}16}]  
\end{array}
\]
\caption{Quivers. On the upper right corner, we assumed that $n$ is even.\label{Qfubar}}

\bigskip

\[
\begin{array}{c|cc}
& \text{before} & \text{after}\\
\hline
\text{O}_-&\incF{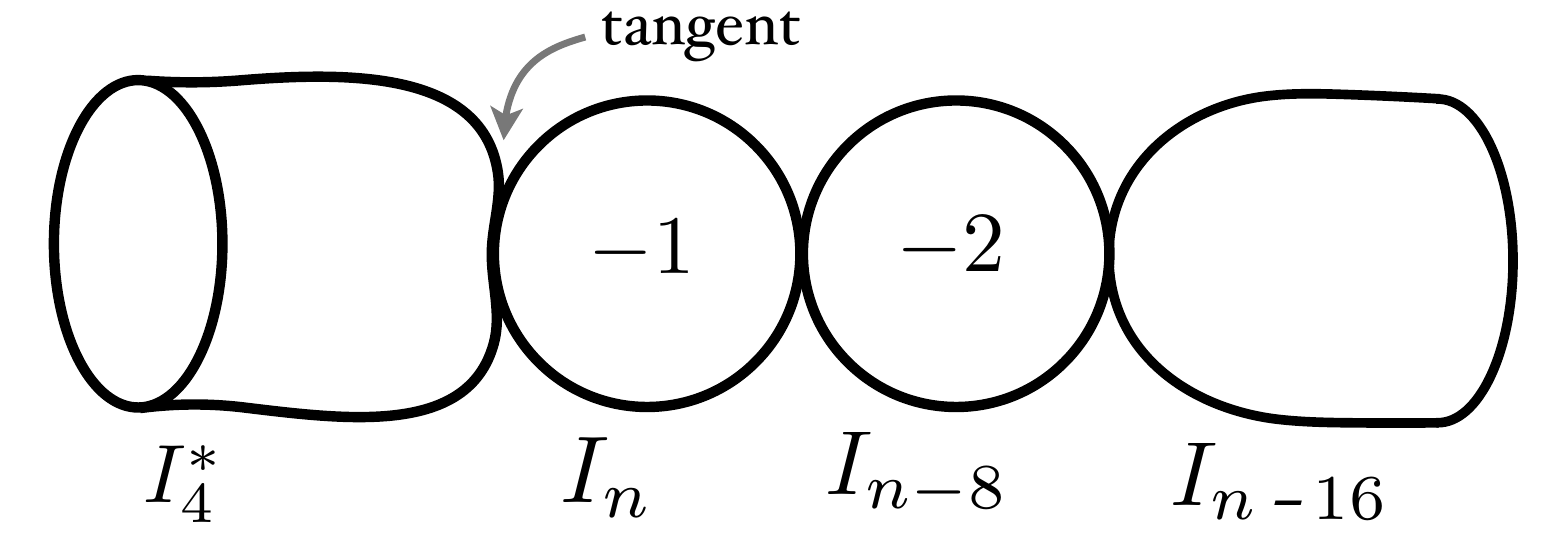}   & \incF{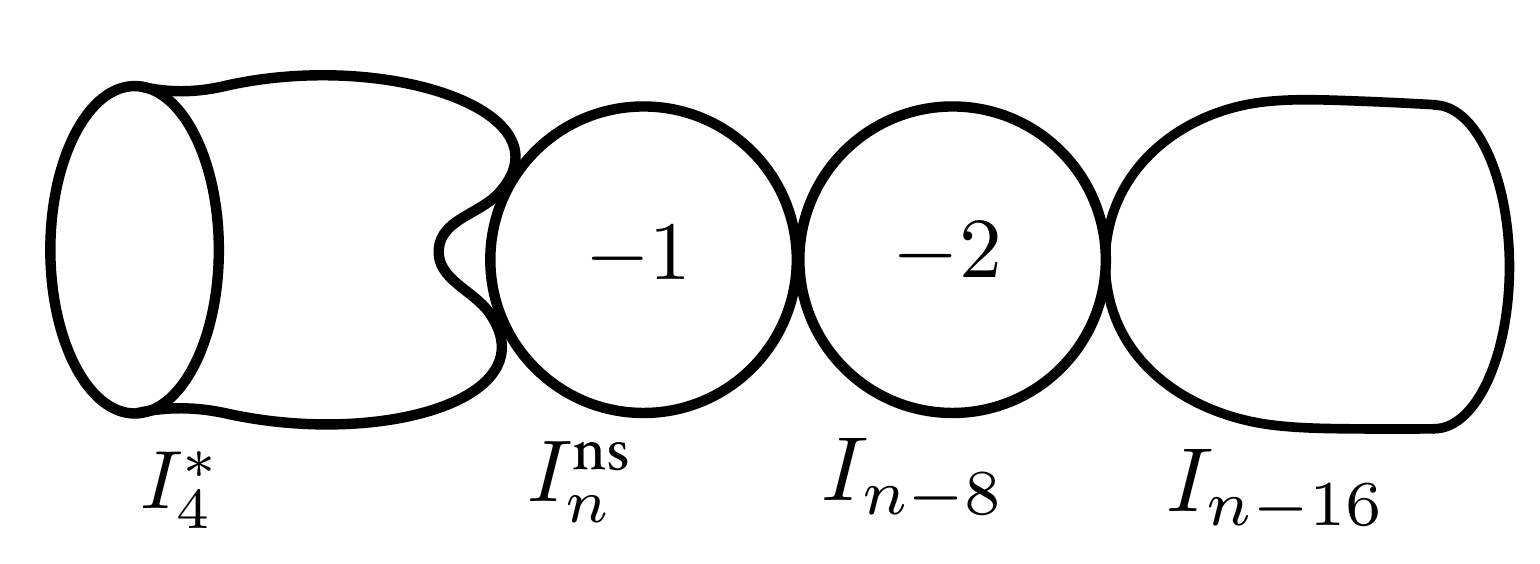} \\
\text{O}_+&\incF{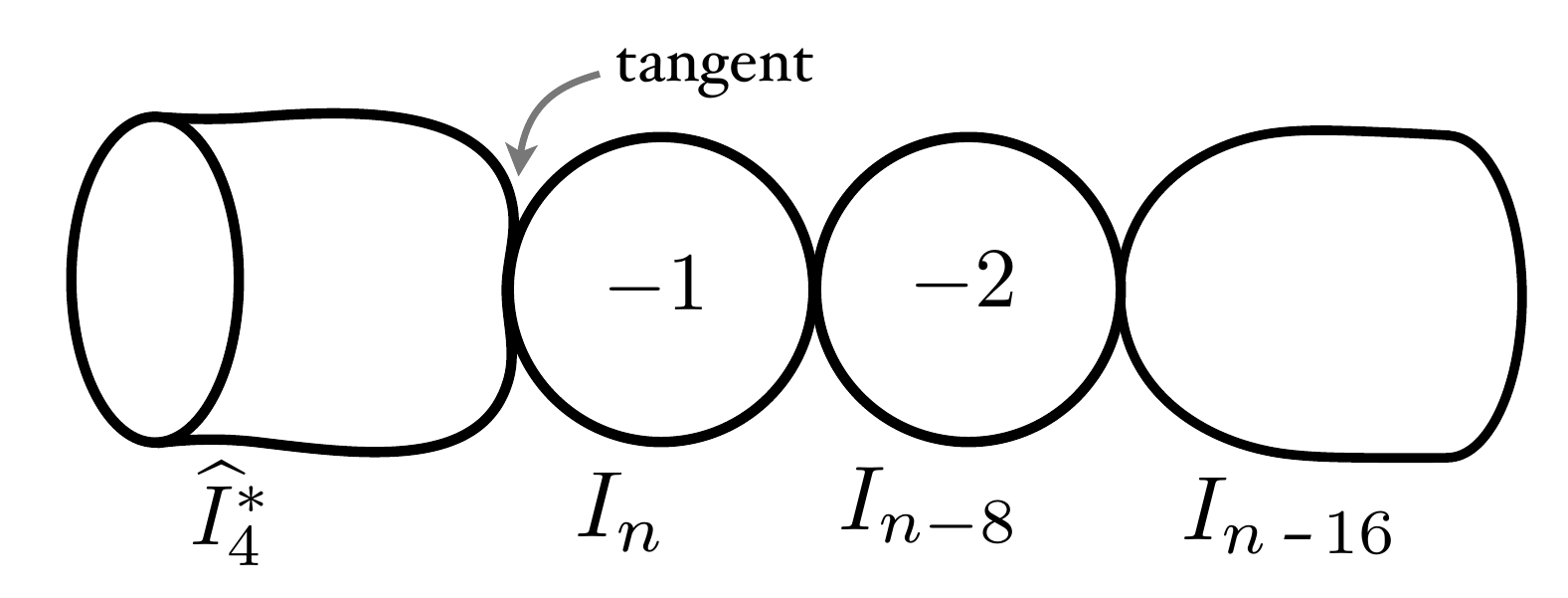}&  \incF{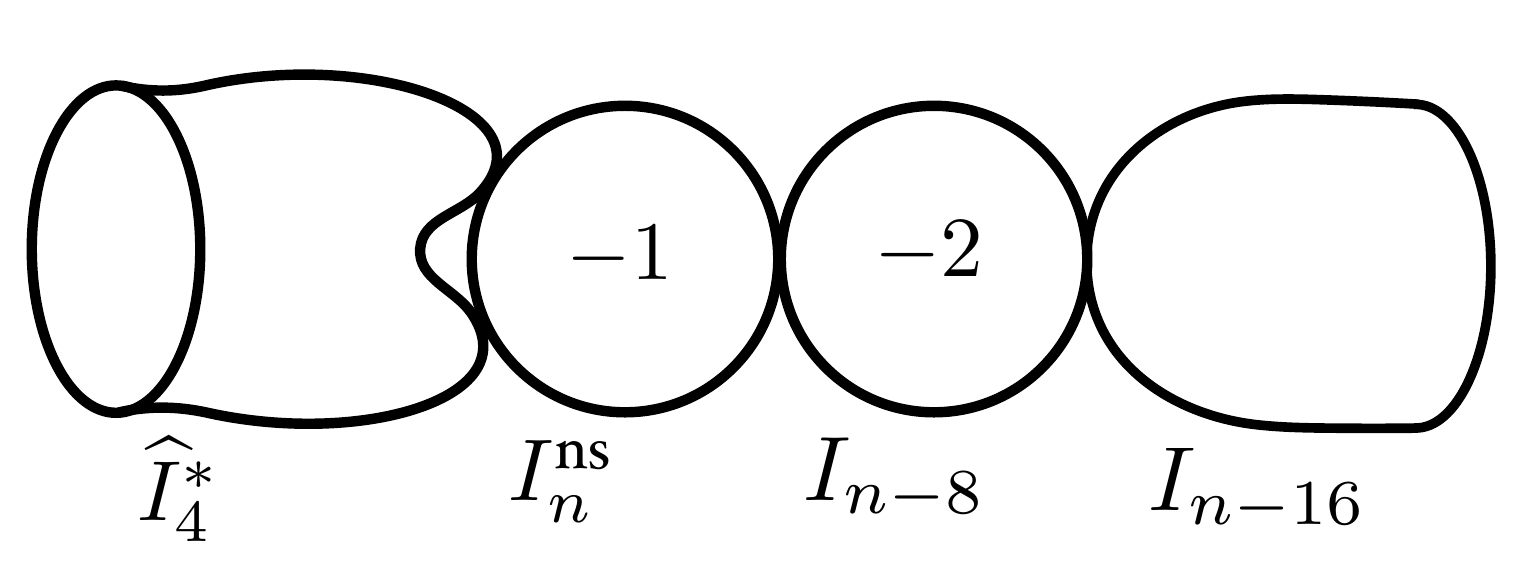} 
\end{array}
\]
\caption{F-theory duals.\label{Ffubar}}
\end{figure}

Let us next consider the IIA configurations shown in Fig.~\ref{fubar}.
The top row and the bottom row are distinguished by the type of the O8-plane;
we add 16 D8-branes for the top row to have the same Romans mass for the both rows. The configurations on the left column contain tangential intersections of the type discussed in section \ref{sub:tangential}. The configurations on the right column are obtained by moving the half-NS5-brane at the intersection of the 6-branes and the 8-branes away from the intersection.
Gauge theoretically, this operation corresponds to giving a vev to hypermultiplets.

Using the discussion in section \ref{sub:tangential} and following  \cite{Brunner:1997gf,Hanany:1997gh}, we find that these configurations realize 6d quivers whose structures are summarized in Fig.~\ref{Qfubar}.
(We did not explicitly write in that figure the standard bifundamental matter hypermultiplets between two consecutive gauge factors.)
The type of the O8-plane is correlated to the type of the two-index tensor representation of the $\su_{n}$ gauge algebra.
Higgsing is done by giving a vev to the hypermultiplet in the antisymmetric or symmetric two-index tensor representations of $\su_{n}$, breaking it to $\sp_{n/2}$ or $\so_{n}$.
Here for simplicity $n$ is assumed to be even in the former case; if $n$ is odd, the gauge algebra is $\sp_{\lfloor n/2\rfloor}$ and one needs to add a flavor to $\su_{n-8}$.

We note that the gauge symmetry $\so_{32}$ on the O8$_-$-plane with 16 D8-branes on top
becomes a flavor symmetry in the theory on the top right of Fig.~\ref{Qfubar}, as expected.
The flavor symmetry is $\su_{16}$ in the theory on the top left instead.
We do not know how to derive this from the perspective of the string theory;
it should be due to the existence of a half-NS5-brane at the intersection of the O8$_-$-plane and the D6-branes.

We can discuss the F-theory duals by T-dualizing the original IIA configurations along the lines of section \ref{sub:tangential}; the results are shown in Fig.~\ref{Ffubar}.
The top row and the bottom row are distinguished by whether we have the ordinary $I^*_4$ singularity
or the frozen $\hat I^*_4$ singularity.
For the left column,
this noncompact divisor of $I_4^*$ or $\hat I_4^*$ is tangent to the divisor with $I_n$ singularity.
To go to the right column, we deform the divisors so that the tangent point is split to two transversal intersection points.
This operation in turn changes the singularity type from $I_n$ to $I_n^\text{ns}$.
The two models on the bottom row realizes 6d quiver gauge theories (the tensor branches of 6d SCFTs) which were not previously possible in an ordinary F-theory compactification without frozen singularities.

Let us name the four divisors in each model as $C_1$, $D_1$, $D_2$, $C_2$ from the left to the right;
$C_{1,2}$ are non-compact and $D_{1,2}$ are compact.
The Green--Schwarz contribution to the anomaly can be written down as follows.

For the top row with the non-frozen $I^*_4$ singularity, we have \begin{equation}
-\frac12\left(- K  \frac{p_1(T)}4  + D_1 \nu( F_1)  + D_2 \nu(F_2) \right)^2
\end{equation} both before and after the Higgsing.
For the bottom row with the frozen $\hat I^*_4$ singularity, we have \begin{equation}
-\frac12\left( - (K+C_1)  \frac{p_1(T)}4 + D_1\nu(F_1)  + D_2 \nu( F_2 )  \right)^2
\end{equation} where we used the fact that the frozen divisor is $C_1$.
After the Higgsing, the Green--Schwarz contribution is \begin{equation}
-\frac12\left( -(K+C_1) \frac{p_1(T)}4 + 2D_1 \nu(F_1)  + D_2 \nu( F_2)  \right)^2
\end{equation}  where the factor in front of $D_1$ is due to the embedding index of $\so_{n} \subset \su_{n}$.\footnote{In particular it explains the superficially funny-looking $\eta_{\mathrm{O}8^+}$ in \cite[(3.23)]{Apruzzi:2017nck}}.
It is a straightforward exercise to show that these Green--Schwarz contributions correctly cancel the gauge squared and the gauge-gravity part of the one-loop anomalies.

The construction discussed here gives a first indication of how the
classification results of \cite{Heckman:2013pva,Heckman:2015bfa} need to be
modified in order to include frozen branes.  We leave a thorough consideration
of the effect of frozen branes on this classification to future work.

\section{Compact models} % (fold)
\label{sec:comp}

In this section we discuss some compact models with O7$_+$-planes in  F-theory language. They are obtained from very classic F-theory models by \emph{flipping} some of the O7$_-$ to O7$_+$.
Our current understanding of the compact models is rather incomplete.
In this paper we will be content with presenting some of the consistent assignments of gauge algebras and hypermultiplet matter content, leaving systematic studies in the future.

\subsection{The $\bF_{-4}$ model and its flip}
\paragraph{Without frozen 7-brane:}

\newcommand{\eff}{\Phi}

\begin{figure}[ht]
	\centering
		\subfigure[\label{fig:F-4}]{\includegraphics[height=3.5cm]{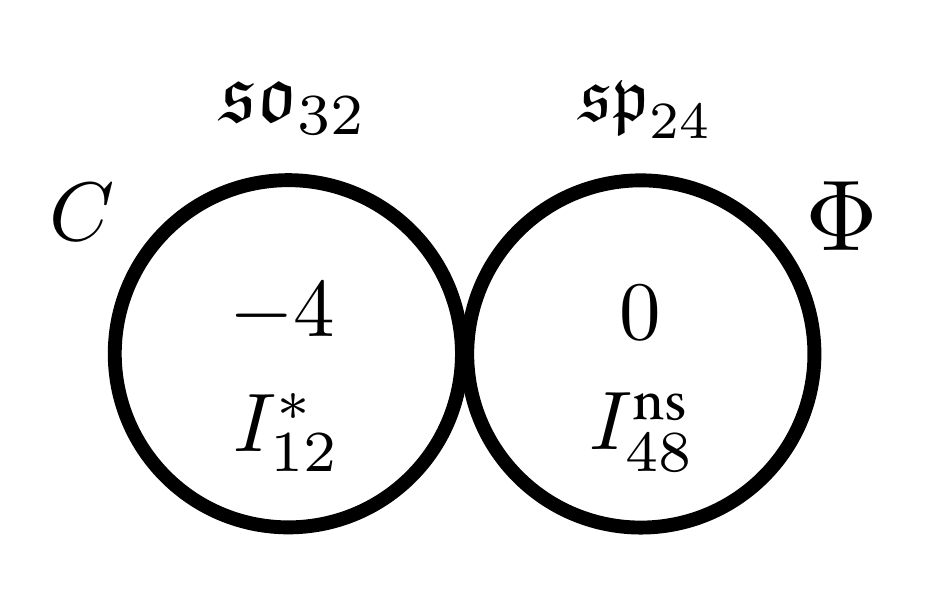}}\qquad
	\subfigure[\label{fig:F-4-flipped}]{\includegraphics[height=3.5cm]{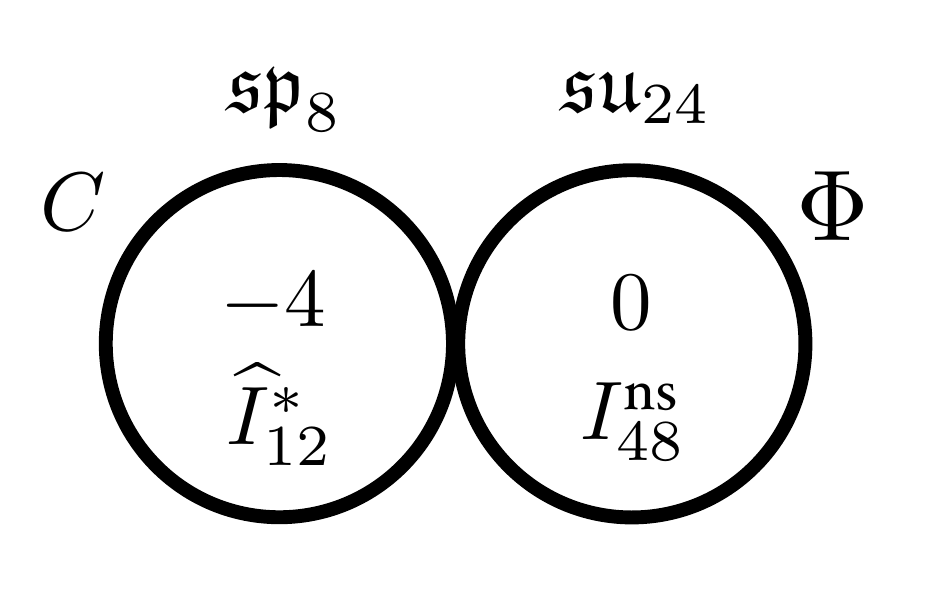}}
	\caption{In \subref{fig:F-4}, the compact model of \cite{Aspinwall:1996vc} on the Hirzebruch surface $\bF_{-4}$. In \subref{fig:F-4-flipped}, a frozen version.}
	\label{fig:F-4s}
\end{figure}

Aspinwall and Gross considered the following model \cite{Aspinwall:1996vc}:
the F-theory base  is the Hirzebruch surface $\bF_{-4}$,
which is a $\CP^1$ bundle over $\CP^1$
such that the base is a $-4$ curve.
We have the $I^*_{12}$ singularity along the $-4$ curve $C$ 
and a fiber $\eff$ hosts an $I_{48}^\text{ns}$ singularity; see Fig.~\ref{fig:F-4}.

This model has the following massless matter content: \begin{itemize}
\item $\mathfrak{so}_{32}$ on $C$ and $\mathfrak{sp}_{24}$ on $\eff$,
\item a half-hypermultiplet in $\rep{32}\otimes \rep{48}$, 
\item a hypermultiplet in $\wedge^2 \rep{48}$, together with
\item one supergravity multiplet, one tensor multiplet and 20 neutral hypermultiplets.
\end{itemize}

Let us remind ourselves how this spectrum can be understood in a dual frame.
We start from the heterotic or type I $\so_{32}$ on a K3, realized as an elliptic fibration over $\CP^1$.
The Green--Schwarz mechanism in ten dimensions requires that the instanton number of the gauge bundle over K3 is 24.
To keep the whole $\so_{32}$ gauge algebra unbroken, we use $24$ point-like instantons.
We then collapse the whole 24 instantons to a point.
This is known to generate $\sp_{24}$ on the heterotic side \cite{Witten:1995gx}.
The spectrum as written above can be found perturbatively on the type I side.

Assuming that the elliptic fiber has small area, we perform fibre-wise the duality between heterotic on $T^2$ and F-theory on an elliptically-fibered K3.
This converts the whole to an elliptically-fibered K3 fibered over $\CP^1$.
The $\so_{32}$ gauge algebra is now realized on the base $C$ as the $I^*_{12}$ singularity,
and the point-like instanton is on the fiber $\eff$ as the $I_{48}^\text{ns}$ singularity.

\paragraph{With a frozen 7-brane:}
Now, let us flip $I^*_{12}$ to $\hat I^*_{12}$.
The anomaly cancellation suggests the following matter content: \begin{itemize}
\item $\mathfrak{sp}_{8}$ on $\frac12C$ and $\mathfrak{su}_{24}$ on $2\eff$,
\item a hypermultiplet in $\rep{16}\otimes\rep{24}$, 
\item two hypermultiplets in $\wedge^2 \rep{24}$, together with
\item one supergravity multiplet, one tensor multiplet and 20 neutral hypermultiplets.
\end{itemize}
This model is  shown in Fig.~\ref{fig:F-4-flipped}. It can be Higgsed to 
\begin{itemize}
\item $\mathfrak{sp}_{8}$ on $\frac12C$ and $\mathfrak{sp}_{12}$ on $2\eff$,
\item a hypermultiplet in $\rep{16}\otimes\rep{24}$, 
\item a hypermultiplet in $\wedge^2 \rep{24}$, together with
\item one supergravity multiplet, one tensor multiplet and 21 neutral hypermultiplets.
\end{itemize}
Here and below, we mean by the sentence ``a gauge algebra $\fg$ on $D$'' that the gauge divisor associated to $\fg$ is $D$, in the language of Sec.~\ref{sec:anom}.

Let us give a derivation of these spectra, using the same duality as in the unfrozen case shown above.
We again start from  the heterotic or type I $\so_{32}$ on a K3, realized as an elliptic fibration over $\CP^1$, but with the generalized Stiefel--Whitney class of $\Spin(32)/\bZ_2$ being nonzero along the fiber, destroying the vector structure \cite{Witten:1997bs}.
The maximal possible gauge algebra is now $\mathfrak{sp}_{8}$.
We now need a gauge configuration of instanton number 12 on the K3 surface,
since the embedding index of $\mathfrak{sp}_{8}\subset \so_{32}$ is two.
We choose to put all 12 point-like instantons at the same place.

The spectrum can be determined perturbatively using the type I description.\footnote{An analysis after a T-dual along one direction in the $T^2$ without vector structure is given around equation \eqref{EEE} of Appendix~\ref{app:one-flipped}.}
We find that when the point-like instanton is on a generic point, the spectrum is as in the Higgsed case above,
while when it is on a singularity of the form $\bC^2/\bZ_2$, the spectrum is the one before the Higgsing.

To go to the F-theory frame, we perform the fiber-wise duality as before.
This time we use the frozen version reviewed in Appendix~\ref{sec:8d}, which relates
heterotic or type I $\so_{32}$ on $T^2$ without vector structure 
to F-theory on K3 with one frozen singularity.
We now have the $\hat I^*_{12}$ singularity on $C$ and the $I^\text{ns}_{48}$ singularity on $\eff$.
The Higgsing distinguishing the two versions
is related to how the residual part of the discriminant with the $I_1$ type singularity intersects with the fiber $\eff$.

\subsection{The unfrozen $\CP^1\times\CP^1$}
\label{sub:cp1cp1}

The next compact model we consider was first considered at a perturbative level by Bianchi and Sagnotti \cite{Bianchi:1990tb} before the second superstring revolution, during which
these models were revisited by many others, including by Gimon and Polchinski \cite{Gimon:1996rq}. In this subsection we will consider its F-theory realization in the case where all O7-planes are O7$_-$; in section \ref{sub:cp1cp1fr} we will consider what happens by changing one or both of them to O7$_+$.

The model is obtained by considering  the $T^2/\bZ_2\times T^2/\bZ_2$ compactification
in type IIB theory, with O7$_-$ at each $\bZ_2$ singularity, together with 16 mobile D7-branes along the first $T^2/\bZ_2$ 
and another 16 mobile D7-branes along the second $T^2/\bZ_2$.
We give the perturbative derivation of the spectrum of these models in Appendix~\ref{app:branes}. The aim here is to understand the spectrum from the F-theory point of view.\footnote{%
Analyses of unfrozen compact models with conformal matter will also be provided in \cite{HayashiJeffersonKimOhmoriVafaToAppear} by other authors, 
where a detailed analysis of the $\CP^1\times \CP^1$ model with four D7-branes per O7$_-$-plane is given, among others.
}

\paragraph{Conformal matter point:}
Since $T^2/\bZ_2\simeq \CP^1$, we take the F-theory base to be $\CP^1\times \CP^1$.
We pick divisors $C$ and $D$ wrapping each of the $\CP^1$ above.
We let each divisor host an $I^*_{12}$ singularity. At the intersection we expect to have the conformal matter theory (see footnote \ref{foot:Dcm})
 $\bD_{32}$, where $\mathfrak{so}_{32}\times\mathfrak{so}_{32}\subset \mathfrak{so}_{64}$ is gauged. 

Let us see this in more detail. We choose coordinates $([s, t], [u, v])$ on $\CP^1\times \CP^1$, and consider bihomogenous
polynomials.  We want to engineer $I^*_{12}$ along $t=0$ and also along
$v=0$.  Doing so is quite constrained.
The equation defining the elliptic fibration was derived in 
\cite{Aspinwall:1996nk}
but we follow the notation of \cite[Eq.~(42)]{Aspinwall:1997ye}:
\begin{equation}\label{eq:det-bis}
y^2 = x^3 + tvp_{3,3}(s, t, u, v)x^2 + t^8v^8x,
\end{equation}
where $p_{3,3}$ is bihomogeneous of degree $(3, 3)$.  (We shall usually
suppress the variables in writing polynomials such as $p_{3,3}$.)

This equation is not in Weierstrass form.
By completing the cube, we find 
\begin{align}
f&=t^2v^2\left(t^6v^6-\frac13p_{3,3}^2\right),\\
g&=t^3v^3p_{3,3}\left(-\frac13t^6v^6+\frac{2}{27}p_{3,3}^2\right),\\
\Delta&=t^{18}v^{18}(2t^3v^3+p_{3,3})(2t^3v^3-p_{3,3}).
\end{align}
By the Kodaira vanishing criteria,
we indeed see $I^*_{12}$ along $t=0$ and $v=0$.
Therefore we have  $\bD_{32}$ conformal matter at $t=v=0$.

The discriminant has components $t=0$ and $v=0$ along which $I^*_{12}$
fibers are located, as well as two components 
\begin{equation}
2t^3v^3=\pm p_{3,3}
\end{equation} 
which comprise the ``residual discriminant'' (the part with no gauge
algebra or type II enhancement).  We denote these by $\Delta_\pm$,
and note that the defining equation of each has bidegree $(3,3)$.
Both of these components
 intersect with $t=0$ at the three points $t=p_{3,3}=0$, 
and similarly intersect with $v=0$ at the three points $v=p_{3,3}=0$. 
The multiplicities of $f$, $g$, and $\Delta$ at all six intersection
points are $(4,6,20)$ so there is conformal matter at those points
as well.

\begin{figure}[ht]
	\centering
		\includegraphics[height=3in]{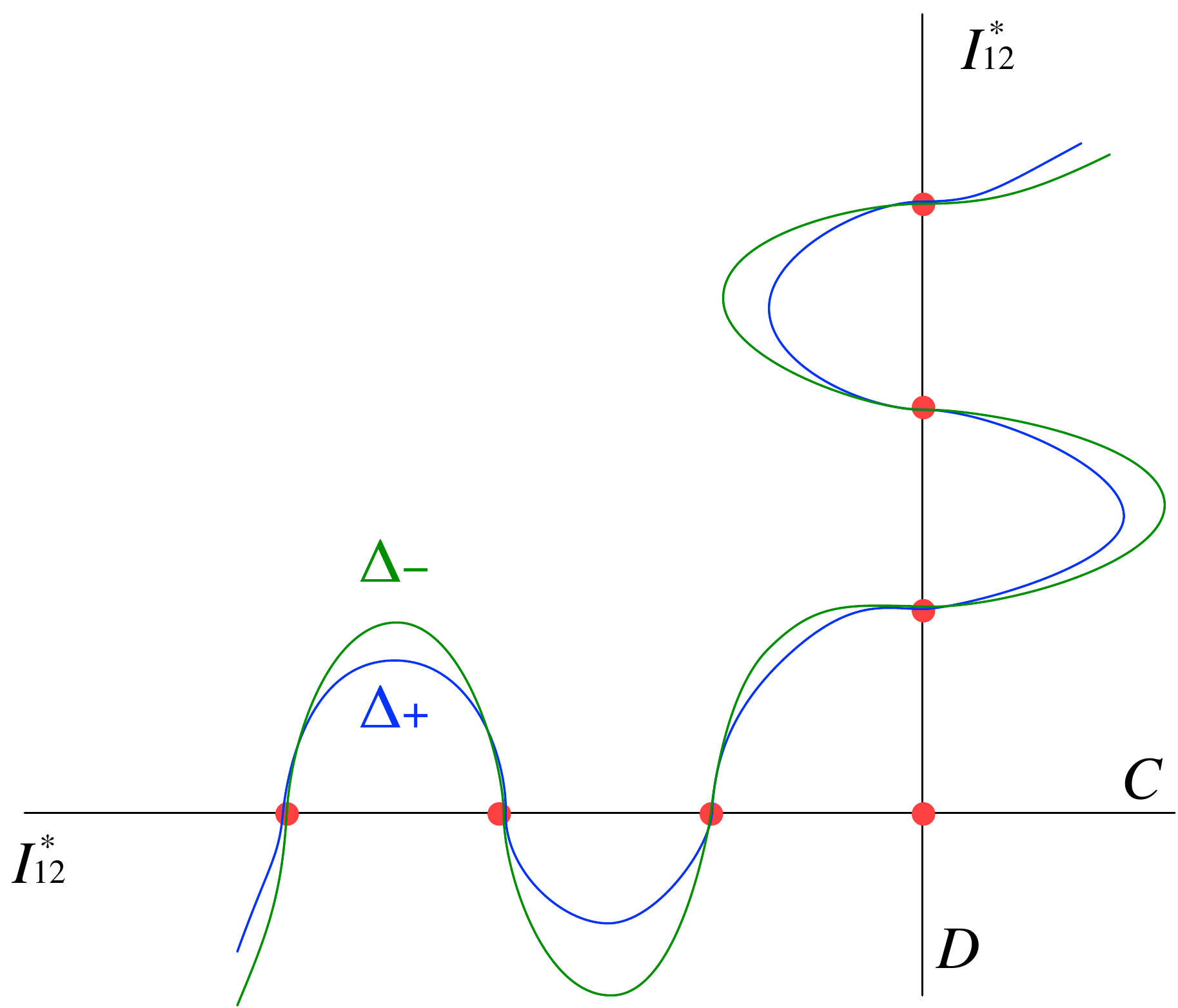}
	\caption{The model with conformal matter points. Note that $\Delta_+$ and $\Delta_-$ have a third order contact with each other at their intersections with $C$ or $D$.}
	\label{fig:No_blowup}
\end{figure}

We can roughly see how this F-theory setup corresponds to the perturbative model reviewed in App.~\ref{app:GP}. The $I^*_{12}$ curves are the counterpart of the O7$_-$-planes with 16 D7s on top. The residual discriminant corresponds to the O7$_-$-planes without D7s. As is customary, such planes are realized in F-theory by a pair of $I_1$ curves.

We illustrate this initial model, which is at its transition point
between tensor and Higgs branches,
in Fig.~\ref{fig:No_blowup}.  We have marked the seven CFT points
with red dots.  The curves $\Delta_+$ and $\Delta_-$
have a third order contact with each other at each point of intersection,
and also pass transversally
through one of the $I_{12}^*$ curves (labeled $C$ and $D$) at each such point.
Key mathematical features not found in the illustration include the
intersection data: $C^2=D^2=0$, $C\cdot D=1$, $K\cdot C = K \cdot D = -2$,
$\Delta_+^2 = \Delta_-^2 = \Delta_+\cdot \Delta_- = 18$, and 
$K\cdot \Delta_+=K\cdot \Delta_-=-12$.  It follows that $\Delta_+$ and 
$\Delta_-$ each have genus $4$.
The matter content is then \begin{itemize}
\item $\mathfrak{so}_{32}$ on $C$ and $\mathfrak{so}_{32}'$ on $D$,
\item the conformal matter $\bD_{32}$ gauged by 
$\mathfrak{so}_{32}\times\mathfrak{so}_{32}'\subset \mathfrak{so}_{64}$
 \item three copies of $\bD_{16}$ gauged by $\mathfrak{so}_{32}$,
 \item three copies of $\bD_{16}$ gauged by $\mathfrak{so}_{32}'$,
 \item one supergravity multiplet, one tensor multiplet and 13 neutral hypermultiplets.
\end{itemize}

\paragraph{Higgsed model:}
To obtain the standard perturbative massless
spectrum of the model, we can Higgs the conformal matter theories. From (\ref{eq:I*I*}) and Fig.~\ref{fig:O6higgs-nong}, we see that at the collision point of two $I^*$ curves there is a hypermultiplet which,
when activated, breaks the global symmetry from\footnote{%
The $\mathfrak{u}_1$ part of both $\mathfrak{u}_{16}$ are known to get Higgsed by the Green-Schwarz mechanism, eating one neutral hypermultiplet each, and becoming massive \cite[Sec.~2]{Berkooz:1996iz}.
Here we follow the older perturbative string terminology. 
}
$\mathfrak{so}_{2n+8}\oplus \mathfrak{so}_{2n+8}'$ to 
$\mathfrak{su}_{n+4}\oplus \mathfrak{su}_{n+4}'$.
 
\begin{figure}[ht]
	\centering
	\subfigure[\label{fig:Blowup}]{\includegraphics[width=7cm]{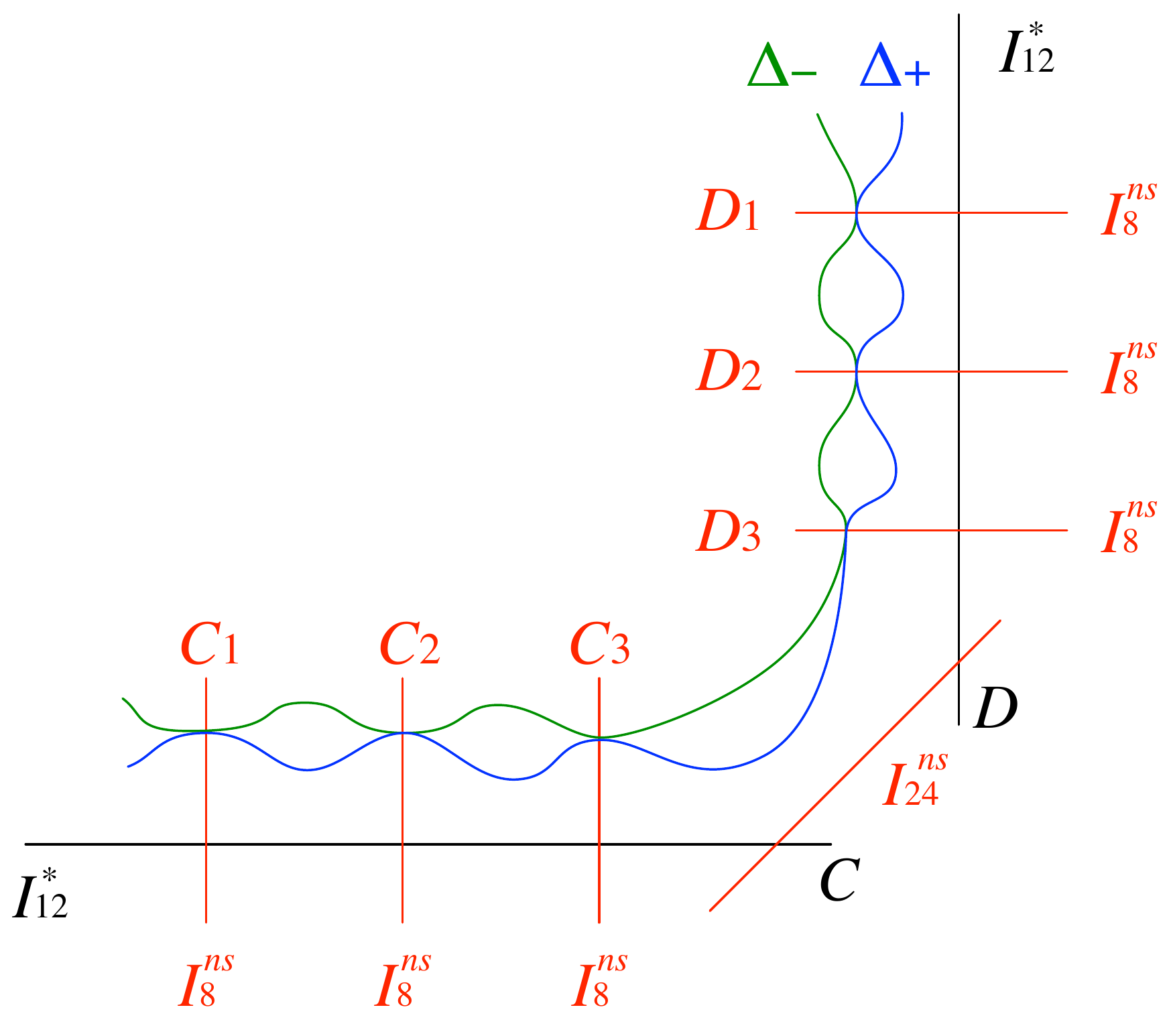}}
	\subfigure[\label{fig:FGP}]{\includegraphics[width=8cm]{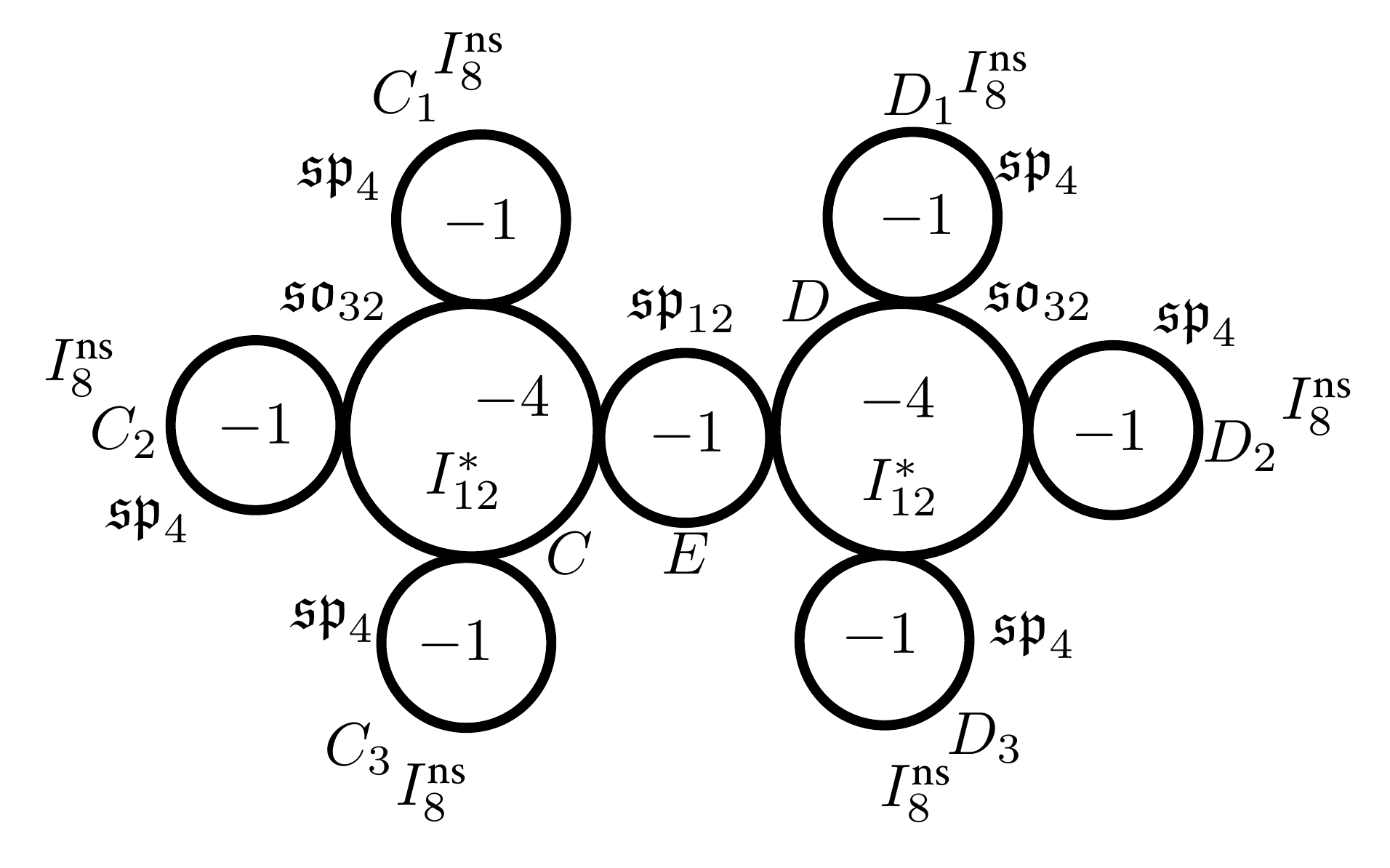}}
	\caption{An F-theory description of the tensor branch of the perturbative model with two O$7_-$ \cite{Bianchi:1990tb,Gimon:1996rq}. In \subref{fig:Blowup} a traditional depiction more similar to the one in Fig.~\ref{fig:No_blowup} is given; 
note that $\Delta_+$ and $\Delta_-$ are tangent at their intersections with $C_i$ and $D_i$.
In \subref{fig:FGP} is a depiction more similar the other figures in the paper, which does not include the residual discriminant.  Upon shrinking $E$ as well as $C_1$, $C_2$, $C_3$, $D_1$, $D_2$, $D_3$, we obtain Fig.~\ref{fig:No_blowup}; by Higgsing the resulting conformal matter theories,
one recover the original perturbative model.}
	\label{fig:FGP-both}
\end{figure}

Doing this for the conformal matter theories in the model of Fig.~\ref{fig:FGP}, one reproduces the perturbative spectrum:
\begin{itemize}
\item $\mathfrak{u}_{16}$ on $C$ and $\mathfrak{u}_{16}'$ on $D$,
\item a hypermultiplet in $\rep{16}\otimes\rep{16}'$,
\item two hypermultiplets in $\wedge^2\rep{16}$,
\item two hypermultiplets in $\wedge^2\rep{16}'$, 
\item one supergravity multiplet, one tensor multiplet and 20 neutral hypermultiplets,
\end{itemize}
as can be found in the original papers \cite{Bianchi:1990tb,Gimon:1996rq,Berkooz:1996iz}, and reviewed in App.~\ref{app:GP}.
The F-theory interpretation of this Higgsed spectrum was given in \cite{Sen:1996vd,Sen:1997kw};
this study eventually led to a refined understanding of the relation between F-theory and O7$_-$ \cite{Sen:1997gv}.\footnote{Let us note that the T-duality between Type IIB on $T^2/\bZ_2\times T^2/\bZ_2$, which we used here,
and Type I on $T^4/\bZ_2$, as originally considered, was first discussed in \cite{Sen:1997pm}.
Let us also mention that when each O7$_-$ has four D7s on top of it,
then the perturbative orientifold construction can be subtly modified so that the system is slightly on the tensor branch side, rather than on the Higgs branch side, of the conformal point, as noticed early in the study of orientifolds \cite{Blum:1996hs,Dabholkar:1996ka}.}

So far we used the process of Fig.~\ref{fig:O6higgs-nong}, which is non-geometric in F-theory, to realize the Higgsed spectrum.
We also expect that giving vevs to other scalars in the same hypermultiplet would have the same effect.
We thus seek a geometric deformation
of the original equation \eqref{eq:det-bis} in which the Kodaira fibers
$I_{12}^*$ become Kodaira fibers $I_{16}^s$.  The deformation involves
a new polynomial $q_{2,2}$ of bidegree $(2,2)$ and takes the
form
\begin{equation}\label{eq:det2}
y^2 + \varepsilon\, q_{2,2}(s,t,u,v)xy 
= x^3 + tvp_{3,3}(s, t, u, v)x^2 + t^8v^8x.
\end{equation}
When we complete the square and then complete the cube, we find the
data for Weierstrass form:
\begin{align}
f&=\left(t^8v^8-\frac13(tvp_{3,3}+\frac14\varepsilon q_{2,2}^2)^2\right),\\
g&=(tvp_{3,3}+\frac14\varepsilon q_{2,2}^2)
\left(-\frac13t^8v^8+\frac{2}{27}(tvp_{3,3}+\frac14\varepsilon q_{2,2}^2)^2\right),\\
\Delta&=t^{16}v^{16}(2t^4v^4+tvp_{3,3}+\frac14\varepsilon q_{2,2}^2)
(2t^4v^4-tvp_{3,3}-\frac14\varepsilon q_{2,2}^2).
\end{align}
This is Kodaira type $I_{16}^s$ on each curve; it is split because 
$(g/f)|_{t=0} = -\frac1{18}\varepsilon q_{2,2}^2|_{t=0}$ is a perfect square,
and likewise for $v=0$.

This time, the intersection of $t=0$ with the residual discriminant
is at two points $t=q_{2,2}$, each of which has multiplicities of
$(f,g,\Delta)$ being $(2,3,18)$.  Such an intersection point is associated
to a matter representation $\Lambda^2$ rather than to conformal matter, 
so the corresponding points should not be blown up.
The same is true of the two points $v=q_{2,2}=0$.

\begin{figure}[ht]
	\centering
		\includegraphics[height=3in]{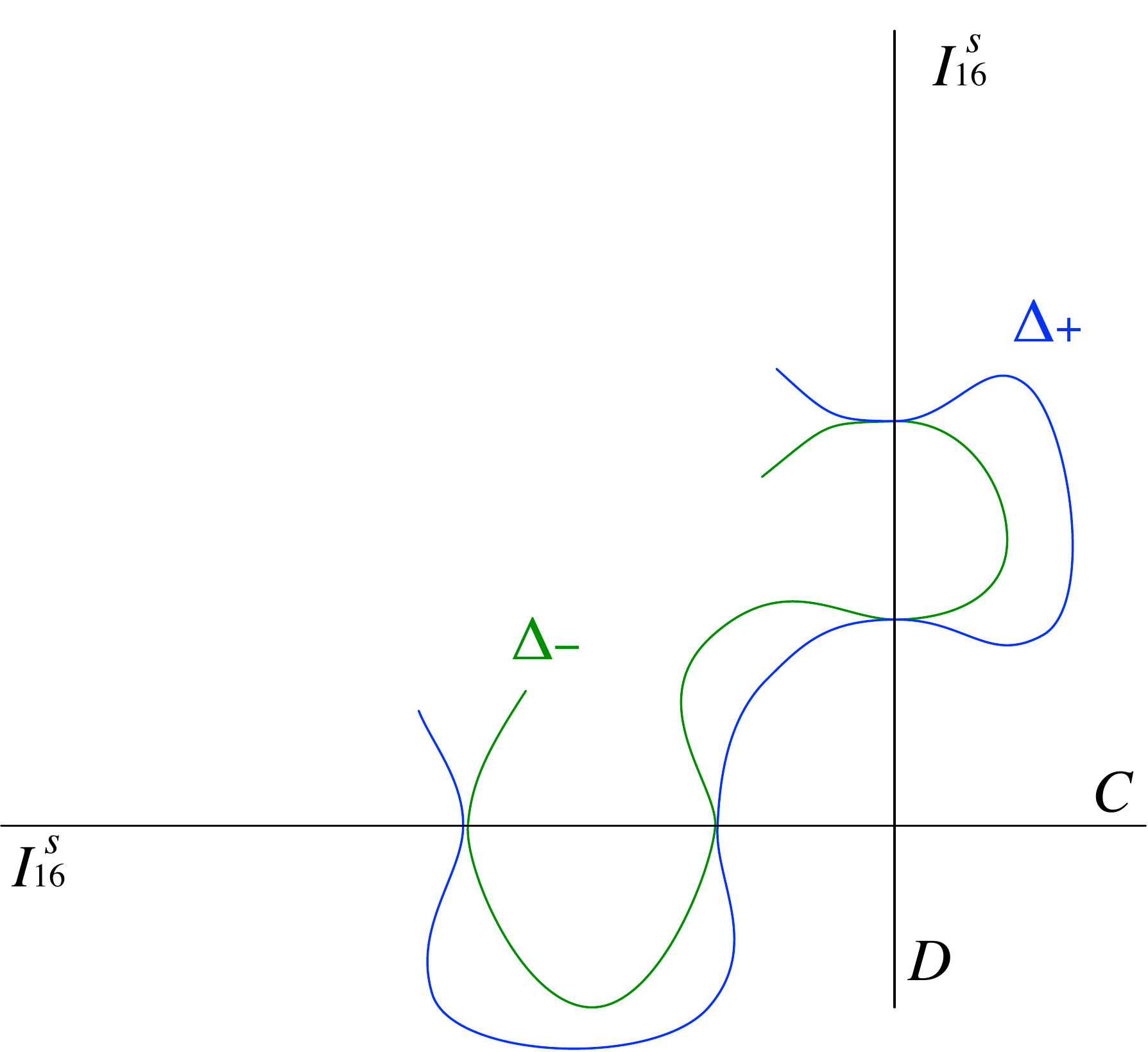}
	\caption{A geometrical realization of the Higgsed model. Note that $\Delta_+$ and $\Delta_-$ have a second order contact with each other at their intersections with $C$ or $D$.}
	\label{fig:First_Higgsing}
\end{figure}

Similarly, the intersection of $t=0$ with $u=0$ is ordinary bifundamental
matter, and this point should not be blown up either.  
The geometry is illustrated in Fig.~\ref{fig:First_Higgsing}. This model reproduces the perturbative spectrum again, but this time with a geometrical Higgsing.

\paragraph{Tensor deformation:}

We will now consider the tensor branch deformation of the model with conformal matter in Fig.~\ref{fig:No_blowup}.

The blowup of the collision point $t=v=0$
is straightforward and produces an exceptional curve $E$ along which
the Kodaira type is $I_{24}^{ns}$.

Let us study the intersection points of $p_{3,3}=0$ with $t=0$ in more detail.
By a change of coordinates, we may
locate one of the intersection points at $t=u=0$.
In that case, we can write $p_{3,3}=u\hat p_{3,2}+t \tilde p_{2,3}$.
Multiplicities of $f$, $g$, and $\Delta$ at $t=u=0$ are easily seen to be 4, 6, and 20 so we have a conformal fixed point and we need to blow up.
To perform the blowup, we work in the affine coordinate chart $v = s = 1$. 
In one coordinate chart of the blowup, we have $t_1 = t$, $u_1 = u/t$, and the Weierstrass coefficients and discriminant become
\begin{align}
f_1 &= t^4_1-\frac13(u_1\hat p + \tilde p )^2,\\
g_1&=(u_1\hat p +\tilde p ) \left(-\frac13t^4_1+\frac2{27}(u_1\hat p +\tilde p )^2\right), \\
\Delta_1 &=t_1^8(2t_1^2+u_1\hat p+\tilde p)(2t_1^2-u_1\hat p-\tilde p).
\end{align}
The exceptional divisor $t_1 = 0$ supports an $I_8$ fiber, since the orders of vanishing are $(0, 0, 8)$,  and there is monodromy: the usual branch divisor $(g_1/f_1)|_{t_1}=0$ vanishes at $u_1 = t_1 = 0$ in this chart and has a single order of vanishing. 
Thus, this is $I_8^\text{ns}$ and the gauge algebra is $\sp_4$. No matter is visible in this chart.
Note that this branch point is the point at which the residual discriminant meets the exceptional divisor. The multiplicities at this point are 2, 3, 10 which is consistent with the enhancement from $A_7$ to $D_8$ which is expected at such a point.
In the other coordinate chart of the blowup, we have $t_2 = t/u, u_2 = u$. 
The Weierstrass coefficients and discriminant become
\begin{align}
 f_2 &= t_2^2 (u_2^4-\frac13(\hat p + t_2\tilde p )^2),\\
 g_2 &=t_2^3 (\hat p + t_2\tilde p ) \left(-\frac13 u_2^4  + \frac2{27}(\hat p + t_2\tilde p )^2 \right),\\
 \Delta_2&=t^{18}_2 u_2^8 (2u_2^2+\hat p + t_2\tilde p)(2u_2^2-\hat p - t_2\tilde p)
\end{align}
and we indeed see the exceptional divisor $u_2=0$ meeting the original $I^*_{12}$ at $t_2=0$.
This intersection point also provides the second branch point defining the monodromy.

This same analysis applies at all six points $t=p_{3,3}=0$ and $v=p_{3,3}=0$
so six additional blowups need to be done.  All in all, we have blown up
$\CP^1\times \CP^1$ at seven points, and we obtain a model with no 
conformal matter and with eight tensor multiplets.  

This model is illustrated in Fig.~\ref{fig:Blowup} and Fig.~\ref{fig:FGP}.  
The curves
$\Delta_+$ and $\Delta_-$ are now simply tangent at each of their points
of intersection, which occur at a point of one of the new exceptional
divisors $C_j$ or $D_j$.  The intersection data this time are:
$C^2=D^2=-4$, $C_j^2=D_j^2=E^2=-1$, $\Delta_+^2=\Delta_-^2=12$;
$K\cdot C = K\cdot D = 2$, $K\cdot C_j = K\cdot D_j = K\cdot E = -1$,
$K\cdot \Delta_+ = K\cdot \Delta_- = -6$.  Note that because of the
tangencies we now have $\Delta_+\cdot\Delta_-=12$.

The massless matter content is:
\begin{itemize}
\item $\mathfrak{so}_{32}$ on $C$, $\mathfrak{sp}_{12}$ on $E$,
 $\mathfrak{so}_{32}'$ on $D$, and a copy of $\mathfrak{sp}_4$ on
each $C_j$ and on each $D_j$,
\item a half-hypermultiplet in $\rep{32}\otimes \rep{24}$,
 a half-hypermultiplet in $\rep{24}\otimes \rep{32}'$, three
half-hypermultiplets in $\rep{32}\otimes \rep{8}$ (corresponding to
$(C,C_j)$) and three half-hypermultiplets in $\rep{8}\otimes\rep{32}'$
(corresponding to $(D_j,D)$),
together with
\item one supergravity multiplet, eight tensor multiplets and 13 neutral hypermultiplets.
\end{itemize}

\subsection{Frozen $\CP^1\times\CP^1$ models} % (fold)
\label{sub:cp1cp1fr}

We will now consider what happens in the  $\CP^1\times\CP^1$ model of section \ref{sub:cp1cp1} when one changes the type of one or both  O7$_-$ to O7$_+$.

\paragraph{With two frozen seven-branes:}

Let us first consider what happens when one changes both O7$_-$ to O7$_+$.

In  the original geometry without blowups in Fig.~\ref{fig:No_blowup}, the two $I^*_{12}$ curves are now changed into $\widehat I^*_{12}$. Since the residual discriminant represents O7$_-$-planes, we do not expect conformal matter at the its intersection with the $\widehat I^*_{12}$ curves, which represent O7$_+$-planes. We have not analyzed this situation before, but we expect it to be similar to the one in Fig.~\ref{fig:smooth-O7} and (\ref{eq:I*hI*}), thus with no conformal matter and a tensor. 

We do not venture to guess the field theory content at this point. It is easier to follow a tensor deformation by blowing up all curves. The result lives again on the geometry illustrated in Fig.~\ref{fig:FGP-both}, $I^*_{12} \to \widehat I^*_{12}$. A possible choice of gauge divisors that cancels all gauge anomalies gives the following model:
\begin{itemize}
\item $(\mathfrak{sp}_{2})_{1,2,3,4}$, supported on $C+ C_1 + C_2+E$, $C+C_1 + C_2 +2C_3+E$, $D + D_1 +D_2 + E$, $D +  D_1 + D_2  + 2D_3+E$ respectively,
\item $\mathfrak{so}_8$ supported on $2E$,
\item hypermultiplets in $\rep{4}_i\otimes \rep{4}_j$ for $i<j$,
\item one supergravity multiplet, 8 tensor multiplets, and 13 neutral hypermultiplets.
\end{itemize}

The $\mathfrak{sp}$ groups living on the $\widehat I^*_{12}$ curves have been shared with the $I^{\rm ns}_8$ curves, just like in Fig.~\ref{fig:smooth-O7}, where O7-planes of different types meet. The $\mathfrak{so}_8$ has appeared on the $E$ curve just like in a usual tensor--Higgs transition. Indeed the field content is related to the one for the perturbative model in App.~\ref{app:two-flipped} by such a transition. 

\paragraph{With one frozen seven-brane:}

We now consider what happens if only one of the $I^*_{12}$ is changed to $\widehat I^*_{12}$ (say $D$). 

As in the previous case, we don't try to write the field content at the singular locus; we instead follow the tensor deformations. 
Here we encountered a problem: we have not found a credible model that cancels all anomalies after blowing up all  those singular points, perhaps because of some global constraint.

What we were able to achieve is the following. 
We consider the geometry in a situation intermediate between Fig.~\ref{fig:First_Higgsing} and Fig.~\ref{fig:Blowup},
namely, we tune the Kodaira type on $D$ to be $\widehat I^*_{12}$,
but we deform the Kodaira type on $C$ to be $I_{16}$.
Then, the intersections of $D$  with others are still conformal points and need to be blown up,
but the intersections of $C$ with others are smooth.
We then only need to blow up the intersection of $C$ and $D$, and the intersections of the residual discriminant and $D$.

\begin{figure}[ht]
	\centering
		\includegraphics[width=7cm]{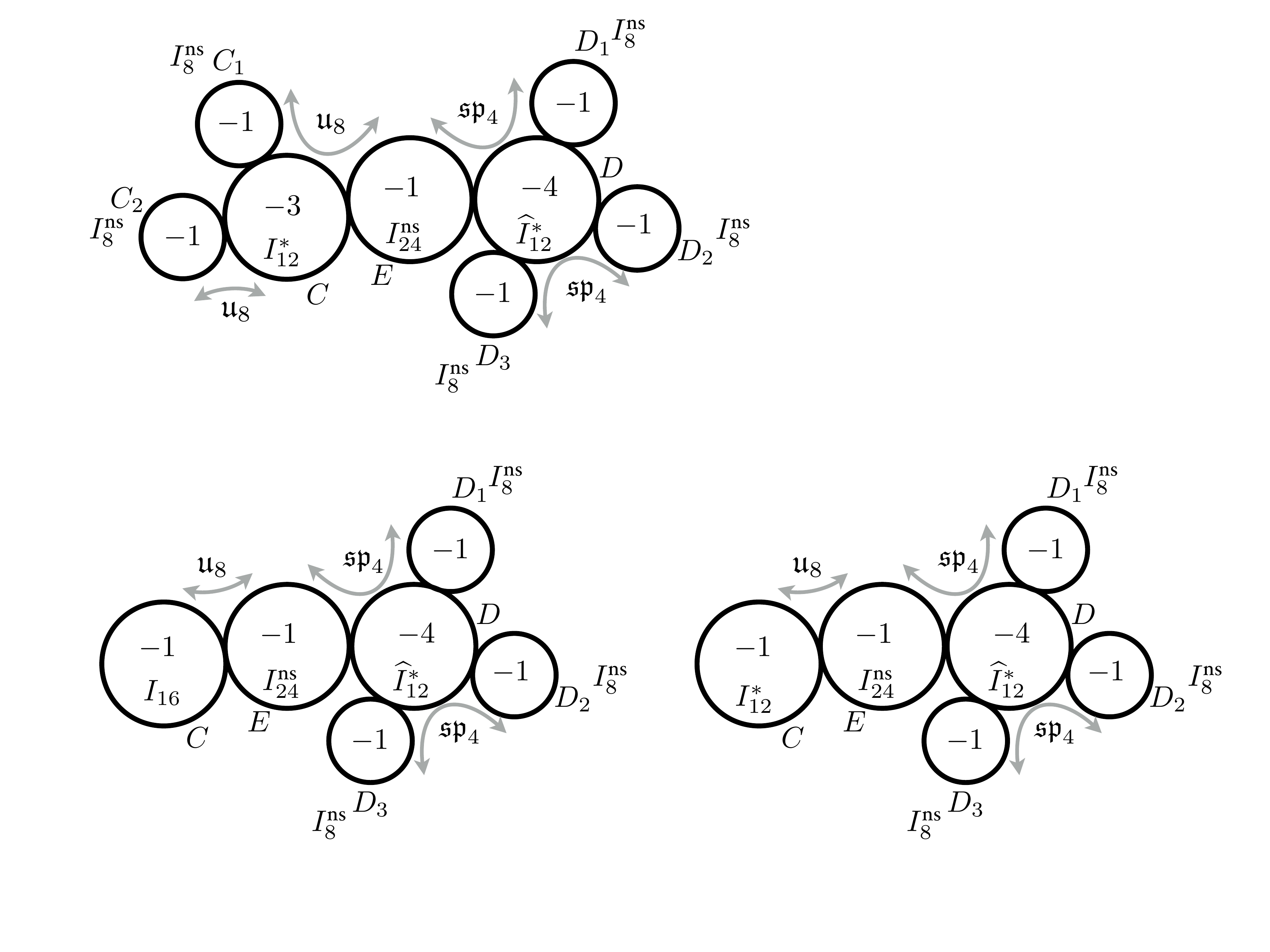}
	\caption{Tensor branch models obtained by changing in the $\mathbb{CP}^1\times \mathbb{CP}^1$ model  one of the $I^*_{12}$ curves to $\widehat I^*_{12}$. 
	As discussed in the main text, we were not able to find a consistent assignment when we have $I^*_{12}$ on $C$ instead.}
	\label{fig:ones-flipped}
\end{figure}

This gives us the geometry in Fig.~\ref{fig:ones-flipped}, and the following spectrum:\footnote{%
As in the previous footnote, we expect the $\mathfrak{u}_1$ part to become massive via the Green-Schwarz mechanism, eating a neutral hypermultiplet.
}
\begin{itemize}
\item $\mathfrak{u}_{8}$ on $2(C+E)$,
\item $\mathfrak{sp}_{4}'$ on $E+ \frac12D+D_1$ and $\mathfrak{sp}_{4}''$ on 
$\frac12D+D_2 + D_3$,
\item hypermultiplets in $\rep{8}\otimes\rep{8}'$, and  $\rep{8}'\otimes \rep{8}''$,  $\rep{8}''\otimes\rep{8}$,
\item two hypermultiplets in $\wedge^2 \rep{8}$,
\item one supergravity multiplet, 5 tensor multiplets and 16 neutral hypermultiplets.
\end{itemize}

This is exactly the spectrum of the perturbative model described in appendix \ref{app:one-flipped}. 
Note that we chose to blow up the intersections of $D$ with other discriminant loci,
while we decided to deform other intersections.
In other words, we chose to go to the tensor branch side for the conformal points on $D$
whereas we went to the Higgs branch side for the conformal points on $C$.
This is in accord with our analysis in Sec.~\ref{ssub:i*i*},
since the perturbative construction naturally gives a tensor at an O7$_+$-O7$_-$ intersection
whereas it gives a hyper at an O7$_-$-O7$_-$ intersection.

% subsection cp1cp1fr (end)

% section comp (end)

\section*{Acknowledgements}
The authors would like to thank O.~Bergman and W.~Taylor for interesting discussions,
and T.~Rudelius for pointing out typos in a previous version of the paper.

The work of L.B.  is partially supported by the Perimeter Institute for Theoretical Physics. Research at Perimeter Institute is supported by the Government of Canada through Industry Canada and by the Province of Ontario through the Ministry of Economic Development and Innovation.

The work of D.R.M.~was supported in part by National Science Foundation grant PHY-1620842 (USA) and by the Centre National de la Recherche Scientifique (France).

Y.T.~is partially supported  by JSPS KAKENHI Grant-in-Aid (Wakate-A), No.17H04837 
and JSPS KAKENHI Grant-in-Aid (Kiban-S), No.16H06335,
and also by WPI Initiative, MEXT, Japan at IPMU, the University of Tokyo.

A.T.~is supported in part by INFN. During this work, his research was also supported by the MIUR-FIRB grant RBFR10QS5J ``String Theory and Fundamental Interactions'', and by the European Research Council under the European Union's Seventh Framework Program (FP/2007-2013) -- ERC Grant Agreement n. 307286 (XD-STRING).

\appendix

\section{Dimension eight}
\label{sec:8d}
There are three families \cite{Witten:1997bs} of vacua with 16 supercharges in dimension eight.
The standard one has gauge algebra of rank 20,
the next one has gauge algebra of rank 12,
and the final one has gauge algebra of rank 4.

The rank-12 case was  found in the perturbative type I frame by Bianchi, Prasidi and Sagnotti in \cite{Bianchi:1991eu} in 1992 and then
in the context of heterotic string by Chaudhuri, Hockney and Lykken in \cite{Chaudhuri:1995fk} in 1995;
the latter construction is known under the name of the CHL string.
An easy generalization of either construction leads to the rank-4 case.
The moduli space of these systems and the possible enhancements of gauge algebras are studied in detail in \cite{Mikhailov:1998si}.

In this appendix, we give an  F-theory description of three cases:
they are models on elliptically-fibered K3 with $0$, $1$, or $2$ frozen seven-branes.

\subsection{IIB with seven-branes}
Let us start by the perturbative IIB setup on the orientifold $T^2/\bZ_2$.
We can either have zero $\Op7$, one $\Op7$ or two $\Op7$: \begin{equation}
\inc{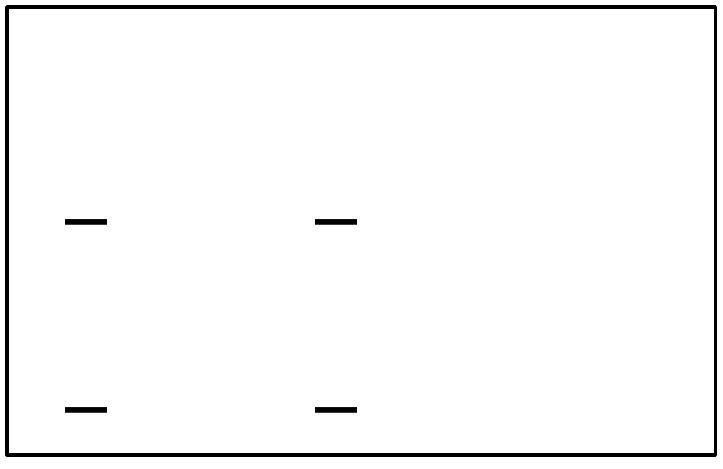},\quad
\inc{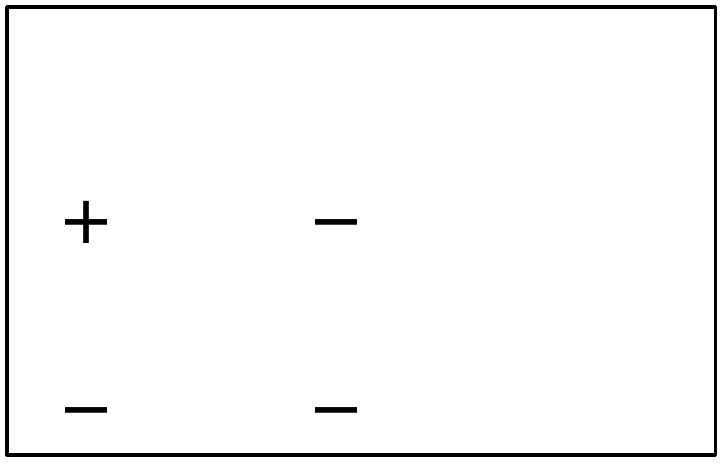},\quad
\inc{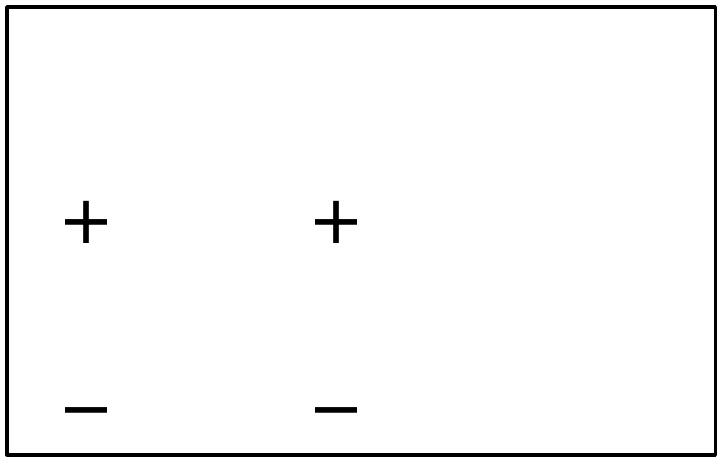}
\end{equation}
with 16, 8 or 0 D7-branes, respectively.
The first one, under T-duality, maps to 2 $\Om8$ in type IIA, and then 1 $\Om9$ in type IIB.
The last one, under T-duality, maps to $\Om8$ and $\Op8$, or to a shift-orientifold of type IIA, and then a shift-orientifold of type IIB, without any D9-brane.

The second one is more peculiar.
One T-duality should combine a pair of two $\Om7$s to $\Om8$, while the other pair of $\Om7$ and $\Op7$ to a shift orientifold.
The resulting geometry is shown below:
\begin{equation}
\inc{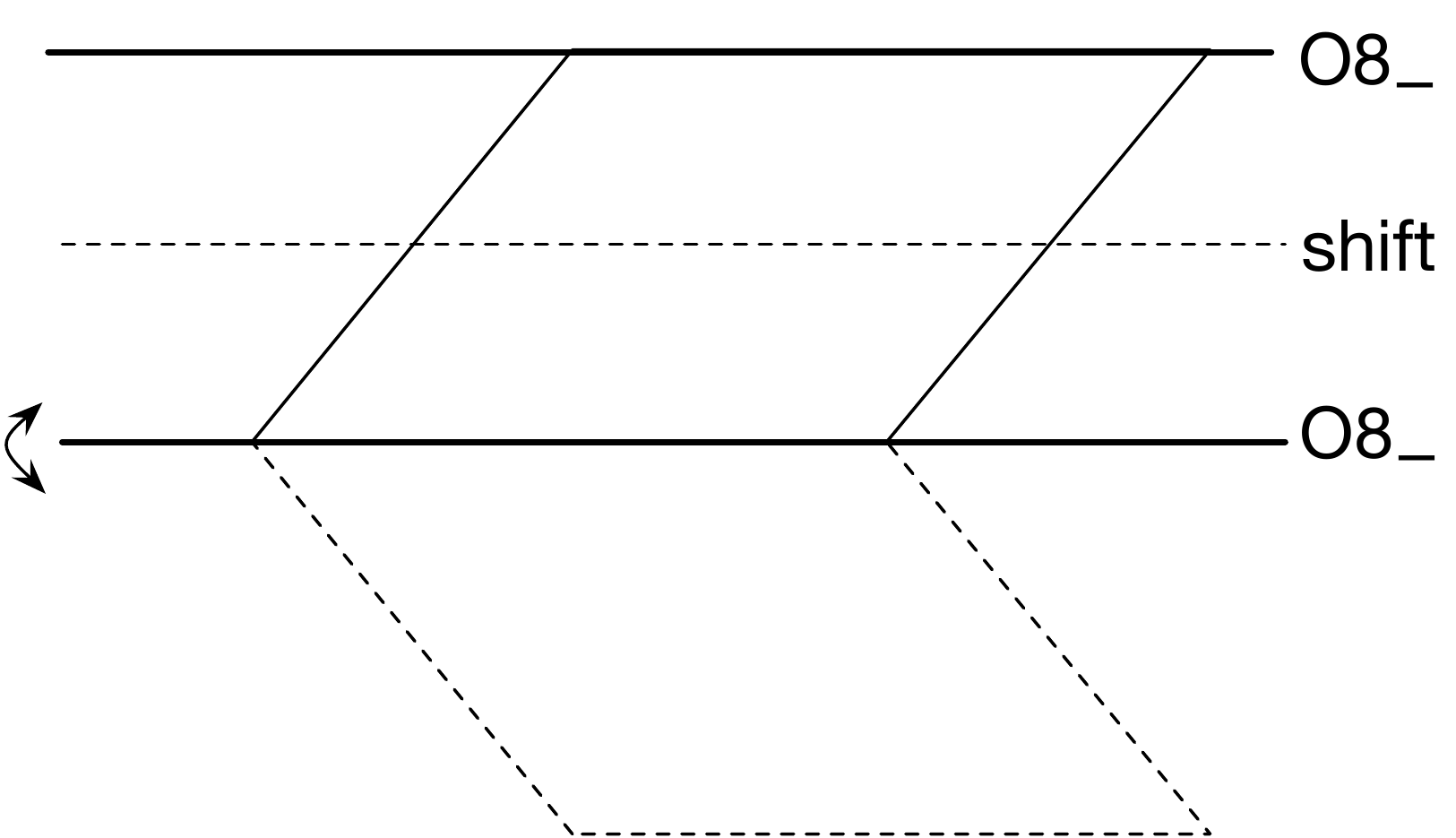}.
\end{equation}
Namely, we consider a $T^2$ whose  complex structure modulus is of the form $\tau\in \mfrac12{+}i\mathbb{R}$, and take the $\bZ_2$ flip along the horizontal axis.
Then we have just one $\Om8$ locus and a shift-orientifold locus.
Another T-duality leads to the $\Spin(32)/\bZ_2$ bundle without vector structure \cite{Witten:1997bs}.

\subsection{F-theory interpretation}
The F-theory representation of the rank-20 case is the standard F-theory compactification on the elliptically-fibered K3 surface.

The F-theory representation of the rank-12 case is given by an elliptic K3 compactification with
a single frozen seven-brane.\footnote{%
Note that this is a substantially different description than the
ones proposed in  \cite{Bershadsky:1998vn} and 
\cite{Berglund:1998va}, where a torsion flux on the base $\CP^1$ was proposed.
It is possible that they are all dual descriptions.}
We use projective coordinates $[z,w]$ on $\CP^1$
and locate the frozen brane at $z=0$:
\begin{equation}
y^2 = x^3 + u_3(z,w)z x^2 + v_4(z,w)z^4x + w_5(z,w)z^7.
\end{equation}
Here we have used the ``Tate form'' \cite{Bershadsky:1996nh,Katz:2011qp}
to present the equation, which involves arbitrary homogeneous polynomials 
$u_3$, $v_4$, and $w_5$ of the labeled degrees.
By a change of variables, the equation can be put 
into Weierstrass form:
\begin{equation}
y^2 = \hat x^3 
+ z^2(- \frac13 u_3^3+z^2v_4 )\hat x +
z^3(\frac2{27}u_3^3- \frac13 z^2 u_3 v_4+z^4 w_5),
\end{equation}
from which we can read off the equation of the discriminant locus
\begin{equation}
\Delta = z^{10}\left(
4u_3^3w_5 - u_3^2v_4^2 - 18z^2 u_3v_4w_5
+ 4z^2v_4^3 + 27 z^4w_5^2 .
\right)
\end{equation}

Generically, in addition to the frozen seven-brane of type $\widehat I_4^*$ 
at $z=0$, which makes no contribution to enhanced gauge symmetry, there
are 14 additional zeros of the discriminant, which correspond to 14 
seven-branes of type $I_1$ (i.e., 14 individual D7-branes)
also contributing no enhanced gauge symmetry.
Tuning the coefficients can lead to enhanced gauge symmetry.

The brane counting becomes clear if we explicitly include a Kodaira fiber
of type $I_0^*$ supporting an $\mathfrak{so}_{8}$ gauge algebra:
this ``uses up'' 6 of the 14 D7-branes, but can be interpreted as an
O7$_-$-plane on top of a stack of 4 D7-branes, which is the quantum splitting
of the O7$_-$-plane \cite{Sen:1996vd}.
Then eight mobile D7-branes remain.

The F-theory representation of the rank-4  8D vacuum
with 16 supercharges involves two frozen seven-branes, which we can
locate at $z=0$ and $w=0$, respectively.
 The equation for these models (in Tate form) is
\begin{equation}
y^2 = x^3 + u_2(z,w)zw x^2 + v_0(z,w)z^4w^4x 
\end{equation}
with frozen brane-locus $\delta=zw$.
 Note that $v_0(z,w)$ is constant,
and the $x^0$ term in the equation vanishes
due to degree considerations.
 This implies that $(x,y)=(0,0)$ is a section
which has order $2$ in the Mordell--Weil group, and suggests
a subtle modification of the F-theory gauge group.\footnote{We are assuming
here that the torsion in the Mordell--Weil group is calculated for
frozen F-theory models in the same way it is calculated for
conventional F-theory models \cite{Aspinwall:1998xj}.
We leave detailed investigations of this for the future.}

\section{$\CP^1\times \CP^1$ model and its flips via branes}
\label{app:branes}

\subsection{Unflipped case}
\label{app:GP}

The original model considered by Bianchi--Sagnotti and Gimon--Polchinski was given in terms of Type I on $T^4/\bZ_2$.
It has $\Om9$ with 16 D9s in the bulk, with $16$ $\Om5$ at the $\bZ_2$ fixed points, and 16 D5s.\footnote{Here the number of D-branes is counted in terms of Type IIB or Type IIA RR-charge, in a way invariant under T-duality.
 In simple orientifold models, this number equals the number of mobile D-branes or the rank of the gauge groups, but in  more complicated models such as those discussed in this note, they can be different. }
 
Let us determine its massless spectrum.
From the bulk closed string modes, we have one supergravity multiplet, one tensor, and four neutral hypermultiplets.
From the $\bZ_2$ twisted closed strings, one neutral hypermultiplet arises from each $\bZ_2$ singularity.

As for the open strings,
$\Om9$ wants to make the gauge algebra on D9 orthogonal.
Therefore the bulk of the 9-brane has $\Spin(32)/\bZ_2$ as the gauge group.
But $\Om5$ wants to make the gauge algebra on D9 symplectic.
This gives a localized $\Spin(32)/\bZ_2$ holonomy around the intersection point, and 
the massless gauge algebra on D9 that can remain is $\mathfrak{u}_{16}$, the intersection of $\sp_{16}$ and $\so_{32}$.
This will keep charged hypermultiplets in $2\cdot \wedge^2\rep{16}$.
One can do the same analysis on the D5-branes, and get the same answer, when all the D5s are on a single $\Om5$.
 Finally, the 5-9 strings give hypermultiplets in $\rep{16}\times \rep{16}$.
The spectrum is then \begin{itemize}
\item gauge algebras $\mathfrak{u}_{16}\times \mathfrak{u}_{16}'$,
\item charged hypermultiplets in $2\cdot \wedge^2\rep{16} \oplus \rep{16} \times \rep{16}' \oplus2\cdot \wedge^2\rep{16}'$,
\item one supergravity multiplet, one tensor multiplet, and 20 neutral hypermultiplets.
\end{itemize}
Anomalies correctly cancel, and $\mathfrak{u}_1$ parts are eaten \cite{Berkooz:1996iz}.

We can take T-duality along two directions and bring this model to
the type IIB $T^4/\bZ_2$ orientifolds with seven-branes, with the structure below: \begin{equation}
\inc{4m}\times \inc{4m}
\end{equation}
where the first $T^2$ has the coordinate $u$, the second has the coordinate $v$, with the orientifolding action sending $u\to -u$ and $v\to -v$ individually.
The spectrum above are when all $16$ D7s along $v$ are on $u=0$ and when all $16$ D7s along $u$ are on $v=0$.

\subsection{Singly-flipped case}
\label{app:one-flipped}
For this, we consider the setup \begin{equation}
\inc{4m}\times\inc{3m}
\end{equation}
with 16 D7s perpendicular to the first $T^2$ and 8 D7s perpendicular to the second $T^2$.

To deduce the open string spectrum on the 8 D7s perpendicular to the second $T^2$, we just T-dualize one direction of the first $T^2$ and apply the rules of \cite{Hanany:1999sj}.
When the 8D7s are on a generic point, one gets $\sp_4$ with a full antisymmetric tensor (both the traceless part and a singlet), and with 16 fundamentals.
If they are on $\Om7$, it gets enhanced to $\mathfrak{u}_8$ with $2\cdot\wedge^2\rep{8}$,
and if they are on $\Op7$, it gets enhanced to $\sp_4\times \sp_4$ with a bifundamental.

For the 16 D7s on the first $T^2$, we can take the T-dual of the second $T^2$: \begin{equation}
\inc{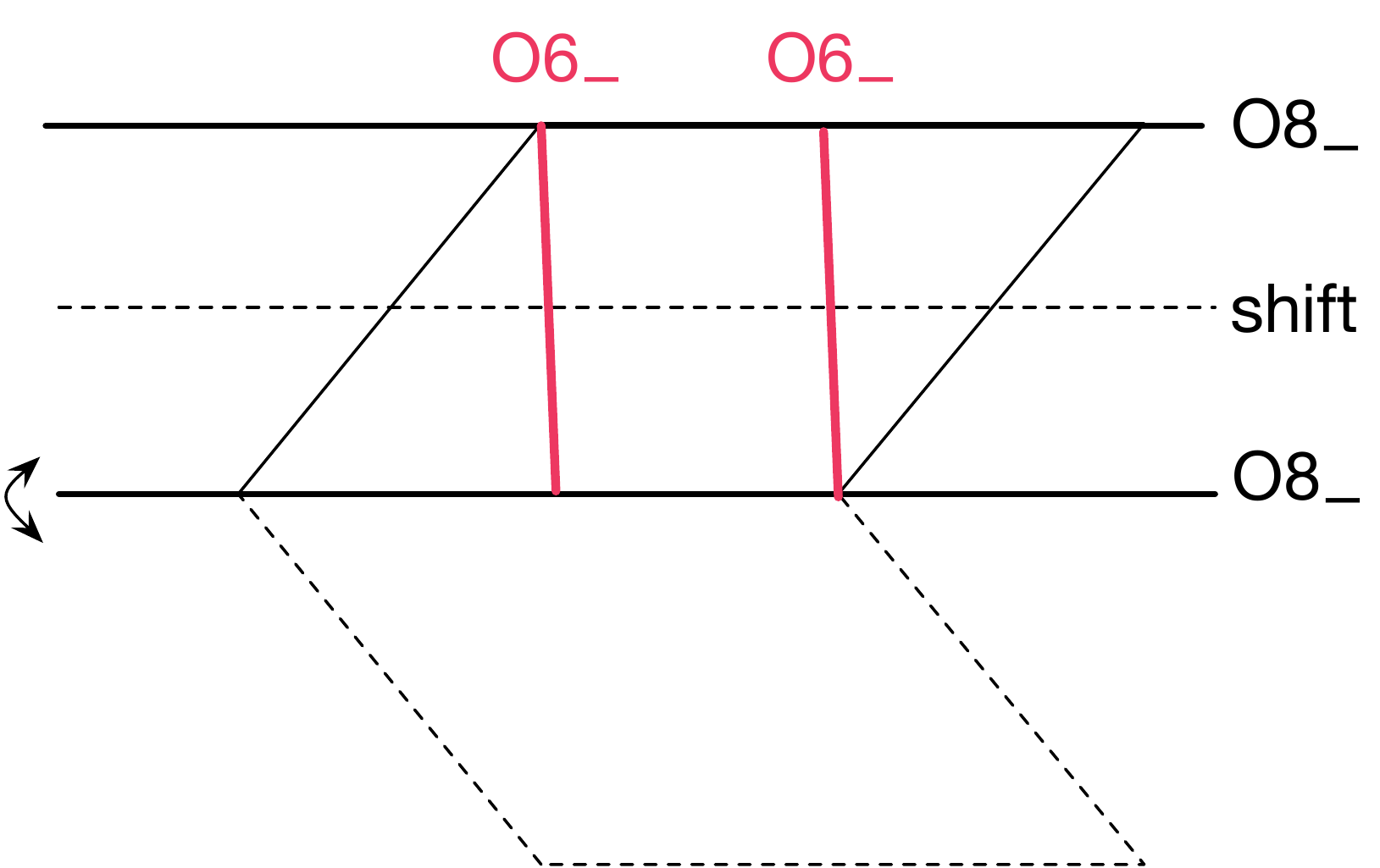}.\label{EEE}
\end{equation}
This T-duality was derived from the worldsheet point of view in \cite{Pradisi:2003ct}.

When 16 D7s are on a single generic point on $T^2$, the T-dual is just 8 D6s suspended between two D8s that are in fact \emph{the same} due to the funny geometry.
 This is $\sp_4$ with one $\rep{asym}$ and  16 flavors.
When they are all on an $\Om7$, this gets enhanced to $\mathfrak{u}_8$ with $2\cdot\wedge^2\rep{8}$.
Although we started from 32 Chan-Paton indices but we got just $\mathfrak{u}_8$.
We give two other explanations to this somewhat unexpected fact: \begin{itemize}
\item If we T-dualize the second torus twice, this describes instantons (or 5-branes) in the $\Spin(32)/\bZ_2$ gauge fields on $T^2$ without vector structure.
As discussed in \cite{Witten:1997bs}, a minimal flat $\Spin(32)/\bZ_2$ configuration without vector structure is in $\SU(2)$ embedded in $\so_{32}$ as $\sp_1\times \sp_{8}$.
Then the instanton needs to be embedded into this $\sp_{8}$; a single such instanton counts as two instantons in the original $\Spin(32)/\bZ_2$.
In other words, two small instantons of $\Spin(32)/\bZ_2$ need to move together.
\item In the original 7-brane description, there are four intersections with transverse O7s; one is with $\Op7$ and three are with $\Om7$.
The former has a monodromy that squares to $-1$ and the latter has a monodromy that squares to $1$.
But one cannot embed them into $\mathrm{O}(1)$: they are not consistent, since the four monodromies need to multiply to one.
To compensate this, one needs an additional flat $\mathrm{SO}(3)$ background on the 7-brane.
\end{itemize}

Summarizing, when 16 D7s perpendicular to the first $T^2$ are on a single $\Om7$ and 8 D7s perpendicular to the second $T^2$ are on a single $\Op7$, the spectrum is \begin{itemize}
\item gauge algebras $\mathfrak{u}_8\times \prod_{i=1,2}(\sp_4)_{i}$,
\item charged hypermultiplets in $2\cdot\wedge^2\rep{8} \oplus (\bigoplus_{i=1,2} \rep{8}\otimes \rep{8}_i )\oplus \rep{8}_1\otimes \rep{8_2}$,
\item one supergravity multiplet, $5$ tensor multiplets, and  $16$ neutral hypermultiplets.
\end{itemize}

\subsection{Doubly-flipped case}
\label{app:two-flipped}

Let us finally consider 
\begin{equation}
\inc{3m}\times \inc{3m}
\end{equation}
with 8 D7-branes on each $T^2$.
Using the analysis as in case II, we see that when 8 D7s are on a single $\Op7$, the gauge algebra is $\sp_2\times \sp_2$.
Considering D7s on both $T^2$, we have $(\sp_2)^4$ in total.
The matter spectrum can be worked out as before:
\begin{itemize}
\item gauge algebras $\prod_{i=1}^4(\sp_2)_i$,
\item charged hypermultiplets in $\bigoplus_{i<j} \rep{4}_i\otimes\rep{4}_j$,
\item one supergravity multiplet, 7 tensor multiplets, and  $14$ neutral hypermultiplets.
\end{itemize}
The anomaly cancels; although there are 8 additional tensors, they do not participate in the gauge anomaly cancellation.
This is as it should be, since they are localized on the intersections of $\Om7$ and $\Op7$, and bifundamentals are supported away from them.

\bibliography{at-yuji}
\bibliographystyle{at}

\end{document}